\documentclass[revtex4,fleqn,usenatbib,useAMS]{emulateapj}

\usepackage{graphicx}	
\usepackage{amsmath}	
\usepackage{amssymb}	
\usepackage{bm}		

\usepackage[T1]{fontenc}

\usepackage{newtxtext,newtxmath}

\shorttitle{The KMOS Cluster Survey (KCS) II}
\shortauthors{Chan et al.}

\begin{document}


\title{The KMOS Cluster Survey (KCS) II -- The Effect of Environment on the Structural Properties of Massive Cluster Galaxies at Redshift $1.39 < z <1.61$\protect\footnote{\MakeLowercase{\MakeUppercase{B}ased on observations obtained at the \MakeUppercase{V}ery \MakeUppercase{L}arge \MakeUppercase{T}elescope  (\MakeUppercase{VLT}) of the \MakeUppercase{E}uropean \MakeUppercase{S}outhern \MakeUppercase{O}bservatory  (\MakeUppercase{ESO}) (program \MakeUppercase{ID}s:  \MakeUppercase{092.A-0210;  093.A-0051;  094.A-0578;  095.A-0137(A); 096.A-0189(A); 097.A-0332(A)}). \MakeUppercase{T}his work is based on observations made with the \MakeUppercase{NASA/ESA HST}, which is operated by the Association of Universities for Research in Astronomy, Inc., under \MakeUppercase{NASA} contract \MakeUppercase{NAS} 5-26555. \MakeUppercase{T}hese observations are associated with program \MakeUppercase{GO} 13687 as well as with the \MakeUppercase{CANDELS} Multi-Cycle Treasury Program and the \MakeUppercase{3D-HST} Treasury Program (\MakeUppercase{GO} 12177 and 12328).}}}

\author{Jeffrey C.C. Chan$^{1,2,3}$, Alessandra Beifiori$^{1,2}$, Roberto P. Saglia$^{2,1}$, J. Trevor Mendel$^{2,1}$,  John P. Stott$^{4,6}$, Ralf Bender$^{1,2}$, Audrey Galametz$^{2,1}$, David J. Wilman$^{1,2}$, Michele Cappellari$^{4}$,  Roger L. Davies$^{4}$, Ryan C. W. Houghton$^{4}$, Laura J. Prichard$^{4}$, Ian J. Lewis$^{4}$, Ray Sharples$^{5}$ and Michael Wegner$^{1}$}

\affil{$^{1}$Universit\"ats-Sternwarte, Ludwig-Maximilians-Universit\"at, Scheinerstrasse 1, D-81679 M\"unchen, Germany\\
$^{2}$Max-Planck-Institut f\"ur extraterrestrische Physik (MPE), Giessenbachstrasse 1, D-85748 Garching, Germany\\
$^{3}$Department of Physics and Astronomy, University of California, Riverside, CA 92521, USA\\
$^{4}$Sub-department of Astrophysics, Department of Physics, University of Oxford, Denys Wilkinson Building, Keble Road, Oxford OX1 3RH, UK\\
$^{5}$Centre for Advanced Instrumentation, Department of Physics, Durham University, South Road, Durham DH1 3LE, UK\\
$^{6}$Department of Physics, Lancaster University, Lancaster LA1 4YB, UK}



\begin{abstract} 
We present results on the structural properties of massive passive galaxies in three clusters at $1.39<z<1.61$ from the KMOS Cluster Survey.  We measure light-weighted and mass-weighted sizes from optical and near-infrared \textit{Hubble Space Telescope} imaging and spatially resolved stellar mass maps.  The rest-frame $R$-band sizes of these galaxies are a factor of $\sim2-3$ smaller than their local counterparts.  The slopes of the relation between the stellar mass and the light-weighted size are consistent with recent studies in clusters and the field.  
Their mass-weighted sizes are smaller than the rest frame $R$-band sizes, with an average mass-weighted to light-weighted size ratio that varies between $\sim0.45$ and $0.8$ among the clusters.  We find that the median light-weighted size of the passive galaxies in the two more evolved clusters is $\sim24\%$ larger than for field galaxies, independent of the use of circularized effective radii or semi-major axes. These two clusters also show a smaller size ratio than the less evolved cluster, which we investigate using color gradients to probe the underlying $M_{*}/L_{H_{160}}$ gradients.
The median color gradients are $\nabla{z-H} \sim-0.4$ mag dex$^{-1}$, twice the local value.  Using stellar populations models, these gradients are best reproduced by a combination of age and metallicity gradients.
Our results favor the minor merger scenario as the dominant process responsible for the observed galaxy properties and the environmental differences at this redshift.  The environmental differences support that clusters experience accelerated structural evolution compared to the field, likely via an epoch of enhanced minor merger activity during cluster assembly.
\end{abstract}





\section{Introduction}
\label{sec:Introduction}
Understanding the formation and evolution of passive galaxies is one of the long-standing problems in astronomy.  In the local Universe, passive galaxies are seen to have regular early-type morphology and are mainly composed of old stellar populations \citep[e.g.][]{Trageretal2000, Thomasetal2005, Gallazzietal2006, Thomasetal2010, McDermidetal2015}. They dominate the massive end of the galaxy population \citep[e.g.][]{Kauffmannetal2003a, Renzinietal2006} and reside at a well-defined region that is separated from star-forming galaxies on the color-magnitude (or stellar mass) space, known as the red sequence \citep[e.g.][]{Boweretal1992, Blantonetal2003, Baldryetal2004, Baldryetal2006}.  

Recent direct look-back studies have revealed that passive galaxies at high redshift show various differences compared to their local counterparts.  Besides having a younger stellar population \citep[e.g.][]{Whitakeretal2013, Bellietal2015, Mendeletal2015, Beifiorietal2017} and higher velocity dispersion \citep[e.g.][]{Cappellarietal2009, vandeSandeetal2013, Bellietal2014a, Bellietal2014b, Beifiorietal2017}, passive galaxies at high redshift are much more compact than those we find in the local Universe.  In the last decade a number of authors first discovered that the sizes of a population of passive galaxies at $z\sim1.5$ in rest-frame UV \citep{Daddietal2005} and rest-frame optical \citep{Trujilloetal2006a} are significantly smaller than their local counterparts.  Initially there were concerns regarding possible biases on the stellar mass and size estimates, owing to uncertainties in the stellar population synthesis models and the imaging depth.  Subsequent dynamical mass measurements from spectroscopic studies have demonstrated that the stellar mass estimates are reliable \citep[e.g.][]{Cappellarietal2009, Bezansonetal2013, vandeSandeetal2013, Bellietal2014a, Shettyetal2014}.  Recent size measurements with the near-IR sensitive \textit{HST} infrared Wide Field Camera 3 (\textit{HST}/WFC3) have confirmed that the measured small sizes are genuine \citep[e.g.][]{Szomoruetal2010,vanderWeletal2014}.  With larger samples it is now established that the massive passive population ($M_{*} \gtrsim 10^{11} M_{\odot}$) have grown by on average a factor of $\sim2$ in size since $z\sim1$ \citep[e.g.][]{Longhettietal2007, Cimattietal2008, vanderWeletal2008, Beifiorietal2014, Chanetal2016} and a factor of $\sim3-4$ since $z \sim2$, from having an effective radius of only $\lesssim1$ kpc \citep[e.g.][]{Trujilloetal2006b, Trujilloetal2007, Toftetal2007, Zirmetal2007, Buitragoetal2008, vanDokkumetal2008, Szomoruetal2012, vanderWeletal2014}.

Despite this discovery, an important component that is still unsettled is the effect of environment on the sizes of these galaxies. 
In the local universe, several studies have found that there is no obvious environmental dependence on passive galaxy sizes \citep[e.g.][]{Maltbyetal2010, Cappellari2013, Huertascompanyetal2013}, although there are reports of a population of dense and compact galaxies in local clusters and the field \citep[e.g.][]{Valentinuzzietal2010b, Poggiantietal2013}. On the other hand, studies at high redshift show contrasting results.  Some works have found that the sizes of passive galaxies are larger in clusters compared to the field \citep[e.g.][]{Cooperetal2012, Zirmetal2012, Papovichetal2012, Strazzulloetal2013, Jorgensenetal2013, Delayeetal2014}, although the magnitude of the effect is not yet clear and might depend on the cluster mass or richness \citep[e.g.][]{Jorgensenetal2014}.  Nevertheless, there were also reports showing cluster passive galaxies have no significant size difference from those in the field \citep[e.g.][]{Maltbyetal2010, Retturaetal2010, Newmanetal2014}, or are even smaller \citep[e.g.][]{Raichooretal2012}.   The apparent discrepancies may in part due to different definitions of environment, the use of different cluster/field samples for comparison and perhaps even related to the mass range of the sample \citep[see][for the size difference in the Coma cluster for different mass ranges]{Cappellari2013}.  \citet{Delayeetal2014} analysed an ensemble of $\sim400$ early-type galaxies (ETGs) in nine clusters at z $\sim$ 1 and showed that the size distributions in clusters and the field peak at the same position, but the size distribution in clusters present a tail of larger galaxies dominated by those with $M_{*} < 10^{11} M_{\odot}$.  
Conversely, \citet{Lanietal2013} reported a significant size increase in cluster with respect to the field at $1<z<2$ for galaxies more massive than $M_{*} > 2 \times 10^{11} M_{\odot}$.

Previous works have also used various definitions of galaxy sizes, the two most used ones being the elliptical semi-major axis $a_{e}$ from S\'ersic profile fitting \citep{Sersic1968}, and the circularized effective radius ($R_{e\rm{-circ}} =  a_{e} \times \sqrt{q}$, where $q$ is the projected axis ratio). The circularized effective radius can better characterise the sizes of early-type galaxies that might exhibit triaxiality \citep[as evident from their distribution of ellipticity or their internal kinematics, e.g.][]{Franxetal1991,Vincentetal2005}, hence is commonly used in works on local early-type galaxies. On the other hand, the semi-major axis is more appropriate for `disk-like' galaxies.

Quantifying the effect of environment can allow us to disentangle the physical processes responsible for the observed size evolution.  Currently the two most plausible explanations are mass-loss driven adiabatic expansion (``puffing-up'') \citep[e.g.][]{Fanetal2008, Fanetal2010, RagoneFigueroaetal2011} and dry minor merger scenarios \citep[e.g.][]{Bezansonetal2009, Naabetal2009, Trujilloetal2011}.  The two scenarios are both able to reproduce the large growth in size, while retaining a minimal increase in stellar mass or the star formation rate, although none can explain the observations fully.  
The ``puffing up'' scenario relies on a rapid mass loss from the centre driven by AGN or supernovae feedback, which then results in a change in the gravitational potential of the galaxy and leads to an expansion in size.  Nevertheless, this scenario fails to explain the observed scatter of the mass-size relation of passive galaxies at high redshift \citep[e.g.][]{Trujilloetal2011} and is difficult to reconcile with the fact that these galaxies may have steep age gradients \citep[e.g.][]{DeProprisetal2015, DeProprisetal2016, Chanetal2016}.  Dry minor mergers, on the other hand, have been shown to be able to explain most of the abovementioned observed properties \citep[see e.g.][]{Trujilloetal2011, Shankaretal2013}.  The progressive mass assembly at outer radii as seen from various studies of the stellar mass surface density profiles further favours this scenario \citep{Bezansonetal2009, vanDokkumetal2010, Pateletal2013}.  Nevertheless, recent works suggest minor mergers can account for the size evolution in field galaxies only up to $z\lesssim1$ \citep[e.g.][]{Kavirajetal2009, LopezSanjuanetal2012, Newmanetal2012, Bellietal2014a}, it is still unclear whether the rate of minor mergers is enough to explain the observed growth at higher redshift. 
A key prediction of the minor merger scenario is that the rate of mergers is environmentally dependent;  they are expected to be more common in higher density environments such as galaxy groups and during cluster formation \citep[e.g.][]{Wetzeletal2008, Linetal2010, Jianetal2012, Wilmanetal2013}.  Although a direct measurement of the merger rates in high-redshift clusters is challenging \citep[see, e.g.][]{Lotzetal2013}, one would expect an environmental dependence on passive galaxy sizes (i.e. larger sizes with higher merger rates), if this is the dominant physical process for the size evolution at high redshift.

Hence in this paper we investigate the structural properties of a sample of passive galaxies in dense environments, as part of the KMOS Cluster Survey (KCS), a guaranteed time observation (GTO) program targeting passive galaxies with the new generation infrared integral field spectrograph, the $K$-band Multi-Object Spectrograph (KMOS), at the Very Large Telescope (VLT) \citep[][Davies, Bender et al., in prep]{Daviesetal2015}. The main goal of KCS is to study the evolution of kinematics and stellar populations in dense environments at high redshift, with a sample of four main overdensities at $1.39 < z < 1.8$ and one lower-priority overdensity at $z = 1.04$.  In paper I of KCS \citep{Beifiorietal2017}, we present the analysis of the fundamental plane for three overdensities at $1.39 < z < 1.61$ in the KCS sample.  The structural and kinematic properties of the galaxies in the overdensity at $z=1.8$ are presented in paper III of KCS \citep{Prichardetal2017}.  The study of the stellar populations of the passive galaxies in the three overdensities in \citet{Beifiorietal2017} will be presented in a forthcoming paper (Houghton et al., in prep).

Here we focus on the structural properties of three overdensities at redshift $1.39 < z < 1.61$, XMMU J2235-2557 at $z=1.39$ \citep{Mullisetal2005}, XMMXCS J2215.9-1738 at $z=1.46$ \citep{Stanfordetal2006} and Cl 0332-2742 at $z=1.61$ \citep{Castellanoetal2007}.  As a pilot study we have examined the structural properties of the overdensity XMMU~J2235-2557 and throughly tested our methodology in \citet{Chanetal2016}.  The interested reader can refer to the paper for detailed explanations of the procedures.

This paper is organised as follows.  A summary of the KCS clusters and data used in this study are described in Section~\ref{sec:Data}.  The selection of passive galaxies is described in Section~\ref{sec:The Red Sequence Sample}.  In Section~\ref{sec:Methods and Analysis} we describe the procedure to derive resolved stellar mass surface density maps, as well as the measurements of the light-weighted and mass-weighted structural parameters.  We present the main results in Section~\ref{sec:Results}. The results are then compared with the field sample, and discussed in Section~\ref{sec:Discussion}.  We conclude our findings in Section~\ref{sec:Conclusion}.  All the measurements are provided in the tables in Appendix~\ref{app:Photometry, Structural parameters and color gradients of the red sequence galaxies in the KCS clusters}.

Throughout the paper, we assume the standard flat cosmology with $H_{0} = 70$~km~s$^{-1}$~Mpc$^{-1}$, $\Omega_{\Lambda} = 0.7 $ and $\Omega_{m} = 0.3$.  With this cosmological model, 1 arcsec corresponds to 8.43 kpc at $z=1.39$, 8.45 kpc at $z=1.45$, and 8.47 kpc at $z=1.61$.  Magnitudes quoted are in the AB system \citep{OkeGunn1983}.  The stellar masses in this paper are computed with a \citet{Chabrier2003} initial mass function (IMF).  


\section{Sample and Data}
\label{sec:Data}
\subsection{The KCS Sample}
\label{sec:The KCS Sample}
The clusters of KCS are selected to have a significant amount of archival data, spanning from multi-band HST imaging to deep ground-based imaging.  They are also selected to have a large number of spectroscopically confirmed galaxy members to enhance the observing efficiency of the KMOS observations.  Here we briefly summarise the properties of the three clusters used in this study.  More details can be found in \citet{Beifiorietal2017}.
 
The cluster \textbf{XMMUJ2235-2257} at $z=1.39$ was discovered in an \textit{XMM-Newton} observation of NGC 7314 \citep{Mullisetal2005}.  The mass of this cluster is estimated to be $M_{200} \sim 7.7 \times 10^{14}$ $M_{\odot}$ \citep[e.g.][]{Stottetal2010, Jeeetal2011}, making it one of the most massive clusters at $z > 1$.  Several works have studied the stellar populations in the massive cluster galaxies using optical colors and agreed on an early formation epoch \citep[e.g.][]{Lidmanetal2008, Rosatietal2009, Strazzulloetal2010}.  The presence of the high mass end in the stellar mass function also indicates that this cluster is already at an evolved mass assembly stage \citep{Strazzulloetal2010}.  \citet{Baueretal2011} found a correlation between the star formation rate (SFR) and projected distance from the cluster centre, suggesting the star formation is shut off within $r <200$~kpc.  All massive galaxies out to $r \sim 1.5$ Mpc have low SFRs, and those in the centre have lower specific SFRs than the rest of the population with the same mass \citep{Grutzbauchetal2012}.  For the structural properties, this cluster has been investigated by \citet{Strazzulloetal2010} and was also included in the cluster samples of \citet{Delayeetal2014}, \citet{DeProprisetal2015} and \citet{Cioccaetal2017}.

The cluster \textbf{XMMXCS~J2215-1738} at $z=1.46$ was discovered in the \textit{XMM} Cluster Survey \citep{Stanfordetal2006}.  Its mass is estimated to be $M_{200} \sim 2.1 \times 10^{14}$ $M_{\odot}$ \citep{Stottetal2010, Jeeetal2011}.  The bimodal velocity distribution of the confirmed cluster members \citep{Hiltonetal2007, Hiltonetal2009, Hiltonetal2010} and the fact that this cluster is under-luminous in X-ray suggest that it is likely not yet virialized \citep{Hiltonetal2007, Maetal2015}.  The cluster shows, in general, a lack of bright galaxies.  Contrary to XMMU~J2235-2557 and most local clusters, the brightest cluster galaxy (BCG) in XMMXCS J2215-1738 is not distinctly bright compared to the other galaxies in the cluster.  This cluster also shows substantial star formation activities, even at the core.  Mid-IR imaging from \textit{Spitzer} revealed eight 24 $\mu$m sources in the core, with three of them within the cluster red sequence.  Most of these objects are dust-obscured star forming galaxies \citep{Hiltonetal2010}.  \citet{Hayashietal2010} identified 44 [O II] emitters with high [O II] SFRs and some of these emitters host AGNs \citep{Hayashietal2011}.  They argued that the cluster has experienced high star-forming activity at rates comparable to the field at $z\sim1.4$.  A search of dust-obscured ultra luminous infrared galaxies (ULIGS) with SCUBA-2 provides further evidence of obscured star formation in the core \citep{Maetal2015}. Recent high-resolution ALMA observations of the cluster core have confirmed the overdensity of dust-obscured star forming galaxies and their spatial distribution imply that these galaxies experienced environmental effects during their infall to the cluster \citep{Stachetal2017, Hayashietal2017}. The existence of substantial star formation together with the hints that this cluster is not virialized suggests that this cluster is dynamically disturbed \citep{Hiltonetal2010} and is not as mature as XMMU~J2235-2257.  This cluster was also included in the cluster sample of \citet{Delayeetal2014}.

The  (proto)cluster \textbf{Cl~0332-2742} at $z=1.61$ is one of the few high redshift clusters detected by clustering in redshift space \citep{Castellanoetal2007}, as opposed to extended X-ray emission (e.g. XMMU~J2235-2557 and XMMXCS~J2215-1738) or red sequence. The structure comprises at least two smaller groups as the cluster members show a bimodal distribution in redshift space, albeit with no clear evidence of spatial separation \citep{Kurketal2009}.  Extended X-ray emission is only detected at one of the substructure that is off-centered from the \citet{Kurketal2009} high-density peak \citep{Tanakaetal2013} and coincides with a concentration of red galaxies. It was confirmed to be a gravitationally bound X-ray group \citep{Tanakaetal2013}.  This suggests that Cl~0332-2742 is a (proto)cluster still in assembly and comprises interacting group structures.  Despite this, Cl~0332-2742 has a well-defined red sequence \citep{Kurketal2009}.  The stacked spectrum of seven red galaxies shows relatively young age ($\sim1$ Gyr), very low specific SFRs and dust extinction \citep{Cimattietal2008}.  Similarly, the members in the \citet{Tanakaetal2013} group have low SFRs, but also have a high AGN fraction: three out of eight of the group members host AGNs. 

In summary, the three overdensities used in this study span a range of environments \citep[see Figure 1 in][]{Beifiorietal2017} and represent clusters in different assembly stage:  from the mature massive cluster XMMU~J2235-2557, to a not yet virialized young cluster XMMXCS~J2215-1738 and to the protocluster Cl~0332-2742.  For simplicity, we will refer to the three overdensities as clusters below.

\subsection{Summary of the \textit{HST} data sets}
We make use of both new and archival deep optical and near-infrared \textit{HST} imaging of the clusters, obtained with \textit{HST}/ACS WFC and \textit{HST}/WFC3 IR.  Table \ref{tab_data_summary} summarises the used \textit{HST} data of the three clusters in various bands.

XMMU~J2235-2557 was observed in June 2005 (as a part of program GTO-10698), July 2006 (GO-10496) and April 2010 (GO/DD-12051). The \textit{HST}/ACS data consist of F775W and F850LP bands (hereafter $i_{775}$ and $z_{850}$) and the WFC3 data comprise four IR bands, F105W, F110W, F125W and F160W (hereafter $Y_{105}$, $YJ_{110}$, $J_{125}$ and $H_{160}$).  The $YJ_{110}$ data are not used due to their short exposure time.  The WFC3 data have a smaller field of view than the ACS data, $145'' \times 126''$, corresponding to a region of up to $\sim550$ kpc from the cluster centre.

The ACS data of XMMXCS~J2215-1738 consist of $i_{775}$ and $z_{850}$ bands, observed during April to August 2006 (GO-10496).  The $i_{775}$ data are not used due to their short exposure time.  The WFC3 data of this cluster come from our cycle 22 observation (GO-13687) observed in June 2015, which is designed for this study and comprises three bands, F125W, F140W, F160W (hereafter $J_{125}$, $JH_{140}$ and $H_{160}$).

Cl~0332-2742 is located at the WFC3 Early Release Science (ERS) field within the GOODS-S field \citep{Windhorstetal2011}, hence \textit{HST} data is publicly available from the CANDELS \citep{Groginetal2011, Koekemoeretal2011} and 3D-HST programs \citep{Brammeretal2012, Skeltonetal2014}. 
We make use of the \textit{HST}/ACS and WFC3 mosaics reduced by the 3D-HST team.  Of all the available mosaics we mainly use the $z_{850}$ and $H_{160}$ band, and the ACS F814W ($i_{814}$) and WFC3 $J_{125}$ photometry from the public released v1.0 3D-HST photometry catalogue \citep{Koekemoeretal2011, Skeltonetal2014}.

\begin{table}
  \caption{Summary of the \textit{HST} imaging of the KCS clusters}
  \centering
  \label{tab_data_summary}
  \begin{tabular}{@{}ccccc@{}}
  \hline
  \hline
 Cluster & Name & Filter &  Rest-frame   & Exposure \\
                &             &           & pivot $\lambda$ (\AA) & time (s)\\
  \hline
  XMMU~J2235\textsuperscript{x}  & $i_{775}$\textsuperscript{+}  & ACS F775W &  3215.2  &  8150   \\
                                        & $z_{850}$\textsuperscript{ } & ACS F850LP  & 3776.1  &  14400  \\   
                                        & $Y_{105}$\textsuperscript{+} & WFC3 F105W  & 4409.5  &  1212  \\   
                                        & $ J_{125}$\textsuperscript{ } & WFC3 F125W & 5217.7  &  1212   \\   
                                        & $H_{160}$\textsuperscript{ } & WFC3 F160W & 6422.5  &  1212  \\   
  XMMXCS~J2215 & $z_{850}$\textsuperscript{ } & ACS F850LP & 3673.2  &  16935  \\   
                                        & $J_{125}$\textsuperscript{ } & WFC3 F125W & 5075.6  &  2662  \\   
                                        & $JH_{140}$\textsuperscript{+} & WFC3 F140W & 5659.8  &  1212  \\   
                                        & $H_{160}$\textsuperscript{ } & WFC3 F160W & 6247.6 &  1312  \\   
  Cl~0332 & $i_{814}$\textsuperscript{ } & ACS F814W & 3108.1  &   - \textsuperscript{o}   \\   
                                        & $z_{850}$\textsuperscript{ } & ACS F850LP & 3462.1  &  - \textsuperscript{o} \\   
                                        & $J_{125}$\textsuperscript{ } & WFC3 F125W & 4783.9  &  1430\textsuperscript{*}  \\   
                                        & $H_{160}$\textsuperscript{ } & WFC3 F160W & 5888.5  &  1518\textsuperscript{*}  \\   
  \hline
  \multicolumn{5}{p{.45\textwidth}}{\textsuperscript{x} The HST imaging data of XMMU~J2235-2557 is also used and described in \citet{Chanetal2016}.} \\
  \multicolumn{5}{p{.45\textwidth}}{\textsuperscript{*} Average exposure time in the section of the GOODS-S field where Cl~0332-2742 resides, derived using the exposure maps in each band.}\\
    \multicolumn{5}{p{.45\textwidth}}{\textsuperscript{+} These filter bands are not used in the analysis in this study, but are included in our photometry catalogue and used in other KCS works.}\\
        \multicolumn{5}{p{.45\textwidth}}{\textsuperscript{o} The exposure time maps of these filter bands are not available from the 3D-HST public release. The interested reader can refer to \citet{Skeltonetal2014} for their depths.}
\end{tabular}
\end{table}

\subsection{\textit{HST} data reduction}
\label{subsec:HST data reduction}
The \textit{HST} reduction of the cluster XMMU~J2235-2557 is described in detail in \citet{Chanetal2016}.  We followed the same procedure for XMMXCS~J2215-1738.  Here we summarise the main steps.  Both clusters are reduced and combined using Astrodrizzle and DrizzlePac (version 1.1.8), an upgraded version of the MultiDrizzle pipeline in the \texttt{PyRAF} interface \citep{Gonzagaetal2012}.  We start with calibrated frames (\texttt{\_flt.fits}) from the Mikulski Archive for Space Telescopes (MAST) archive.  All the \texttt{flt} files are first examined to check the quality of the cosmic rays and bad pixels identification by \texttt{calacs} (for ACS data) and \texttt{calwfc3} (for WFC3 data) pipeline.  Occasionally there can be hot stripes that span across the FOV (e.g. satellite trails) which are not fully flagged.  We mask these regions generously in the data quality array of the \texttt{flt} files.  For exposures taken in multiple visits, the relative WCS offsets between exposures are corrected using the \texttt{tweakreg} task in DrizzlePac before drizzling.

For XMMU~J2235-2557 the ACS and WFC3 images have been drizzled to pixel scales of 0.05 and 0.09 arcsec pixel$^{-1}$ respectively \citep[see][for details]{Chanetal2016}.  For the new Cycle 22 data for XMMXCS~J2215-1738 we adopt a pixel scale of 0.03 and 0.0642 arcsec pixel$^{-1}$ for the ACS and WFC3 images to better match the 3D-HST mosaics (0.03 and 0.06 arcsec pixel$^{-1}$).  Note that the choice of pixel scale does not affect the result in our case, as we have tested extensively with the data of XMMXCS~J2215-1738.  For the drizzling, we use a \texttt{pixfrac} of 0.8, a square kernel, and produce weight maps using both inverse variance map (\texttt{IVM}) and error map (\texttt{ERR}) weighting for different purposes.  The \texttt{IVM} weight maps, which contain all background noise sources except Poisson noise of the objects, are used for object detection, while the \texttt{ERR} weight maps are used for structural analysis as the Poisson noise of the objects is included.  The full-width-half-maximum (FWHM) of the PSF is $\sim$0.11 arcsec for the ACS $z_{850}$ data and $\sim$0.18 arcsec for the WFC3 $H_{160}$ data, as measured from characteristic PSFs of each band constructed by median-stacking bright unsaturated stars. For XMMU J2235-2557 and XMMXCS J2215-1738, 15 (12) stars are used in the stack for ACS $z_{850}$ and 4 (5) stars are used in the stack for the WFC3 $H_{160}$ bands, due to the smaller FOV of WFC3.  We follow \citet{Casertanoetal2000} and apply a scaling factor to the weight maps to account for correlated noises from the drizzle process.  Initial WCS calibrations of the drizzled ACS images are derived using GAIA in the Starlink library \citep{Berryetal2013} with Guide Star Catalog II (GSC-II) \citep{Laskeretal2008}.  For the WFC3 images, we derive their WCS by comparing the coordinates of unsaturated stars on the WFC3 images to the WCS calibrated ACS $z_{850}$ images.

PSF matching is crucial in photometry as well as our resolved stellar mass measurements, as the measured flux of the galaxy has to come from the same physical projected region.  We use the \texttt{psfmatch} task in IRAF to PSF-match the $z_{850}$ image to the resolution of the $H_{160}$ images of XMMU~J2235-2557 and XMMXCS~J2215-1738, which we used to derive photometry and resolved stellar mass measurements.  For Cl~0332-2742, the 3D-HST releases have already provided PSF-matched images in multiple bands that are matched to the $H_{160}$ band.  For XMMU~J2235-2557 and XMMXCS~J2215-1738, the ratios of the growth curves of the matched PSF fractional encircled energy deviate by $< 2.5\%$ from unity.  For Cl~0332-2742, the resultant growth curves after PSF matching are consistent within $1\%$ \citep[see][]{Skeltonetal2014}.

\subsection{Construction of photometric catalogues}
\label{sec:Construction of photometric catalogues}
Before deriving the catalogues, because of the relative small FOV of the ACS and WFC3 images, we first register the images of XMMU~J2235-2557 and XMMXCS~J2215-1738 to a larger image mosaic to improve their absolute WCS accuracies.  We utilise the $K_{s}$-band HAWK-I images for these two clusters.  For XMMU~J2235-2557, the images were taken as part of the first HAWK-I science verification run\footnote{Based on data products from observations made with ESO Telescopes at the La Silla Paranal Observatory under programme \MakeUppercase{ID 060.A-9284(H)}.}, in October 2007 \citep{Lidmanetal2008,Lidmanetal2013}, whereas for XMMXCS~J2215-1738 the images were obtained under ESO program ID 084.A-0214(A) in October 2009 (C. Lidman, private communication). These large-scale images cover a $\sim10' \times 10'$ region (for XMMXCS~J2215-1738) and a $\sim13' \times 13'$ region (for XMMU~J2235-2557), much larger than the ACS and WFC3 images.  For Cl~0332-2742 this step is not needed as the 3D-HST mosaics have already high absolute astrometric accuracy.

For each cluster we use the $H_{160}$ image, the reddest available band, as the detection image for SExtractor \citep{BertinArnouts1996} to construct photometric catalogues.  We derive multiband photometry using SExtractor in dual image mode with the $H_{160}$ image as the detection band.  {\tt MAG\_AUTO} are used for galaxy magnitudes and aperture magnitudes ($1''$ in diameter) are used for color measurements. Point sources ({\tt class\_star} $\geq 0.9$) are removed from the catalogues.  Galactic extinction is corrected using the dust map of \citet{Schlegeletal1998} and the recalibration $E(B-V)$ value from \citet{Schlaflyetal2011}.   Since the 3D-HST photometry catalogue \citep{Skeltonetal2014, Momchevaetal2016} does not provide aperture magnitude measurements for the ACS bands as well as using a $0.7''$ aperture for the WFC3 bands, we run SExtractor on Cl~0332-2742 to have consistent photometric measurements for all three KCS clusters.

We then cross-match our photometric catalogue with existing catalogues from the literature. For XMMU~J2235-2557, we cross-match our SExtractor catalogue to the catalogue from \citet{Grutzbauchetal2012} to identify spectroscopically confirmed cluster members from previous literature \citep[mostly from][]{Mullisetal2005, Lidmanetal2008, Rosatietal2009}. 12 (out of 14) spectroscopically confirmed cluster members are within the WFC3 FOV and are identified.  Similarly, we cross-match our catalogue of XMMXCS~J2215-1738 with the photometric and spectroscopic redshift catalogue from \citet{Hiltonetal2009,Hiltonetal2010}.  52 objects (out of 64) of the \citet{Hiltonetal2009} catalogue, and 26 (out of 44) spectroscopically confirmed objects from \citet{Hiltonetal2010} are detected.  Most of the undetected objects are out of the WFC3 FOV or are deblended to be multiple objects with our higher resolution \textit{HST}/WFC3 imaging.

For Cl~0332-2742, we cross-match our catalogue with the 3D-HST photometry catalogue (v4.1.5) of the GOODS-S field \citep{Skeltonetal2014, Momchevaetal2016}.  A redshift and spatial area selection was first applied to the 3D-HST catalogue to select objects that are plausibly cluster members for the KMOS observation \citep[see][for a description]{Beifiorietal2017}.  We select objects that are within a region of 10\arcmin~in diameter and are within $\pm 3000$ km s$^{-1}$ of the cluster redshift using the spectroscopic, grism and photometric redshift information from the 3D-HST catalogue \citep{Momchevaetal2016} as well as spectroscopic redshifts from our own KMOS observation.  This region encloses the \citet{Tanakaetal2013} group and the upper main parts of the \citet{Kurketal2009} structures, where the most massive galaxies reside.  We have estimated using monte-carlo methods that the redshift selection (for those within our magnitude limits, see Section~\ref{sec:The Red Sequence Sample} below) is $\sim85\%$ complete.  Our catalogue includes all the 37 $UVJ$ passive objects from this selection of the 3D-HST catalogue.  

\section{The Red Sequence Sample}
\label{sec:The Red Sequence Sample}
\begin{figure*}
  \centering
  \includegraphics[scale=0.60]{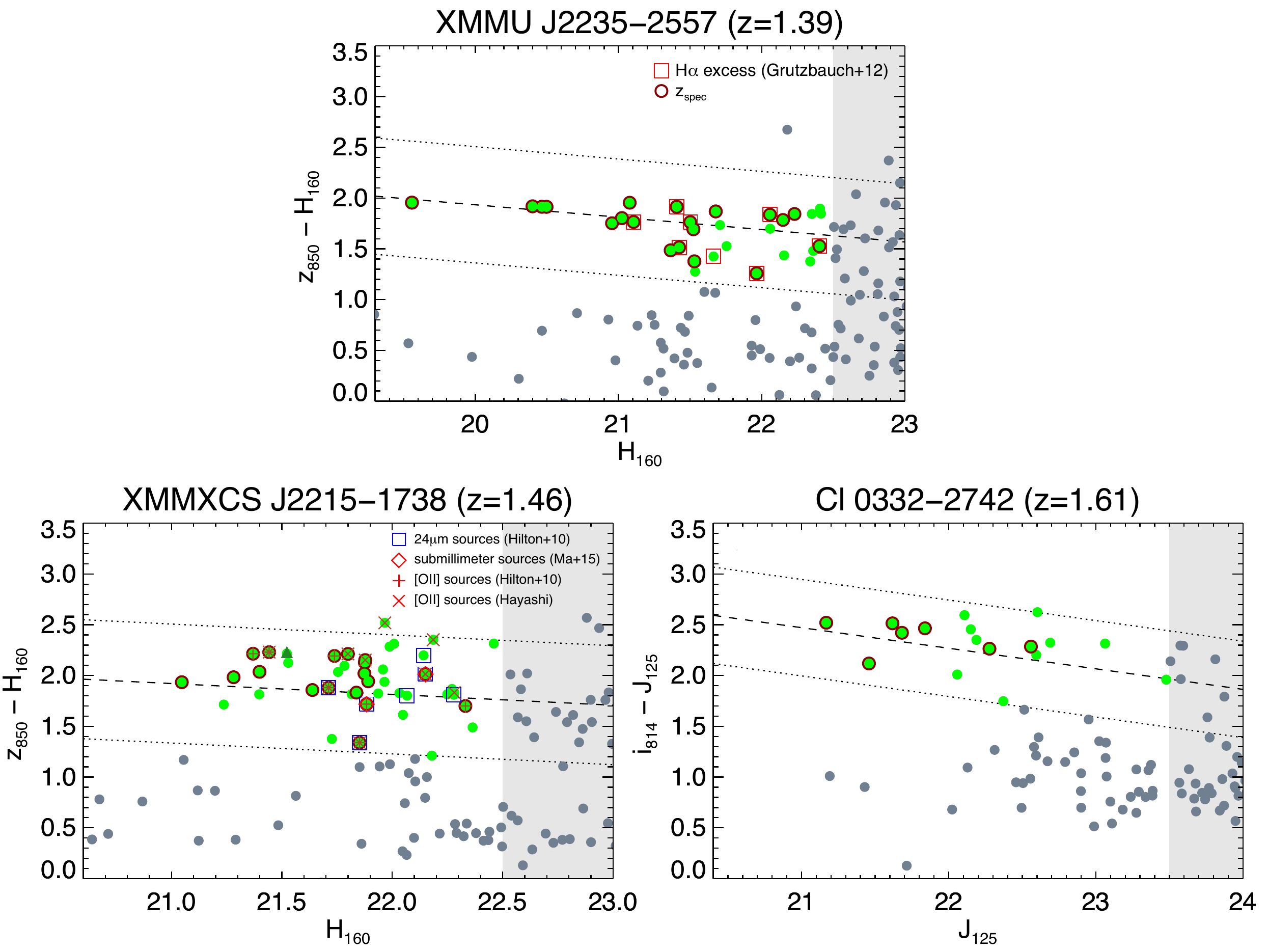}  
  \caption[Color-magnitude diagram of the three KCS clusters used in this study]{Color-magnitude diagram of the three KCS clusters used in this study.  For XMMU~J2235-2557 and XMMXCS~J2215-1738, the $H_{160}$ magnitudes are \textit{HST}/WFC3 \texttt{MAG\_AUTO} magnitudes while the $z_{850} - H_{160}$ colors are from $1''$ aperture magnitudes.  For Cl~0332-2742, the $J_{125}$ magnitudes and $i_{814}  - J_{125}$ colors are from total magnitudes of the 3D-HST photometric catalogue \citep{Momchevaetal2016}.  The dashed line in each panel corresponds to the fitted red sequence and the dotted lines are $\pm 2\sigma$.  The scatter is measured through the galaxy number distribution obtained from marginalising over the magnitude.  Green circles correspond to objects that are selected to be red sequence objects and are brighter than the magnitude limit ($H_{160} < 22.5$ for XMMU~J2235-2557 and XMMXCS~J2215-1738, $J_{125} < 23.5$ for Cl~0332-2742) as denoted by the grey shaded area.  Objects that are spectroscopically confirmed cluster members (combining KCS observations and previous literature) are circled in dark red.  For XMMU~J2235-2557, H$\alpha$ excess emitters from \citet{Grutzbauchetal2012} are shown as red squares.  These H$\alpha$ excess objects have a narrow $H$-band flux that is 3 times greater than the noise in the broad $H$-band (continuum) image and an equivalent width $> 20$~\AA.  For XMMXCS~J2215-1738, the $24\mu$m sources from \citet{Hiltonetal2010} (blue squares), submillimeter sources at 450 and $850\mu$m from \citet{Maetal2015} (red diamonds) and [O II] sources from \citet{Hiltonetal2010} and \citet[][private communication]{Hayashietal2010, Hayashietal2011, Hayashietal2014} (red crosses) are shown.  The dark green triangle indicates the most massive galaxy in this cluster (see Section~\ref{subsec:Stellar mass -- light-weighted size relations} for details).  Only objects that fulfil the selection with redshift and area are plotted (see Section~\ref{sec:Construction of photometric catalogues} for details).}
  \label{fig_color_mag_plot}
\end{figure*}

The goal of this study is to investigate the structural properties of passive galaxies in dense environments, hence we first need to select a clean sample of passive cluster galaxies by removing star-forming galaxies in the clusters and field galaxies. We identify passive galaxies in the KCS clusters through a red sequence and color-color selection.  Figure~\ref{fig_color_mag_plot} shows the color-magnitude diagram of the detected sources in the three clusters that fulfil the selection described in Section~\ref{sec:Construction of photometric catalogues}.  We use the PSF-matched $1''$ aperture $z_{850} - H_{160}$ colors for the cluster XMMU~J2235-2557 and XMMXCS~J2215-1738, while for the Cl~0332-2742 we use the $i_{814}  - J_{125}$ colors from the 3D-HST catalogue \citep{Momchevaetal2016}, in order to match the selection for the KMOS observations \citep[see][for more details]{Beifiorietal2017}.  We have checked that using the $i_{814}  - J_{125}$ colors (and the color-color selection) will result in the same selection as with $z_{850} - H_{160}$ colors.

Objects that are within 2$\sigma$ from the fitted color-magnitude relation in each cluster are selected as the initial red sequence sample, with the exception that some red objects that are slightly above 2$\sigma$ are also selected for completeness.  A magnitude cut is applied for each cluster: $H_{160} < 22.5$ for XMMU~J2235-2557 and XMMXCS~J2215-1738,  $J_{125} < 23.5$ for Cl~0332-2742.  This magnitude cut corresponds to a completeness of $\sim95\%$ for XMMU~J2235-2557 and XMMXCS~J2215-1738, and $\sim97\%$ for Cl~0332-2742. 

Also shown on Figure~\ref{fig_color_mag_plot} are the $24\mu$m and submillimeter detections from \citet{Hiltonetal2010} and \citet{Maetal2015}.  Using the template library of \citet{CharyElbaz2001} and \citet{DaleHelou2002},  these sources have very high derived SFR $ \gtrsim 100 M_{\odot}$~yr$^{-1}$ \citep{Hiltonetal2010, Maetal2015} despite their red color, indicating that they are dusty-starburst galaxies.  This demonstrates that the red sequence method alone suffers from contamination.  One way to distinguish between `genuine' passive galaxies and dusty star forming galaxies is through color selection techniques, such as the $UVJ$ classification \citep[e.g.][]{Labbeetal2005, Williamsetal2010, Brammeretal2011}.  Hence, on top of the red sequence selection we perform a color-color selection to reduce contamination.

\begin{figure*}
  \centering
  \includegraphics[scale=0.555]{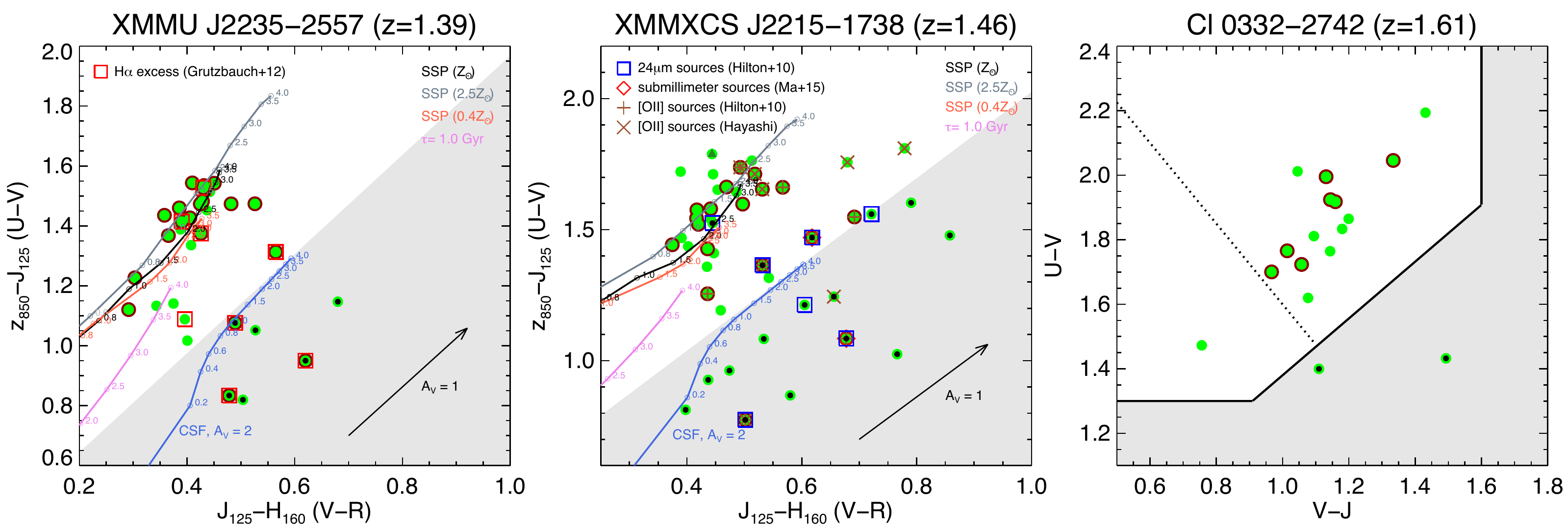}  
  \caption[Color-color selection for the three KCS clusters]{Color-color selection for the three KCS clusters.   Only red sequence selected galaxies are shown.  Objects that are spectroscopically confirmed cluster members (combining KCS observations and previous literature) are circled in dark red.  For XMMU~J2235-2557 (left) and XMMXCS~J2215-1738 (middle), the selection is performed using $z_{850} - J_{125}$ versus $J_{125} - H_{160}$ color, which correspond to rest-frame $(U-V)$ and $(V-R)$.  Different colored lines correspond to the evolution tracks of \citet{BruzualCharlot2003} models with different properties:  a constant SFH with $A_V =2$ mag of extinction (CSF, blue), SSPs with sub-solar, solar and super-solar metallicities with no dust ($0.4 Z_{\odot},Z_{\odot}, 2.5Z_{\odot}$ with light red, black, dark grey, respectively),  an exponentially declining SFH with no dust ($\tau = 1$ Gyr, violet). 
The empty circles represent the model colors at the specified ages (in Gyr).  The dust vector indicates an extinction of $A_V$ = 1 mag, assuming the \citet{Calzettietal2000} extinction law.  Different properties available from the literature (H$\alpha$, $24\mu$m, [O II] and submillimeter sources) are plotted. The dark green triangle indicates the most massive galaxy in XMMXCS~J2215-1738 (see Section~\ref{subsec:Stellar mass -- light-weighted size relations} for details). For Cl~0332-2742 (right), the $UVJ$ color selection is used.  The rest-frame $(U-V)$ and $(V-J)$ color are taken from the 3D-HST photometric catalogue \citep{Skeltonetal2014, Momchevaetal2016}.  The dashed line corresponds to the division between `young' and `old' passive galaxies based on their colors, adopted from \citet{Whitakeretal2013}.   The grey shaded region in each panel is where the star-forming galaxies reside. Objects that are excluded are labeled with a black dot at the centre. See text for details.}
  \label{fig_color_color}
\end{figure*}

For Cl~0332-2742, a $UVJ$ classification was performed from the rest-frame $(U-V)$ and $(V-J)$ color provided in the 3D-HST photometric catalogue. The right panel of Figure~\ref{fig_color_color} shows the $UVJ$ digram of the red sequence of this cluster.   We remove the two objects that are not in the quiescent region.  

For XMMU~J2235-2557 and XMMXCS~J2215-1738, since rest-frame $J$-band magnitudes are not available, we construct a color-color selection using the $z_{850} - J_{125}$ and the $J_{125} - H_{160}$ color, shown in the left and middle panel of Figure~\ref{fig_color_color}.  These two colors correspond roughly to the rest-frame $(U-V)$ and $(V-R)$ color (hereafter $UVR$ selection).  We identify the region occupied by star forming galaxies in the $UVR$ plane by computing evolutionary tracks with different star formation histories for each cluster redshift using \citet{BruzualCharlot2003} stellar population models (hereafter BC03).  The evolutionary track of a constant star forming population (CSF, blue) is clearly separated from various passively evolving populations (SSP with different metallicities and an exponentially declining $\tau$ model with $\tau = 1$ Gyr).

We exclude galaxies that are within the star forming region spanned by the CSF track with various dust extinction values, as shaded in grey on Figure~\ref{fig_color_color}.  The edge of the shaded star forming region is parallel to the dust vector.  Objects that are within this region are presumably dusty-star forming galaxies or interlopers that are not at the cluster redshift.  In addition, we have also excluded the $24~\mu$m and submillimeter sources that have high derived SFR ($\gtrsim100 M_{\odot}$ yr$^{-1}$) from \citet{Hiltonetal2010} and \citet{Maetal2015}, marked with a black dot on Figure~\ref{fig_color_color}.  Additional information would be required to identify star forming objects that have lower SFR.  For completeness we also plot the matched [O II] sources from \citet{Hiltonetal2010} and \citet[][private communication]{Hayashietal2010, Hayashietal2011, Hayashietal2014} in Figure~\ref{fig_color_color}.

We have also verified that all the $UVJ$-passive galaxies in Cl 0332-2742 fall into the $UVR$ passive regions of this cluster.  Since with $UVR$ we only removed galaxies that are in the region occupied by star forming galaxies with constant star formation rate, the $UVR$ selection we used is a less stringent selection compared to the $UVJ$.

In summary,  the red sequence and color selection result in a sample of \textbf{25} objects in the cluster XMMU~J2235-2557, \textbf{29} objects in XMMXCS~J2215-1738 and \textbf{15} objects in Cl~0332-2742.  The photometric catalogues of the three clusters are provided in Table~\ref{tab_photopara_xmmu}, Table~\ref{tab_photopara_xmmxcs} and Table~\ref{tab_photopara_cl}, respectively.  Note that we have revised the sample of XMMU~J2235-2557 described in \citet{Chanetal2016} in this paper based on the additional color-color selection as well as new redshift information from recent KCS observations \citep{Beifiorietal2017}.  Hence, compare to Table F1 in \citet{Chanetal2016}, Table~\ref{tab_photopara_xmmu} comprises the new $z_{850} - J_{125}$ and the $J_{125} - H_{160}$ color we used for the selection and more updated spectroscopic member information. Objects that are spectroscopically confirmed non-members are excluded from this sample.  For the same reason the sample here is slightly different from the passive sample for KMOS observations described in \citet{Beifiorietal2017}.

\section{Analysis}
\label{sec:Methods and Analysis}

\subsection{Light-weighted structural parameters}
\label{subsec:Light-weighted structural parameters}
We derive the light-weighted structural parameters of the red-sequence sample using the same procedure described in \citet{Chanetal2016}.   Two-dimensional single S\'ersic profile fitting \citep{Sersic1968} is performed on individual galaxies in each \textit{HST} band independently.  The parameters are derived using a self-modified version of GALAPAGOS (based on v.1.1) \citep{Bardenetal2012} with GALFIT (v.3.0.5).


The S\'ersic profile can be characterised by five independent parameters:  the total luminosity $L_{\rm{tot}}$, the S\'ersic index $n$, the effective semi-major axis $a_e$, the axis ratio $q$ ($=b/a$, where $a$ and $b$ is the major and minor axis respectively) and the position angle $P.A.$.  All five parameters as well as the centroid ($x,y$) of the galaxy are left to be free parameters in the fitting process. The fitting constraints are set to be: $0.2 < n < 8$, $0.3 < a_{e} < 500$ (pix), $0 < $mag$ < 40$, $0.0001 < q < 1$, $-180^{\circ} < P.A. < 180^{\circ}$.  The sky level is fixed to the value determined by GALAPAGOS.  The S\'ersic model is convolved with the PSF constructed from stacking bright unsaturated stars in the images (see Section~ \ref{subsec:HST data reduction} for details on the PSF derivation).

We modify GALAPAGOS to use the RMS maps derived from \texttt{ERR} weight maps output by Astrodrizzle as input for $\chi^2$ fitting.  The version of GALAPAGOS code we used relies only on the internal error estimation in GALFIT.  The RMS maps that we generate from \texttt{ERR} weight maps are a more realistic representation of the noise than the internal error estimation in GALFIT (see Section~\ref{subsec:HST data reduction}), as they include pixel-to-pixel exposure time differences originating from image drizzling and dithering patterns in observations, as well as a more accurate estimation of shot noise.

We then perform quality checks on the fitted structural parameters and derive uncertainties on top of the error output by GALFIT using simulated galaxies.  We randomly drop on average a set of 20000 simulated galaxies (one at a time) with surface brightness profiles described by a S\'ersic profile on the ACS and WFC3 images of the three clusters, and recover their parameters with our pipeline.  For each galaxy we then add the corresponding dispersion in quadrature to the error output by GALFIT \citep[see][for details of the simulation]{Chanetal2016}.  The best-fitting light-weighted structural parameters and the corresponding uncertainties of XMMXCS~J2215-1738 and Cl~0332-2742 are provided in Table~\ref{tab_strpara_xmmxcs} and Table~\ref{tab_strpara_cl}, respectively.  For the parameters of XMMU~J2235-2557 the interested reader can refer to Table~F1 in \citet{Chanetal2016}. \\

\subsection{Stellar mass-to-light ratio -- color relation and integrated stellar masses}
\label{subsec:Stellar mass-to-light ratio-color relation and integrated stellar masses}
We estimate the stellar mass-to-light ratios $M_{*}/L$ and stellar masses $M_{*}$ of the galaxies using an empirical relation between the observed color and the stellar mass-to-light ratio.  
At $z\sim1.5$, the $z_{850} - H_{160}$ color is the perfect proxy for the $M_{*}/L$ as it straddles the $4000 \AA$ break and has a wide dynamic range.

\begin{figure*}
  \centering
  \includegraphics[scale=0.81]{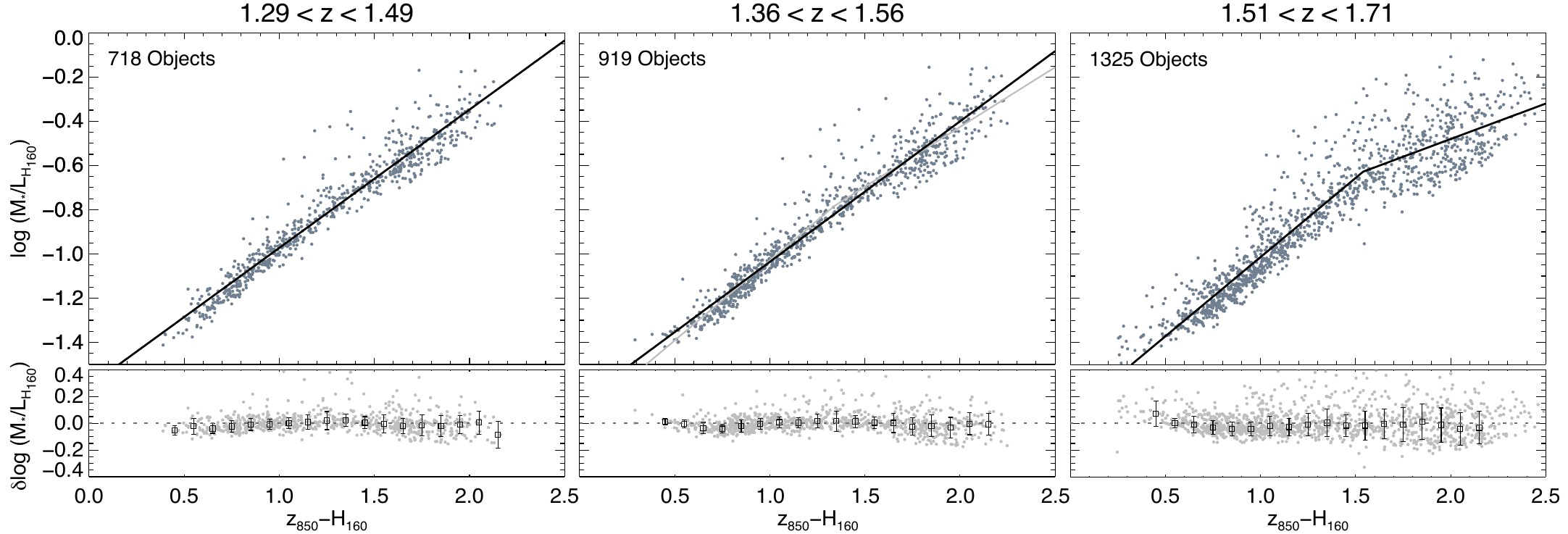}
  \caption[Relations between stellar mass-to-light ratio and $z_{850}-H_{160}$ color at redshifts of the three KCS clusters]{Relations between stellar mass-to-light ratio and $z_{850}-H_{160}$ color at the redshift of the three KCS clusters.  Gray points in each panel are galaxies from the NMBS catalogue that satisfy the selection criteria.  The black line is the best-fit relation.  The light gray line in the XMMXCS~J2215-1738 panel illustrates the effect of using the bilinear fit instead of a single linear fit.  The bottom part in each panel shows the residuals of the relation $\delta \log(M_{*}/L) = $ data - linear fit in color bins of 0.1 dex.  The empty squares are the median residual in bins of 0.1 mag.}
  \label{fig_ml_color_relation}
\end{figure*}

Figure~\ref{fig_ml_color_relation} shows the stellar mass-to-light ratio -- color relations for the three clusters respectively.  The relations are derived using the public catalogue from the NEWFIRM medium band survey (NMBS) in the COSMOS field \citep{Whitakeretal2011}, which comprises photometries in 37 bands, spectroscopic redshifts for a subset of the sample, photometric redshifts derived with \texttt{EAZY} \citep{Brammeretal2008} and stellar population parameters derived with \texttt{FAST} \citep{Krieketal2009}.  The stellar masses used are estimated using BC03 models with exponentially declining SFHs and a \citet{Chabrier2003} IMF.  We derive the $M_{*}/L$-color relation for each cluster in the observer frame (observed $z_{850} - H_{160}$ color), contrary to the typical approach which interpolates the data to obtain rest-frame colors \citep[e.g.][]{Szomoruetal2013}, to reduce the number of interpolations required for our data.  

For each cluster, we select NMBS galaxies within a redshift window of $\pm0.1$ of the cluster redshift and apply a magnitude cut as for our red sequence selection.  With these criteria we select 718 objects for $1.29 < z < 1.49$, 919 objects for $1.36 < z < 1.56$ and 1325 objects for $1.51 < z < 1.71$.  We rerun \texttt{EAZY} for these objects using the redshifts and photometries in the NMBS catalogue to obtain the best-fit SEDs for computing the observed frame $z_{850} - H_{160}$ colors and luminosities $L_{H_{160}}$.  A more detailed description of deriving the stellar mass-to-light ratio -- color relation can be found in \citet{Chanetal2016}.   \\


\noindent For XMMU~J2235-2557:
\begin{equation}
   \log(\frac{M_{*}/L_{H_{160}}} {M_\odot/L_\odot}) =  (0.625 \pm 0.004)  \>(z_{850} - H_{160}) - (1.598 \pm 0.005)
\end{equation}

\noindent For XMMXCS~J2215-1738:
\begin{equation}
   \log(\frac{M_{*}/L_{H_{160}}} {M_\odot/L_\odot}) =  (0.635 \pm 0.002) \>(z_{850} - H_{160}) - (1.671 \pm 0.002)
\end{equation}

\noindent For Cl~0332-2742:\\
if $(z_{850} - H_{160}) \leq 1.54$: 
\begin{equation}
   \log(\frac{M_{*}/L_{H_{160}}} {M_\odot/L_\odot}) =  (0.716 \pm 0.007) \>(z_{850} - H_{160}) - (1.730 \pm 0.005)
\end{equation}
if $(z_{850} - H_{160}) > 1.54$:
\begin{equation}
   \log(\frac{M_{*}/L_{H_{160}}} {M_\odot/L_\odot}) =  (0.320 \pm 0.011) \>(z_{850} - H_{160}) - (1.121 \pm 0.023)
\end{equation}

The relation is linear in XMMU~J2235-2557 and XMMXCS~J2215-1738, while in Cl~0332-2742 a bilinear function is preferred.  
Using a two-component function is common in fitting $M_{*}/L$-color relation \citep[e.g.][]{Moketal2013}, primarily due to the difference in $M_{*}/L$ of the blue and red stellar population.  We have also tried to use a bilinear fit for the other two clusters, but the results are consistent with single linear fits.  We have checked that using non-parametric regression methods, such as the constrained B-Splines (cobs) and the local regression (locfit) implemented in R, will not change our relations.  The fitting uncertainties are estimated from bootstrapping with 1000 realizations.  The global scatter of the fits are $\sim 0.06$ dex for XMMU~J2235-2557, $\sim0.06$ dex for XMMXCS~J2215-1738 and $\sim0.10$ dex for Cl~0332-2742 respectively.  The uncertainty in log($M_{*}/L$) is generally $\sim0.1$ in each bin and the bias is negligible.  

We estimate the integrated stellar masses ($M_{*}$) of the cluster galaxies using the $M_{*}/L$-color relations, $z_{850} - H_{160}$ aperture colors and total luminosities $L_{H_{160}}$ from the best-fit 2D GALFIT S\'ersic models. 
The typical uncertainty of the mass estimates is $\sim0.1-0.15$ dex.  
We compared our stellar masses with masses derived from SED fitting from \citet{Strazzulloetal2010, Delayeetal2014, Santinietal2015, Momchevaetal2016} for a subset of our sample. The mass estimates from the two methods are consistent with each other, with a median difference of $\lesssim 0.1$ dex \citep[see][for details]{Beifiorietal2017}..

\subsection{Resolved stellar mass surface density maps and mass-weighted structural parameters}
\label{subsec:Resolved stellar mass surface density maps and mass-weighted structural parameters}
Because of the varying color gradients in the passive galaxies \citep[e.g.][]{Guoetal2011, Chanetal2016}, the luminosity-weighted size is dependent on the filter band of the image and is not always a reliable proxy of the stellar mass distribution.  This may complicate the interpretation of the size evolution or the comparison between different environments.  One way to resolve this is to measure characteristic sizes of the stellar mass distribution (i.e. mass-weighted sizes) instead of using the wavelength dependent luminosity-weighted sizes. Recently a number of works attempted to reconstruct stellar mass profiles taking into account the $M_{*}/L$ gradients.  Two techniques have been primarily used: resolved spectral energy distribution (SED) fitting \citep[e.g.][]{Wuytsetal2012, Langetal2014} and the use of $M_{*}/L$ - color relation \citep[e.g.][]{BelldeJong2001, Belletal2003}.  In \citet{Chanetal2016} we construct resolved stellar mass surface density maps (hereafter referred to as mass maps) of individual galaxies in XMMU~J2235-2557 using the $M_{*}/L$-color relation and color maps derived from the $z_{850}$ and $H_{160}$ images.  In this paper we extend this method to two additional clusters in KCS.  Here we review only the key processing steps.

We first resample the PSF-matched $z_{850}$ image to the same grid as the $H_{160}$ image using SWarp \citep{Bertinetal2002}.  
For each galaxy, we run the Voronoi binning algorithm \citep{Cappellarietal2003} on the sky-subtracted PSF-matched $z_{850}$ band galaxy postage stamp to group pixels to a target S/N level of 10 per bin.  The same binning scheme is then applied to the sky-subtracted $H_{160}$ postage stamp, which has a higher S/N.  
Binned $z_{850} - H_{160}$ color maps are obtained by converting the ratio of the two images into magnitudes.  Binned $M_{*}/L$ maps are then derived by converting the color in each bin to a mass-to-light ratio with the derived color - $M_{*}/L$ relation for each cluster respectively.  For areas with insufficient S/N, (i.e. $<$ 1.5 times of our target S/N
), we fix the $M_{*}/L$ to the annular median of $M_{*}/L$ bins at the last radius with sufficient S/N, as determined from the one-dimensional S/N profiles of the galaxy in the $z_{850}$ band. 
To cope with a ``discretization effect" that arises from the binning procedure, for each galaxy we perform the abovementioned binning procedure 10 times, each with a different randomised set of initial Voronoi nodes.  We then median-stack the resulting $M_{*}/L$ maps.  The mass maps are then constructed by directly combining the median-stacked $M_{*}/L$ map and the original (i.e. unbinned) $H_{160}$ images, in order to preserve the WFC3 spatial resolution.  Similarly, we also generate mass RMS maps for each galaxy from the \texttt{ERR} weight maps output by Astrodrizzle.

We then measure mass-weighted structural parameters from the resolved stellar mass surface density maps, following a similar procedure as with the light-weighted structural parameters.  All five parameters of the S\'ersic profile ($M_{*, tot}$, $n_{mass}$, $a_{e,mass}$, $q_{mass}$ and $P.A._{mass}$) and the centroid are left to be free parameters.  The sky level (i.e. the background mass level in mass maps) is fixed to zero.  We use the same GALFIT constraints as for the light-weighted structural parameters, except allowing a larger range for the S\'ersic indices: $0.2 < n < 15.0$ since mass profiles are expected to be more centrally peaked compared to light profiles \citep{Szomoruetal2013}.  Again we derive the uncertainties of the structural parameters using simulated galaxies \citep[see][for details]{Chanetal2016}.

While the fitting process is straightforward for most of the galaxies, we found that for a couple of objects the fits do not converge, or have resultant sizes smaller than half of the PSF HWHM, which are unreliable \citep[see the discussion in Appendix A3 of][]{Chanetal2016}.  We remove these objects from the mass parameter sample.  Most of them initially have small light-weighted sizes.  5 objects (out of 25) in XMMU~J2235-2557 and 9 objects (out of 29) in XMMXCS~J2215-1738 are discarded, among them two objects in XMMU~J2235-2557 and four objects in XMMXCS~J2215-1738 that are spectroscopically confirmed.  All of the objects in Cl~0332-2742 are well-fitted.   The mass-weighted structural parameters of the three clusters are also provided in  Table~\ref{tab_strpara_xmmxcs} and Table~\ref{tab_strpara_cl} and Table~F1 in \citet{Chanetal2016}, respectively.

\section{Results}
\label{sec:Results}
\ifx true false
\fi

In this section we derive stellar mass -- size relations of the passive galaxies in the KCS clusters.  As we discussed in Section~\ref{sec:Introduction}, previous studies have used different definitions of galaxy size to derive stellar mass -- size relations.  Hence, we have derived relations using both circularised effective radii ($R_{e\rm{-circ}} =  a_{e} \times \sqrt{q}$) and elliptical semi-major axes ($a_{e}$) as galaxy sizes.  To compare with the literature, in this section we will mainly focus on the result of stellar mass -- size relations derived using $R_{e\rm{-circ}}$, as using semi-major axes instead do not change the conclusion.

\subsection{Stellar mass -- light-weighted size relations}
\label{subsec:Stellar mass -- light-weighted size relations}
We first compare the stellar mass -- light-weighted size relation of the KCS clusters with other clusters as well as field galaxies from the literature.  The mass -- size relations of XMMU~J2235-2557 and XMMXCS~J2215-1738 in rest-frame UV have been studied in \citet{Strazzulloetal2010} and \citet{Delayeetal2014}.


\subsubsection{Comparison to local samples}
\label{subsubsec:Comparison to local samples}
Figure \ref{fig_light_masssize} shows the $H_{160}$ band (rest-frame $R$-band) mass -- light-weighted size relations of the three KCS clusters.  Also shown in Figure \ref{fig_light_masssize} is the local mass -- size relation of the SDSS passive sample \citep[single S\'ersic fit relation,][]{Bernardietal2014}  and the ATLAS\textsuperscript{3D} sample \citep[the peak ridge-line of the distribution for ETGs with stellar mass larger than $3 \times 10^{10} M_{\odot}$,][]{Cappellarietal2013b} for comparison. To be consistent with the relation of the KCS clusters and the \citet{Bernardietal2014} local relation, we have circularized the sizes used in the relation of \citet{Cappellarietal2013b}.   Although both relations were derived for galaxies regardless of their local density, a number of studies have established that there is no obvious environmental dependence on passive galaxy sizes in the local universe \citep{Guoetal2009, Weinmannetal2009, Tayloretal2010, Huertascompanyetal2013, Cappellari2013}. We also plot the zone of exclusion as defined by local mass -- size and mass -- velocity dispersion relations from \citet{Cappellarietal2013b}, \citet{Cappellari2016}.  We have also confirmed that the wavelength-dependent size-correction for this comparison is negligible \citep[see][for a discussion on the wavelength-dependence relation of XMMU~J2235-2557]{Chanetal2016}.  Both local relations are based on $r$-band photometry \citep{Cappellarietal2013a, Bernardietal2014}, which roughly corresponds to the observed $H_{160}$ band at redshifts of $1.39 -1.61$.

The $H_{160}$ band sizes of the passive galaxies in XMMU~J2235-2557 are on average $\sim42\%$ smaller than expected from the \citet{Bernardietal2014} relation (i.e. the average deviation of the sample from the local relation), with $\langle \log(R_{e\rm{-circ}} / R_{\rm{Bernardi}}) \rangle = -0.24$ ($\sim41\%$ smaller for the spectroscopic confirmed members).  These galaxies are on average $\sim22\%$ smaller than expected from the ATLAS\textsuperscript{3D} relation. Note that part of the difference between the two local relations is due to sample selection and how the masses and sizes are measured; the ATLAS\textsuperscript{3D} sizes are measured from the multi-Gaussian expansion models and the masses are dynamical masses determined from JAM models \citep[see,][for details]{Cappellarietal2013a}.  The difference between dynamical masses and stellar masses could add a systematic offset to the comparison and potentially make our sample less different from the local sample \citep[see, e.g. Section 4][]{Beifiorietal2017}.  There are galaxies in XMMU~J2235-2557 whose sizes are $\sim70\%$ smaller than those of their average local counterparts.  As one can see from Figure~\ref{fig_light_masssize}, the BCG also has the largest size ($\sim24$ kpc) and lies on the local relation.  This is consistent with previous works showing BCGs as a population have had very little evolution in mass or size since $z \sim 1$ \citep[e.g.][]{Stottetal2010, Stottetal2011}.  

For XMMXCS~J2215-1738, the sizes of the passive galaxies are on average $\sim55\%$ smaller than the \citet{Bernardietal2014} relation, with $\langle \log(R_{e\rm{-circ}} / R_{\rm{Bernardi}}) \rangle = -0.34$ ($\sim65\%$ smaller for the spectroscopic confirmed members). They are on average $\sim40\%$ smaller than expected from the ATLAS\textsuperscript{3D} relation.  This suggests that the sizes in XMMXCS~J2215-1738 are on average smaller compared to those in XMMU~J2235-2557.  The galaxy with the smallest size is $\sim86\%$ smaller than its average local counterpart.  As we mentioned in Section~\ref{sec:The KCS Sample}, the BCG in XMMXCS~J2215-1738 is not exceptionally bright compared to other galaxies.  From Figure~\ref{fig_light_masssize} it is clear that this atypical BCG is not the most massive object in the cluster and has a relatively small size ($\sim 1.0$ kpc), and is even below the zone of exclusion.  On the other hand, the most massive galaxy in this cluster, although not spectroscopically confirmed, has a redder color but is 0.5 mag less bright compared to the BCG (marked with a triangle in Figure~\ref{fig_color_mag_plot} and~\ref{fig_color_color}).  Both galaxies are off-centered, which is probably related to the fact that XMMXCS~J2215-1738 is not virialized \citep[e.g.][]{Hiltonetal2010, Maetal2015}.

The average $H_{160}$ band size of the passive galaxies in Cl~0332-2742 is the smallest among the three clusters, as expected from the size evolution.  The galaxies are on average $\sim69\%$ smaller than expected from the \citet{Bernardietal2014} relation, with $\langle \log(R_{e\rm{-circ}} / R_{\rm{Bernardi}}) \rangle = -0.51$ ($\sim69\%$ smaller for the spectroscopic confirmed members), and on average $\sim59\%$ smaller than expected from the ATLAS\textsuperscript{3D} relation.  Half of the galaxies are below the zone of exclusion, which is expected in the case of size evolution.  The smallest galaxy is $\sim 82\%$ smaller than expected from the local relation.  The most massive object also has the largest size among the sample ($\sim 6.2$ kpc).  This object (ID 11827) locates at the west part of the structure and is the brightest group galaxy (BGG) in the \citet{Tanakaetal2013} group.

To measure the slope of the mass -- size relation, we fit the relation using \texttt{LINMIX\_ERR}, an IDL routine using Bayesian inference approach to linear regression with Markov Chain Monte Carlo \citep[MCMC;][]{Kelly2007} with the following linear regression:
\begin{equation}
  \label{eqt-fit}
   \log(R_{e\rm{-circ}}/\textrm{kpc}) =  \alpha +  \beta~ (\log(M_{*} /M_\odot) - 11.0) + N(0, \epsilon)
\end{equation}
\noindent where $N(0,\epsilon)$ is a normal distribution with mean 0 and dispersion $\epsilon$.  The $\epsilon$ represents the intrinsic random scatter of the regression.  The normalisation of the stellar masses in equation~\ref{eqt-fit} is chosen to be $\log(M_{*}/M_\odot) = 11.0$, which is close to the average mass in the three clusters, to minimise the uncertainty in $\alpha$ and facilitate comparison with the literature.  We did not consider the small covariance between the $R_{e\rm{-circ}}$ and $M_\odot$ while fitting the relations.  Moreover, the BCGs (as well as the BGG in the \citet{Tanakaetal2013} group of Cl~0332-2742) have been excluded in the fitting process, as they may have experienced a different evolutionary path \citep[e.g.][]{Stottetal2011}.

The best-fit intercept $\alpha$, slope $\beta$ and scatter $\epsilon$ for both the entire red-sequence selected sample (case A) and only the spectroscopically confirmed members (case B) of the three clusters are summarised in Table~\ref{tab_bestfit}.  For comparison to previous literature, we also fit only massive objects with $\log(M_{*}/M_\odot) \geq 10.5$ (case C \& D) , the limiting mass adopted in \citet{Delayeetal2014}, to ensure the mass range of the fitted data is comparable.  Since all objects in Cl~0332-2742 have $\log(M_{*}/M_\odot) \geq 10.5$, we only give the result of case C \& D.  We have also derived relations using semi-major axes as galaxy sizes, the fitted parameters are presented in Appendix~\ref{The fitted parameters of mass -- size relations using semi-major axes}.

The measured slopes $\beta$ for the red-sequence selected samples (A \& C) as well as the spectroscopic confirmed members (B \& D) in the three clusters are consistent within $1\sigma$ respectively.

Using the fitted relations, the average size of XMMU J2235-2557, XMMXCS J2215-1738 and Cl~0332-2742 at $\log(M_{*}/M_{\odot}) = 11$ are $\sim40\%, \sim51\%, \sim70\%$ smaller compared to the \citet{Bernardietal2014} relations, respectively ($\sim21\%, \sim35\%, \sim60\%$ compared to ATLAS\textsuperscript{3D}).

Among the three clusters, XMMXCS~J2215-1738 has the steepest relations for the full sample fit (A \& B) as well as the massive sample fit (C \& D).  Comparing the fits in different mass ranges (A \& C), the fits with only massive objects always have a steeper slope, which probably hints that the mass -- size relation of passive galaxies are also curved similar to the local mass -- size relation \citep{Hydeetal2009}. 

For completeness, if we fit the entire massive red sequence sample (C) of all three clusters simultaneously, we measure a typical slope of $\beta = 0.79 \pm 0.14$ and an intercept of $\alpha = 0.32 \pm 0.03$.

\subsubsection{Comparison to high redshift samples}
\label{subsubsec:Comparison to high redshift samples}
\citet{Delayeetal2014} studied the mass -- size relation of a sample of red sequence galaxies in nine clusters at $0.89 < z < 1.5$ in the rest-frame $B$-band for $\log(M_{*}/M_\odot) > 10.5$.  They reported a typical slope of $\beta = 0.49 \pm 0.08$ for the seven clusters up to $z\sim1.2$, and relatively shallow slopes of XMMU~J2235-2557 ($\beta = 0.22 \pm 0.32$) and XMMXCS~J2215-1738 ($\beta = 0.31 \pm 0.32$) in their sample.  Their relations are systematically flatter by more than $1\sigma$ compared to both our full sample (A) and massive sample (C) fit.  Our relations also show smaller intrinsic scatter compared to \citet{Delayeetal2014}.  On the other hand, their intercepts (0.44 for XMMU~J2235-2557 and 0.43 for XMMXCS~J2215-1738 after converted to our definition) are roughly consistent with our measurements. 

The discrepancies are primarily driven by the sample selection.  It is known from local studies that the galaxy properties related to the stellar populations (such as age and color) tend to vary along lines of nearly constant velocity dispersion, which traces equal mass concentration, hence the slope of the mass -- size relation depends strongly on the sample selection \citep[e.g.][]{Cappellari2016}. Samples that are more passive are expected to be steeper, while more shallow values are obtained with less stringent criteria for being passive \citep[see Section 4.3 in][for more details]{Cappellari2016}.  With only the red sequence selection (i.e. without the color-color selection), our relation of XMMU~J2235-2557 would have a flatter slope and a larger scatter: $\beta = 0.48 \pm 0.21$ (instead of $0.71$), $\epsilon = 0.25$.  Other effects such as the method of computing stellar masses and also the band which the relations are measured in ($z_{850}$ vs. our $H_{160}$) may also play a role. For example, if we keep our selection and use masses scaled with \texttt{MAG\_AUTO} as in \citet{Delayeetal2014}, the relation of XMMU~J2235-2557 is slightly flattened: $\beta = 0.69 \pm 0.24$.


Our derived slopes are also consistent with recent work by \citet{Sweetetal2017}, who studied a mass - size relation for SPT-CL J0546-5345 at $z=1.067$ and found a slope of $\beta = 0.74 \pm 0.06$.

The abovementioned published relations are for passive galaxies in high redshift clusters.  For field galaxies, the slopes of our massive sample fit (C) are consistent with previous works by \citet{Newmanetal2012} ($\beta = 0.62 \pm 0.09$, for $\log(M_{*}/M_\odot) > 10.7$ galaxies at $1.0 < z < 1.5$) and \citet{Cimattietal2012} ($\beta = 0.50 \pm 0.04$, for $\log(M_{*}/M_\odot) > 10.5$ galaxies at $z > 0.9$).  Our result is also consistent with the recent study by \citet{vanderWeletal2014}, who found the slope of the mass -- size relation at $z \sim 1.25$ and $z \sim 1.75$ to be $\beta = 0.76 \pm 0.04$ for $UVJ$ passive galaxies ($\log(M_{*}/M_\odot) > 10.3$) in CANDELS.   Note that \citet{vanderWeletal2014} used semi-major axis $a_{e}$ as sizes instead of $R_{e\rm{-circ}}$.  We find that using semi-major axis instead of circularized effective radius does not have a huge impact on the measured slopes, for example for XMMU~J2235-2557 the slope is only slightly flatter ($\beta = 0.55 \pm 0.15$ (A) and $\beta = 0.66 \pm 0.15$ (C)) if $a_{e}$ is used (see also Appendix~\ref{The fitted parameters of mass -- size relations using semi-major axes} for the fitted parameters of the relations using semi-major axes as sizes).  Given the uncertainties, we conclude that there is no evidence that the slope of the mass -- size relation for this mass range depends on the environment.

\subsubsection{Caveats - Progenitor bias}
\label{subsubsec: Caveats - Progenitor bias}
The scenario described above does not include the continual addition of newly quenched young galaxies onto the red sequence, i.e. the progenitor bias \citep[e.g.][]{vanDokkumetal2001}.  The progenitor bias complicates the interpretation of the evolution of red sequence galaxies.  Several studies have shown that it has non-negligible effect on the size evolution \citep[e.g.][]{Sagliaetal2010, Valentinuzzietal2010b, Carolloetal2013, Poggiantietal2013, Beifiorietal2014, Jorgensenetal2014, Shankaretal2015, Fagiolietal2016}, as young galaxies have preferentially larger sizes \citep[see, e.g.][for a clear color-size correlation at the low mass end]{Morishitaetal2017}.  Unfortunately additional information such as age is required to correct for the progenitor bias. Although the mean age for each KCS cluster is available \citep{Beifiorietal2017}, we do not have precise ages for individual galaxies to implement a correction for the three clusters. However, we can take a different approach and use age information of the local sample to select likely descendant of our KCS cluster galaxies and examine how this will affect the above size comparison to local passive galaxies.  

\begin{figure*}
  \centering
  \includegraphics[scale=0.80]{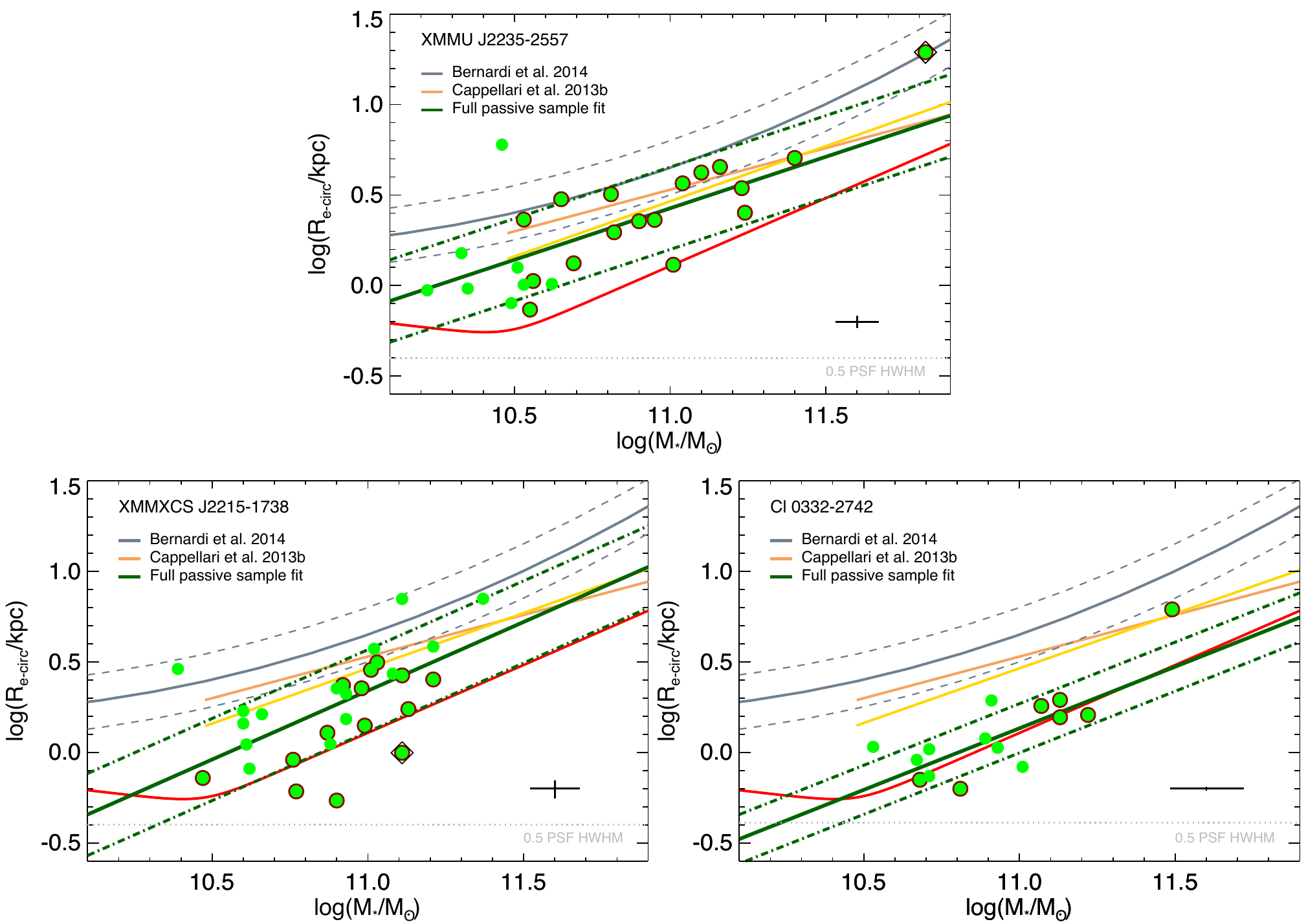}
  \caption[Stellar mass -- light-weighted size relations of the red sequence galaxies in the KCS clusters]{Stellar mass -- light-weighted size relations of the red sequence galaxies in the KCS clusters.  The green line in each panel is a linear fit to the full passive sample (Case A for XMMU~J2235-2557 and XMMXCS~J2215-1738, Case C for Cl~0332-2742), while the dot-dashed lines represent $\pm 1\epsilon$ fitted intrinsic scatter.  The dark grey line corresponds to the local $r$-band mass -- size relation from \citet{Bernardietal2014}. The light brown line corresponds to the local mean mass -- size relation for the ATLAS\textsuperscript{3D} sample (the peak ridge-line of the distribution for ETGs with stellar mass larger than $ 3 \times 10^{10} M_{\odot}$) from \citet{Cappellarietal2013b}.  Note that the offset between the two local relations is largely due to how the masses and sizes are measured. The ATLAS\textsuperscript{3D} masses are dynamical masses determined from JAM models. See text for details.  The golden line corresponds to the mass -- size relation for the ATLAS\textsuperscript{3D} sample with our progenitor biased correction.  Individual objects are shown in green and spectroscopically confirmed objects are circled with dark red.  The BCGs in XMMU~J2235-2557 and XMMXCS~J2215-1738 are indicated with black diamonds.  The cross shows the typical uncertainty of the sizes and the median uncertainty of the integrated mass in our sample.  The red line shows the zone of exclusion for local galaxies from Equation 4 of \citet{Cappellarietal2013b}, \citet{Cappellari2016}. The light grey dotted line corresponds to 0.5 PSF HWHM, the limit where we can measure reliable sizes as derived from our simulations (see Appendix A3 in \citet{Chanetal2016} for details).}
  \label{fig_light_masssize}
\end{figure*}

We attempt to correct for the progenitor bias on the ATLAS\textsuperscript{3D} sample using the stellar ages derived by \citet{McDermidetal2015}.  We remove galaxies in the sample with ages that are too young to be descendants of passive galaxies at the cluster redshift, following the procedure used in \citet{Beifiorietal2014}, \citet{Chanetal2016} and \citet{Beifiorietal2017}.  The fitted relations of this progenitor bias corrected ATLAS\textsuperscript{3D} sample, using the same method as the KCS sample, are shown in each panel in Figure \ref{fig_light_masssize} as a golden line.  As expected, the relation with the progenitor bias corrected sample has a steeper slope compared to the original sample, as young(er) passive galaxies are preferentially larger and less massive \citep[e.g][]{Cappellari2016}.  A similar effect on the slope after correcting progenitor bias is also seen in \citet{Beifiorietal2017}.  The size difference between the ATLAS\textsuperscript{3D} sample and our KCS sample is reduced with the correction applied.  Note that the difference quoted below should only be compared to the uncorrected ATLAS\textsuperscript{3D} sample and understood in relative terms due to issues of the masses and sizes described in Section~\ref{subsubsec:Comparison to local samples}.  The sizes in XMMU J2235-2557 are on average $\sim3.5\%$ smaller than the progenitor bias corrected sample, as opposed to $\sim22\%$ without the correction (compared to the average deviation we measured in Section~5.1.1). The sizes in XMMXCS J2215-1738 and Cl 0332-2742 are $\sim28\%$ and $\sim51\%$ smaller than this sample, respectively ($\sim40\%$ and $\sim59\%$ before the correction). This comparison is consistent with all the abovementioned previous studies, showing that neglecting progenitor bias can lead to an overestimation of the size evolution.

Another way to look at this would be to compare the KCS sample to the superdense galaxies (SDGs) found in local clusters, as these galaxies are primarily the descendant of high redshift passive galaxies. \citet{Valentinuzzietal2010a} reported that $\sim22\%$ of massive cluster galaxies in the WIde-field Nearby Galaxy-cluster Survey (WINGS) local cluster sample are superdense massive galaxies, which have sizes (and masses) comparable to passive galaxies observed at high redshift.  These SDGs are also found to be more abundant in local clusters than the field \citep{Poggiantietal2013}.  Taking the characteristic value of the \citet{Valentinuzzietal2010a} sample (V-band $\langle  R_{e\rm{-circ}} \rangle = 1.61$ kpc at $\langle \log(M_{*}/M_\odot) \rangle = 10.94$) and applying a wavelength-dependent size-correction as in \citet{Chanetal2016}, the characteristic size of XMMU J2235-2557 at this mass as determined from the fitted relation is even larger than the median value of these SDGs. The sizes in XMMXCS J2215-1738 are comparable to the SDG sample, while those in Cl 0332-2742 are $\sim40\%$ smaller.

\begin{table}
  \centering
  \noindent
  \caption{Best-fit parameters of the stellar mass -- light-weighted size relations of the three KCS clusters}
  \label{tab_bestfit}
  \begin{tabular}{@{}p{0.3cm}lccc@{}}
  \\
  \multicolumn{5}{c}{XMMU~J2235-2557\textsuperscript{$\star$}} \\  
  \hline
  \hline
  \multicolumn{5}{l}{Stellar mass -- light-weighted size relation} \\
  \hline
  Case &  Mass range &  $\alpha \pm \Delta \alpha$  & $\beta \pm \Delta \beta$ & $\epsilon$ \\
  \hline
A & $10.0 \leq M_{*} \leq 11.5$              & $ 0.427 \pm  0.061 $ &  $ 0.569  \pm 0.156 $  &  0.228  \\    
C & $10.5 \leq M_{*} \leq 11.5$              & $ 0.423  \pm 0.051 $  &  $ 0.713  \pm 0.168  $   & 0.180 \\  
D & $10.5 \leq M_{*} \leq 11.5$ (spec)  & $ 0.426  \pm 0.056 $  &  $ 0.642  \pm  0.211 $   & 0.196 \\  
  \hline
  \multicolumn{5}{p{.45\textwidth}}{\textsuperscript{$\star$} Since all the spectroscopic members in XMMU~J2235-2557 are $\log(M_{*}/M_\odot) \geq 10.5$, case B is identical to case D.} \\
  \\
  \multicolumn{5}{c}{XMMXCS~J2215-1738} \\  
  \hline
  \hline
  Case &  Mass range &  $\alpha \pm \Delta \alpha$  & $\beta \pm \Delta \beta$ & $\epsilon$ \\
  \hline
  A & $10.0 \leq M_{*} \leq 11.5$              & $ 0.341 \pm 0.054$  &  $0.760 \pm 0.228$  &  0.227 \\    
  B & $10.0 \leq M_{*} \leq 11.5$ (spec)  & $ 0.255 \pm 0.078$   &  $1.125 \pm 0.435$  &  0.211 \\    
  C & $10.5 \leq M_{*} \leq 11.5$              & $ 0.345 \pm  0.050 $  &  $1.087 \pm  0.265 $   & 0.187 \\     
  D & $10.5 \leq M_{*} \leq 11.5$ (spec)  & $  0.271 \pm  0.086 $  &  $1.968 \pm  0.860 $   & 0.175 \\    
  \hline
  \\
  \multicolumn{5}{c}{Cl~0332-2742*} \\  
  \hline
  \hline
  Case &  Mass range &  $\alpha \pm \Delta \alpha$  & $\beta \pm \Delta \beta$ & $\epsilon$ \\
  \hline
  C & $10.5 \leq M_{*} \leq 11.5$              & $ 0.134 \pm 0.058$  &  $0.680 \pm 0.316$  &  0.136 \\     
  D & $10.5 \leq M_{*} \leq 11.5$ (spec)  & $ 0.086 \pm 0.104$   &  $1.022 \pm 0.644$  &  0.169 \\      
   \hline
   \hline
    \multicolumn{5}{p{.45\textwidth}}{\textsuperscript{*} Since all the selected galaxies in Cl~0332-2742 have $\log(M_{*}/M_\odot) \geq 10.5$, only case C \& D are applicable.} \\
\end{tabular}
\end{table}

\begin{table}
  \centering
  \noindent
  \caption{Best-fit parameters of the stellar mass -- mass-weighted size relations of the three KCS clusters}
  \label{tab_bestfitmass}
  \begin{tabular}{@{}p{0.3cm}lccc@{}}
  \\
  \multicolumn{5}{c}{XMMU~J2235-2557\textsuperscript{$\star$}} \\  
  \hline
  \hline
  \multicolumn{5}{l}{Stellar mass -- mass-weighted size relation} \\
  \hline
  Case &  Mass range &  $\alpha \pm \Delta \alpha$  & $\beta \pm \Delta \beta$ & $\epsilon$ \\
  \hline
 A & $10.0 \leq M_{*} \leq 11.5$              & $ 0.159 \pm 0.083 $  &  $ 0.403 \pm 0.211 $  &  0.280 \\        
 C & $10.5 \leq M_{*} \leq 11.5$              & $ 0.144  \pm 0.066 $  &  $ 0.551 \pm 0.211  $  & 0.218 \\       
 D & $10.5 \leq M_{*} \leq 11.5$ (spec)  & $ 0.148  \pm 0.075 $  &  $ 0.526 \pm 0.267 $  & 0.239 \\          
  \hline
  \multicolumn{5}{p{.45\textwidth}}{\textsuperscript{$\star$} Since all the spectroscopic members in XMMU~J2235-2557 are $\log(M_{*}/M_\odot) \geq 10.5$, case B is identical to case D.} \\
  \\
  \multicolumn{5}{c}{XMMXCS~J2215-1738} \\  
  \hline
  \hline
   Case &  Mass range &  $\alpha \pm \Delta \alpha$  & $\beta \pm \Delta \beta$ & $\epsilon$ \\
   \hline
 A & $10.0 \leq M_{*} \leq 11.5$              & $ -0.031 \pm 0.052 $  &  $ 0.319 \pm 0.231 $  &  0.168 \\          
 B & $10.0 \leq M_{*} \leq 11.5$ (spec)  & $ -0.059 \pm 0.084 $  &  $ 1.241 \pm 0.842 $  &  0.158 \\          
 C & $10.5 \leq M_{*} \leq 11.5$              & $ -0.038 \pm 0.046 $  &  $ 0.622 \pm 0.258  $  & 0.133 \\           
 D & $10.5 \leq M_{*} \leq 11.5$ (spec)  & $  -0.055 \pm 0.086 $  &  $ 1.250 \pm 0.862  $  & 0.159 \\          
   \hline
  \\
  \multicolumn{5}{c}{Cl~0332-2742*} \\  
  \hline
  \hline
   Case &  Mass range &  $\alpha \pm \Delta \alpha$  & $\beta \pm \Delta \beta$ & $\epsilon$ \\
   \hline
 C & $10.5 \leq M_{*} \leq 11.5$              & $ 0.009  \pm  0.045$  &  $0.483 \pm 0.256$  &  0.112 \\          
 D & $10.5 \leq M_{*} \leq 11.5$ (spec)  & $ 0.008  \pm  0.102$  &  $0.754 \pm 0.614$  &  0.170 \\          
   \hline
   \hline
 \multicolumn{5}{p{.45\textwidth}}{\textsuperscript{*} Since all the selected galaxies in Cl~0332-2742 have $\log(M_{*}/M_\odot) \geq 10.5$, only case C \& D are applicable.} \\
\end{tabular}

\end{table}

\subsection{Stellar mass -- mass-weighted size relations}
\label{subsec:Stellar mass -- mass-weighted size relations}
With the mass-weighted sizes, we are able to derive the stellar mass -- mass-weighted size relations of the three clusters.  Here we investigate how using mass-weighted sizes can affect the mass -- size relations.

\begin{figure*}
  \centering
  \includegraphics[scale=0.80]{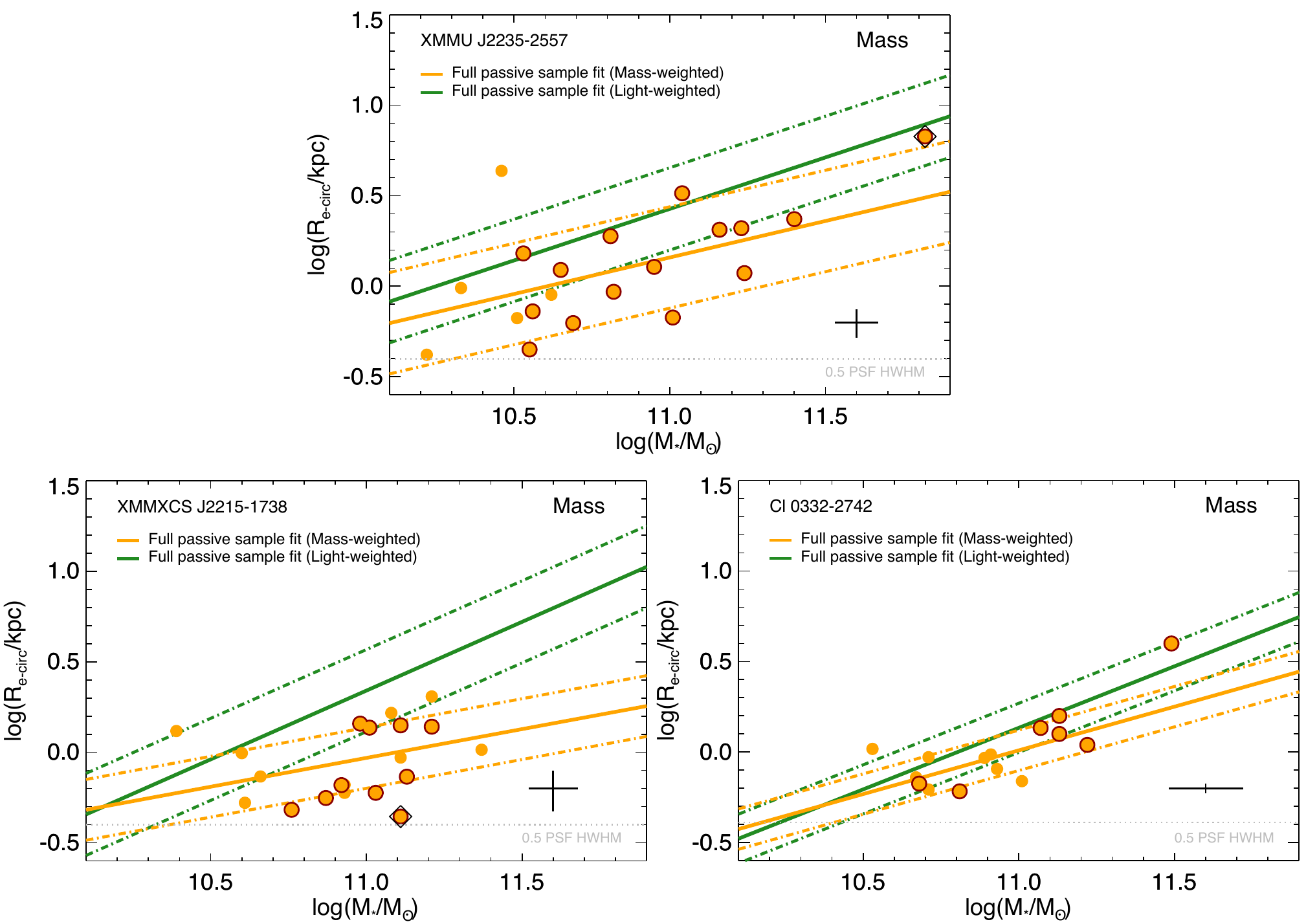}  
  \caption[Stellar mass -- mass-weighted size relations of the red sequence galaxies in the KCS clusters]{Stellar mass -- mass-weighted size relations of the red sequence galaxies in the KCS clusters.  The orange line in each panel is a linear fit to the full passive sample (Case A for XMMU~J2235-2557 and XMMXCS~J2215-1738, Case C for Cl~0332-2742),  while the dot-dashed lines represent $\pm 1\epsilon$ fitted intrinsic scatter.  The green line is the same mass -- light-weighted size relation fit in Figure~\ref{fig_light_masssize} for comparison.  Individual objects are shown in orange and spectroscopically confirmed objects are circled with dark red.  The BCGs in XMMU~J2235-2557 and XMMXCS~J2215-1738 are indicated with black diamonds.  The cross shows the typical uncertainty of the sizes and the median uncertainty of the integrated mass in our sample.}
  \label{fig_mass_masssize}
\end{figure*}

Figure~\ref{fig_mass_masssize} shows the stellar mass -- size relations of the clusters using mass-weighted size (hereafter mass-weighted relations).  The relations are fitted in the same way as the mass -- light-weighted size relations (hereafter light-weighted relations) using equation~\ref{eqt-fit}.  The results are summarised in Table~\ref{tab_bestfitmass}.  We have also fitted the relations using semi-major axes as galaxy sizes, the results are again presented in Appendix~\ref{The fitted parameters of mass -- size relations using semi-major axes}.  In Figure~\ref{fig_mass_masssize} we have also over-plotted the best-fit of the light-weighted relations in each panel for comparison.

The mass-weighted sizes are on average smaller than the light-weighted sizes, as is evident from comparing the intercepts of the fitted mass-weighted relations to those from the light-weighted relations.
 
Indeed for XMMU~J2235-2557, comparing the two sizes of each galaxy we find that the mass-weighted sizes are on average $\sim45\%$ smaller than the $H_{160}$ light-weighted sizes, with a median difference of $\langle \log(R_{e\rm{-circ},mass}/ R_{e\rm{-circ}})\rangle = -0.26$.  The 1$\sigma$ scatter $\sigma_{\log(R_{e\rm{-circ},mass}/ R_{e\rm{-circ}})} $ is $\sim0.11$.  

For XMMXCS~J2215-1738, the mass-weighted sizes are on average $\sim55\%$ smaller than the $H_{160}$ sizes, with a median difference of $\langle \log(R_{e\rm{-circ},mass}/ R_{e\rm{-circ}})\rangle = -0.34$ and a scatter of $\sigma_{\log(R_{e\rm{-circ},mass}/ R_{e\rm{-circ}})} \sim0.14$.  For some galaxies the mass-weighted size can be $\sim87\%$ smaller than its light counterpart.

For Cl~0332-2742, the mass-weighted sizes are similar to the $H_{160}$ sizes, with an average of only $\sim20\%$ decrease. The median difference is $\langle \log(R_{e\rm{-circ},mass}/ R_{e\rm{-circ}})\rangle = -0.095$,  much smaller than the other two clusters.  The scatter is also smaller, with $\sigma_{\log(R_{e\rm{-circ},mass}/ R_{e\rm{-circ}})} \sim0.065$.

The general trend of mass-weighted sizes being smaller than the light-weighted sizes suggests that the mass distribution is more concentrated than the light distribution.  The ${M_{*}/L}$ ratio at the inner part of the galaxy is hence higher compared to the outskirts, implying the existence of a ${M_{*}/L}$ gradient.  This trend of mass-weighted sizes being smaller is in qualitative agreement with \citet{Szomoruetal2013}, who computed mass-weighted sizes using 1D surface brightness profiles for passive field galaxies in CANDELS at a similar redshift.

On the other hand, the slope of the relations are consistent with the light-weighted relations, given the large uncertainties in the measured values.  For completeness, we measure a typical slope of $\beta = 0.49 \pm 0.13$ and $\alpha = 0.04 \pm 0.03$ if the entire massive red sequence sample (C) for all three cluster is fitted simultaneously.  At face value there might be a hint of a slight change in the slope if mass-weighted sizes are used (0.49 vs 0.79),  although at least part of it is due to the effect of the discarded objects.  Recall that we remove objects that have mass sizes smaller than the PSF size or problematic fits.  If we fit the \textit{light-weighted} relations for the entire mass range including only objects that have reliable mass-weighted sizes, this will give a slope of $\beta = 0.47 \pm 0.19$ (A) and $\beta = 0.61 \pm 0.23$ (D) for XMMU~J2235-2557, $\beta = 0.57 \pm 0.22$ (A) and $\beta = 1.35 \pm 0.84$ (B) for XMMXCS~J2215-1738, which slightly reduces the difference between the slopes of the mass-weighted relations to the light-weighted ones.

\section{Discussion}
\label{sec:Discussion}

\subsection{Environmental dependence of structural properties of massive passive galaxies}
\label{sec:Environmental dependence of photometric properties of massive passive galaxies at high redshift}
In this section, we compare the structural properties of the massive passive KCS galaxies ($\log(M_{*}/M_\odot) \geq 10.5$) to passive field galaxies at similar redshifts. 

Below we use the sample from \citet{Langetal2014} as our field comparison sample.  \citet{Langetal2014} derived both $H_{160}$ light-weighted and mass-weighted structural properties for a mass-selected sample ($\log(M_{*}/M_\odot) > 10$) spanning a redshift range $0.5< z<2.5$ in all five CANDELS fields.

We select a subsample of massive passive galaxies with the same mass cut ($\log(M_{*}/M_\odot) \geq 10.5$) from the \citet{Langetal2014} sample following the $UVJ$ passive criteria to match the KCS sample (hereafter L14 field sample).  Problematic objects with mass-weighted or light-weighted structural parameters that hit the boundary of the allowed ranges (e.g. $n = 8.0$) are removed from the sample.  We also noticed and removed an excess of $UVJ$ passive objects with extremely small ($q_{mass} < 0.1$) mass-weighted axis ratios, which are not present in the light-weighted axis ratio distributions or the KCS sample. Since Cl~0332-2742 is in GOODS-S, we have also removed our cluster galaxies in Cl~0332-2742 from the L14 field sample.  A total of 1055 objects are selected, among them 226 are in the redshift range comparable to the three KCS clusters ($1.3<z<1.7$).

\subsubsection{Size distributions in different environments}
\label{subsec:Size distributions in different environment}
We first compare the size distributions in different environments.  As we discussed in the introduction, recent works have found differences between the size distributions of massive passive galaxies in clusters and the field at high redshift, although the extent is still under debate \citep[e.g.][]{Cooperetal2012, Zirmetal2012, Papovichetal2012, Lanietal2013, Strazzulloetal2013, Jorgensenetal2013, Delayeetal2014}. On the other hand, no such difference can be seen in the local universe \citep[e.g.][]{Maltbyetal2010, Huertascompanyetal2013, Cappellari2013}.

Similar to Section~\ref{sec:Results}, we have derived the size distributions using both circularised effective radii ($R_{e\rm{-circ}}$) and elliptical semi-major axes ($a_{e}$) as galaxy sizes.  Here we will first present the result of using circularised effective radii, followed by the one using semi-major axes.

We note that, between different studies differences on the order of $\sim10\%$ are difficult to consistently reproduce \citep{Cappellarietal2013a}, hence one must first ensure the sizes in both samples are comparable. Through a direct comparison with our derived sizes, we found that the light-weighted and mass-weighted sizes from both samples are highly consistent (see Appendix~\ref{Mass-weighted size comparison with Lang et al. 2014} for an example of the comparison), although different methods have been used.

To compare the observed size distributions, we first mass-normalise the measured sizes ($R_{e\rm{-circ},\rm{MN}}$ or $a_{e,\rm{MN}}$), following the definition in \citet{Newmanetal2012} and \citet{Delayeetal2014}. For the case of $R_{e\rm{-circ},\rm{MN}}$, it is defined as:
\begin{equation}
  \label{eqt-MN}
  R_{e\rm{-circ},\rm{MN}} = R_{e\rm{-circ}}/ (M_{*} /10^{11} M_\odot)^{\beta}
\end{equation}
where $\beta$ is the slope of the mass-size relation and $R_{e\rm{-circ},\rm{MN}}$ is the mass-normalised size at $\log(M_{*}/M_\odot) =11.0$.  Using mass-normalised sizes removes the correlation between stellar mass and size, and hence allows us to compare the size distribution of two samples that do not share the same mass distribution. We compute both the light-weighted and mass-weighted mass-normalised size distributions of each cluster using the best-fit slope (case C) of the mass -- size relations in Section~\ref{subsec:Stellar mass -- light-weighted size relations} and~\ref{subsec:Stellar mass -- mass-weighted size relations} (see Table~\ref{tab_bestfit} and \ref{tab_bestfitmass} for the light-weighted and mass-weighted slopes).  For the case of $a_{e,\rm{MN}}$ we replace $R_{e\rm{-circ}}$ in Equation~\ref{eqt-MN} with $a_e$ and use the slopes of the mass-size relation derived using $a_e$ for the normalisation.

\begin{figure*}
  \centering
  \includegraphics[scale=0.425]{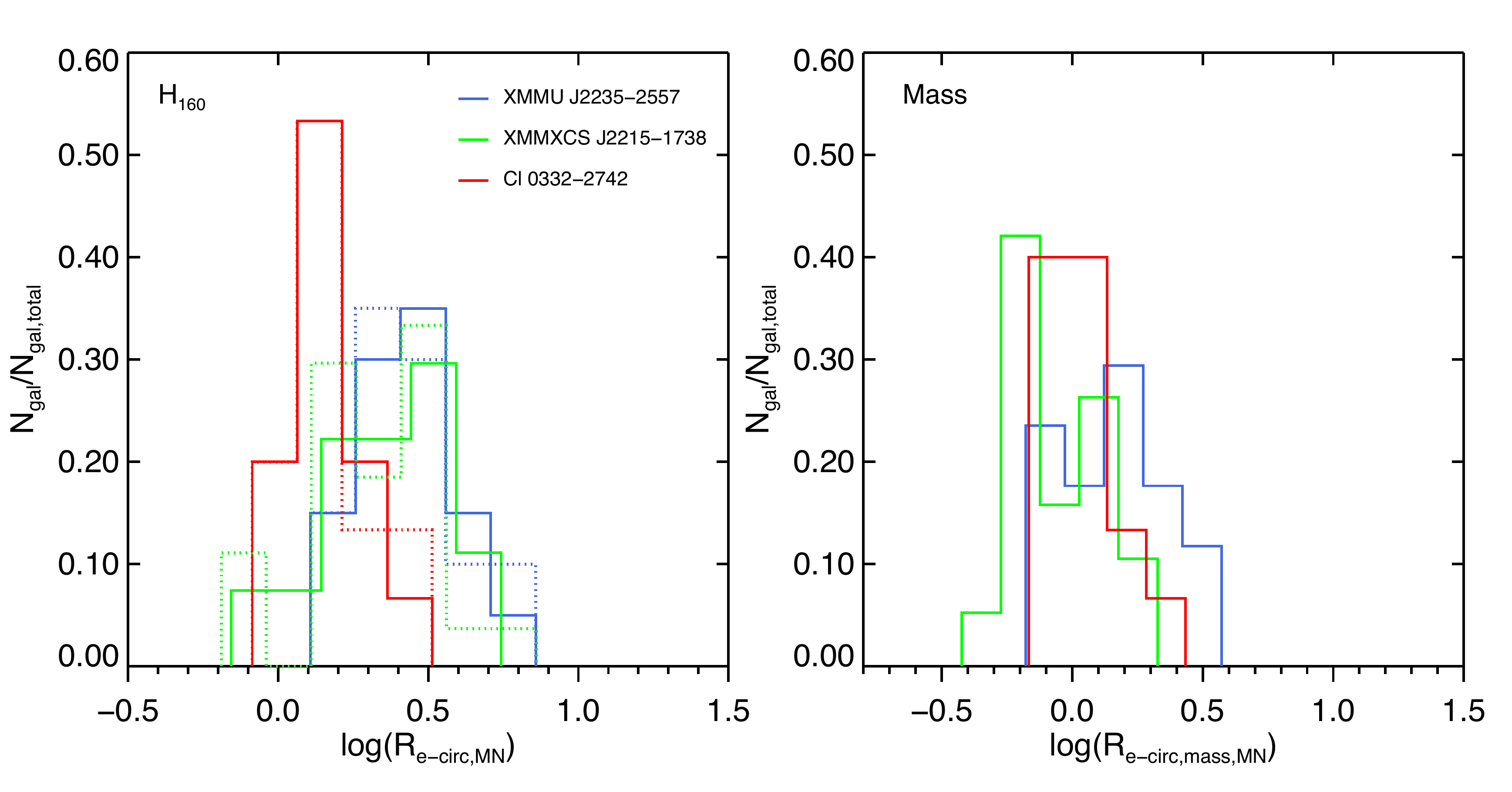}   
  \caption[Mass-normalised size distributions of the KCS clusters]{Mass-normalised size distributions of the KCS clusters.  Left:  The light-weighted mass-normalised size distributions.  Distribution of XMMU~J2235-2557 is shown in blue, the one of XMMXCS~J2215-1738 is shown in green and the one of Cl~0332-2742 is shown in red, respectively.  The solid histograms are computed with the slope ($\beta$) of the fitted mass -- light-weighted size relation of each cluster, while the dotted histograms are computed with the slope adopted from \citet{vanderWeletal2014} for all three clusters ($\beta = 0.76$).  Right:  The mass-weighted mass-normalised size distributions.  The histogram of each cluster is computed with the slope ($\beta$) of the fitted mass -- mass-weighted size relation respectively.} 
  \label{field_histo_kcsonly}
\end{figure*}

Figure~\ref{field_histo_kcsonly} shows the mass-normalised circularised effective radius distributions of the three KCS clusters for galaxies with $\log(M_{*}/M_\odot) \geq 10.5$.  XMMU~J2235-2557 and XMMXCS~J2215-1738 show comparable light-weighted mass-normalised size distributions.  On the other hand,  the size distribution in the higher redshift cluster Cl~0332-2742 is distinct from the other two, as discussed in Section~\ref{subsec:Stellar mass -- light-weighted size relations}. We checked that this difference is not due to the applied best-fit slope.  In the left panel of Figure~\ref{field_histo_kcsonly} we show also the size distributions computed using the slope from \citet{vanderWeletal2014} ($\beta = 0.76$) as dotted histograms, which are very similar to the ones computed with the best-fit slope.   
On the right panel of Figure~\ref{field_histo_kcsonly} we show the mass-weighted mass-normalised size distributions of the three clusters.  The differences between Cl~0332-2742 and XMMU~J2235-2557/XMMXCS~J2215-1738 seem to be reduced.  The median of all three distributions are consistent within the errors.

We then compare our sample to the L14 field sample.  The disparity between the Cl~0332-2742 and other two clusters in properties and redshift can make a cluster - field comparison of the whole redshift range problematic.   Hence we split the cluster - field comparison into two redshift ranges:   XMMU~J2235-2557 \& XMMXCS~J2215-1738 with the L14 field subsample at $1.3 < z < 1.5$,  and Cl~0332-2742 alone with the L14 field subsample at $1.5 < z < 1.7$.

We follow \citet{Newmanetal2012} to fit each size distribution with a skew normal distribution, which takes into account the asymmetry in the size distributions to estimate the mean sizes of the distributions:


\begin{equation}
  \label{eqt-SN}
  P(\log(R_{e\rm{-circ},\rm{MN}}), s, \phi, \omega) =  \frac{1}{\omega\pi} e^{-\frac{\kappa^2}{2}}  \int_{-\infty}^{s \kappa} e^{-\frac{t^2}{2}} dt
\end{equation}
where $\kappa = \frac{\log(R_{e\rm{-circ},\rm{MN}})-\phi}{\omega} $.  

The mean of the best-fit distribution $\overline{\log(R_{e\rm{-circ},\rm{MN}})}$ is given by $\overline{\log(R_{e\rm{-circ},\rm{MN}})} = \phi + \omega (s/ \sqrt{1+s^2}) \sqrt{2/\pi}$, and $s$ is the `shape' parameter that governs the skewness.  We have performed the same fitting to the $a_e,\rm{MN}$ distributions.   The mean of the best-fit distributions ($\overline{\log(R_{e\rm{-circ},\rm{MN}})}$ and $\overline{\log(a_{e,\rm{MN}})}$) and the median of the original (not fitted) distributions ($\langle \log(R_{e\rm{-circ},\rm{MN}}) \rangle$ and $\langle \log(a_{e,\rm{MN}}) \rangle$) discussed below are given in Table~\ref{tab_sizeMN}. To evaluate the fits we have applied both Kolmogorov-Smirnov (KS) and Anderson-Darling (AD) tests on the results, with the null hypothesis that they come from a common distribution.  The resulting $p$-values are also given in Table~\ref{tab_sizeMN}.  We found that in some cases, especially for the mass-weighted size distributions in the field, both the KS and AD tests indicate they are not good fits, which are probably due to the double-peaked features or excesses at large or small sizes.  We have excluded those fits from Table~\ref{tab_sizeMN}.  Due to low number statistics we will only compare the mean and median of the size distributions later on.

\begin{figure*}
  \centering
  \includegraphics[scale=0.425]{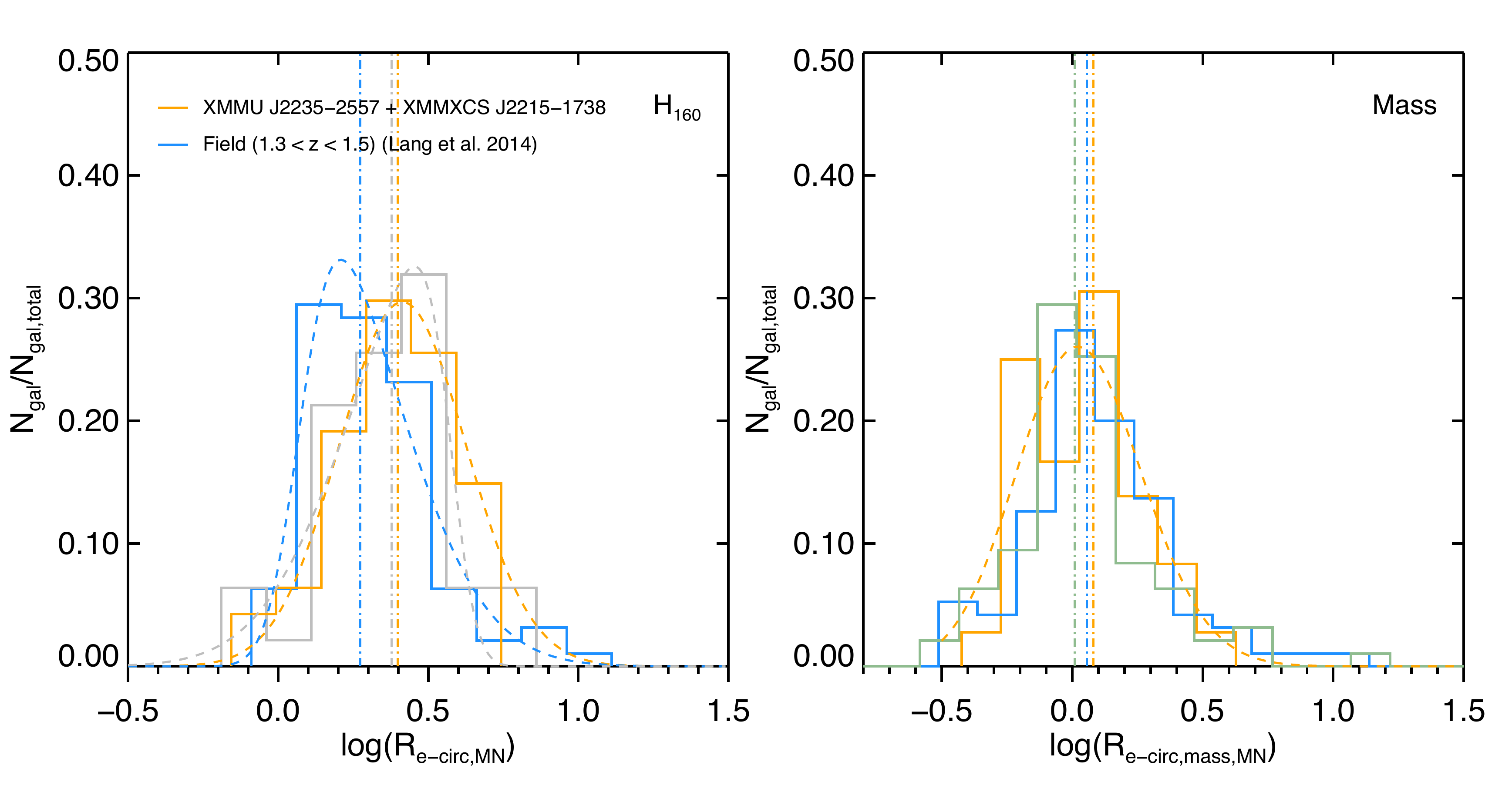} 
  \caption[Comparison of the mass-normalised size distributions of XMMU~J2235-2557 and XMMXCS~J2215-1738 with the field]{Comparison of the mass-normalised size distributions of XMMU~J2235-2557 and XMMXCS~J2215-1738 with the field.  Left:  The light-weighted mass-normalised size distributions.  The combined size distribution of XMMU~J2235-2557 $+$ XMMXCS~J2215-1738 is shown in orange.  The size distribution of the L14 field sample with a redshift range of $1.3<z<1.5$ is shown in blue.  The grey histogram is the size distribution of XMMU~J2235-2557 $+$ XMMXCS~J2215-1738 computed with the slope adopted from \citet{vanderWeletal2014}.   Right:  The mass-weighted mass-normalised size distributions.  The blue histogram shows the size distribution of the L14 field sample computed with an assumed slope of $\beta = 0.76$, identical to the mass -- light-weighted size relations.  The light green histogram shows the size distribution with an assumed slope identical to the mass -- mass-weighted size relations of the clusters ($\beta=0.49$).  The colored dash-dotted lines show the median sizes for each size distribution, while the dashed lines show the best-fit skew normal distributions respectively.} 
  \label{field_histo_xmm}
\end{figure*}

Figure~\ref{field_histo_xmm} shows the comparison of the combined mass-normalised circularised effective radius distributions of XMMU~J2235-2557 \& XMMXCS~J2215-1738 with the L14 field sample at $1.3 < z < 1.5$.  The light-weighted distributions of the field sample are computed with the slope from \citet{vanderWeletal2014} ($\beta = 0.76$).  Although this slope is computed with passive galaxies in a wider redshift bin of $1.0 < z < 1.5$, the slope of the mass -- light-weighted size relations in the field is found to be an invariant with redshift \citep{vanderWeletal2014}.  We also plot the size distributions of the clusters computed using this slope in grey for illustrative purposes.

From Figure~\ref{field_histo_xmm}, it is clear that the mode of the light-weighted size distribution of the two clusters is offset to larger sizes compared to the field.  The median as well as the mean of the best-fit distribution of the clusters is larger than those of the field, suggesting the median sizes in the clusters is $\sim33\%$ larger than the field ($\sim30\%$ from the best-fit mean).  Using $\beta = 0.76$ instead of the best-fit slope would give a consistent $\overline{\log(R_{e\rm{-circ},\rm{MN}})}$ to the field, although the difference in the median remains unchanged.  Contrary to \citet{Delayeetal2014}, we do not see a tail of large-size cluster galaxies in the distribution with respect to the field.  This may be due to the small sample that we have (a total of 47 galaxies in XMMU~J2235-2557 and XMMXCS~J2215-1738) or the fact that our color-color selections remove dusty star-forming galaxies, which would predominantly have large sizes.   

On the right panel of Figure~\ref{field_histo_xmm} we show the comparison of the mass-weighted circularised effective radius distribution of clusters and the field.  Since there is no available estimate of the slope of the mass -- mass-weighted size relations in the field,  we assume two different slopes:  a) same slope as the light-weighted relation ($\beta = 0.76$) and b) same slope as we found in the clusters ($\beta = 0.49$).  The two cases are shown as blue and light green, respectively.  

We found that the difference between the size distributions in clusters and the field is reduced when mass-weighted sizes are used, independent of the assumed value of $\beta$.  As an additional check, we use the KS and AD tests to evaluate whether the size distributions in clusters and field are different.  The results are given in Table~\ref{tab_sizeKS}.  For the light-weighted size distributions,  we see mild significance from the $p$-values derived from the KS and AD tests to reject the null hypothesis that they come from the same distribution.  The $p$-value of the KS test for the light-weighted size distributions of the two clusters and the field is $0.02$ ($p \simeq 0.01$ for $\beta=0.76$).  Similar values are also seen for the AD tests. While for the mass-weighted size distributions this is not true,  we derive a $p$-value of $0.87$ for the mass-weighted size distribution ($p \simeq 0.61$ for $\beta =0.49$).

\begin{figure*}
  \centering
  \includegraphics[scale=0.425]{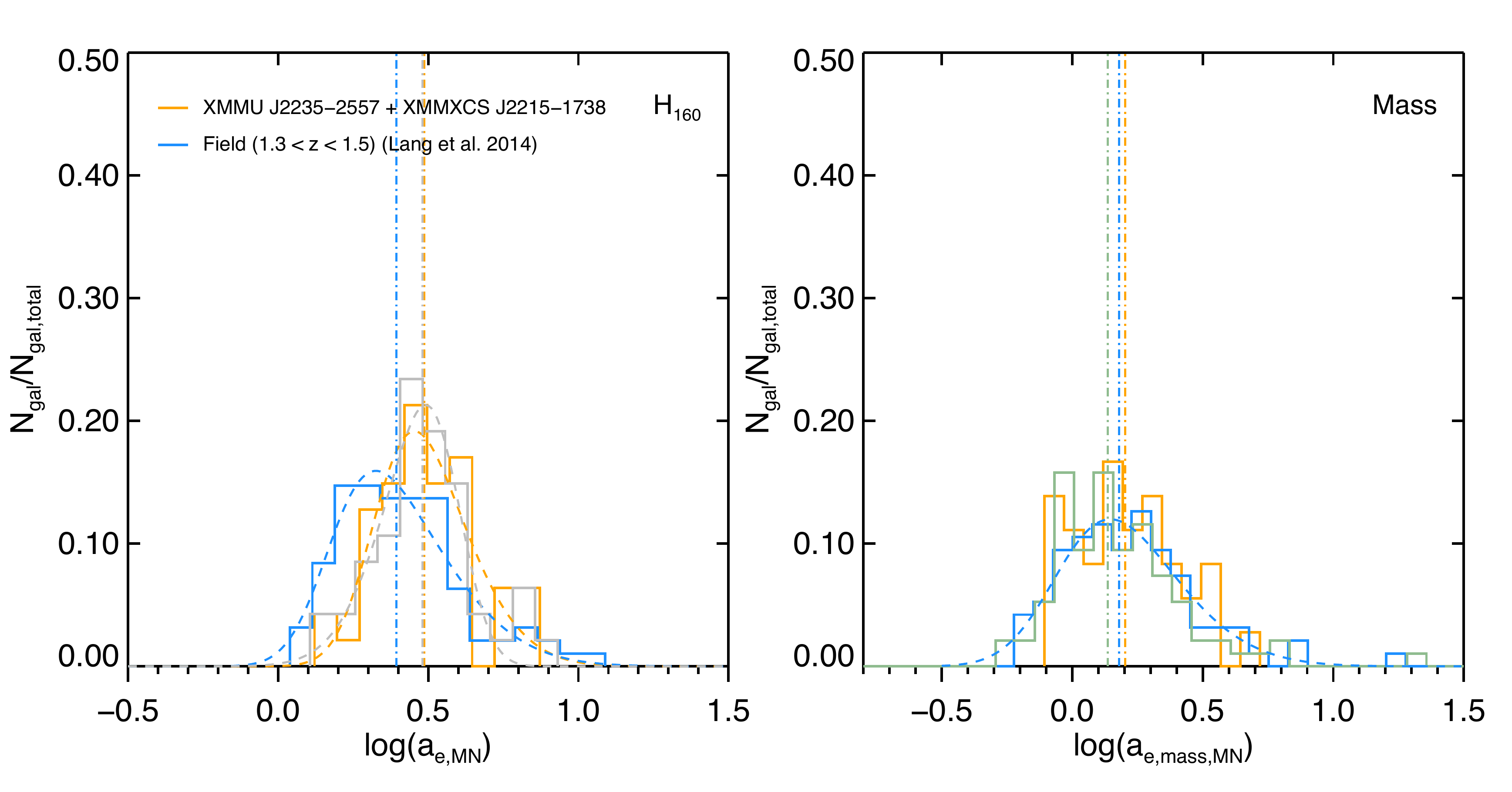}
  \caption[Comparison of the mass-normalised size distributions of XMMU~J2235-2557 and XMMXCS~J2215-1738 with the field (SMA)]{Comparison of the mass-normalised semi-major axis distributions of XMMU~J2235-2557 and XMMXCS~J2215-1738 with the field.  Same as Figure~\ref{field_histo_xmm}, but using semi-major axis ($a_e$) as galaxy size instead of the circularized radius.}  
  \label{field_histo_xmm_sma}
\end{figure*}

Using $a_{e,\rm{MN}}$ instead of $R_{e\rm{-circ},\rm{MN}}$ shows similar results. Figure~\ref{field_histo_xmm_sma} shows the comparison of the combined mass-normalised semi-major axis distributions of XMMU J2235-2557 \& XMMXCS J2215-1738 with the L14 field sample at $1.3 < z < 1.5$.  The mode, median as well as the mean of the best-fit light-weighted size distribution of the clusters are also larger than those of the field, albeit with a smaller difference.  The median sizes in the clusters is $\sim24\%$ larger than the field ($\sim25\%$ from the best-fit mean). The KS and AD test also result in low $p$-values to reject the null hypothesis that they come from the same distribution.  Again, we see that the difference is reduced when mass-weighted sizes are used. The KS and AD tests show large $p$-values for the mass-weighted $a_{e,\rm{MN}}$  distributions.

\begin{figure*}
  \centering
  \includegraphics[scale=0.425]{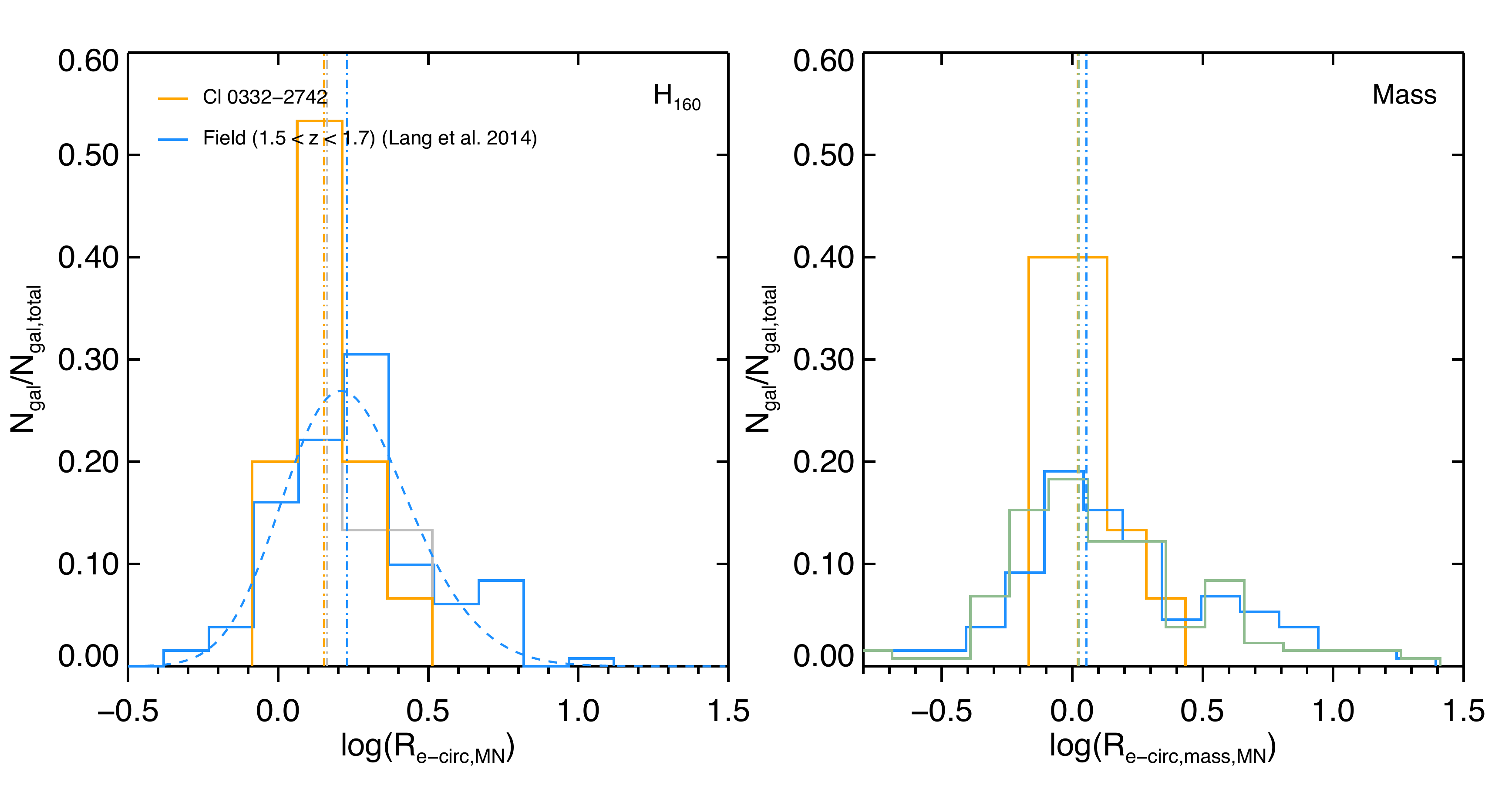}
  \caption[Comparison of the mass-normalised size distributions of Cl~0332-2742 with the field]{Comparison of the mass-normalised size distributions of Cl~0332-2742 with the field.  Left:  The light-weighted mass-normalised size distributions.  The size distribution of Cl~0332-2742 is shown in orange.  The size distribution of the L14 field sample with a redshift range of $1.5<z<1.7$ is shown in blue.  The grey histogram is the size distribution of Cl~0332-2742 computed with the slope adopted from \citet{vanderWeletal2014}.   Right:  The mass-weighted mass-normalised size distributions.  The blue histogram shows the size distribution of the L14 field sample computed with an assumed slope of $\beta = 0.76$, identical to the mass -- light-weighted size relations.  The light green histogram shows the size distribution with an assumed slope identical to the mass -- mass-weighted size relations of the clusters ($\beta=0.49$).  The colored dash-dotted lines show the median sizes for each size distribution, while the dashed lines show the best-fit skew normal distributions respectively.} 
  \label{field_histo_cl}
\end{figure*}

Figure~\ref{field_histo_cl} shows the comparison of the higher redshift cluster, Cl~0332-2742, with the L14 field sample at $1.5< z < 1.7$.  We did not attempt to fit a skew normal distribution for Cl~0332-2742 due to the small number of galaxies.  Comparing the median sizes of the distributions, passive galaxies in Cl~0332-2742 seem to have comparable if not smaller sizes compared to the field galaxies.  This is true for both light-weighted and mass-weighted sizes.  The KS test presents a small value of $p \simeq 0.03$ for the light-weighted distributions of Cl~0332-2742 and the field, but not for the AD test ($p_{AD} \simeq 0.08$) or if $\beta=0.76$ is used ($p_{KS} \simeq 0.11$).  Both the KS and AD tests do not present a small $p$-value for the mass-weighted size distributions.

\begin{figure*}
  \centering
  \includegraphics[scale=0.425]{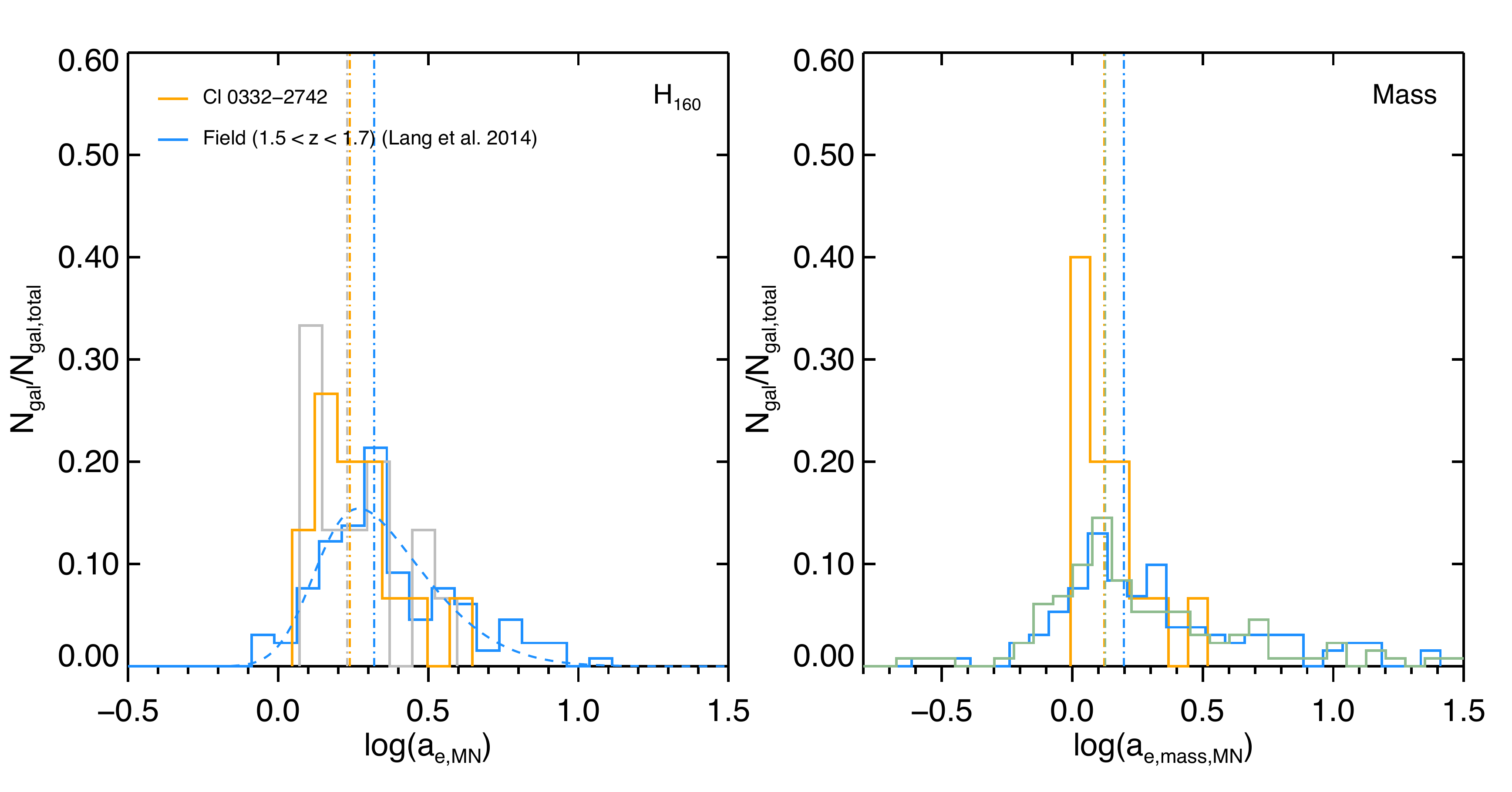}
  \caption[Comparison of the mass-normalised size distributions of Cl~0332-2742 with the field (SMA)]{Comparison of the mass-normalised semi-major axis distributions of Cl~0332-2742 with the field.  Same as Figure~\ref{field_histo_cl}, but using semi-major axis ($a_e$) as galaxy size instead of the circularized radius.}  
  \label{field_histo_cl_sma}
\end{figure*}

Using $a_{e,\rm{MN}}$ gives consistent results as $R_{e\rm{-circ},\rm{MN}}$ for Cl~0332-2742. The comparison of the combined mass-normalised semi-major axis distributions of Cl~0332-2742 with the L14 field sample at $1.5 < z < 1.7$ is shown in Figure~\ref{field_histo_cl_sma}.  The KS and AD tests do not suggest that the size distributions of the Cl~0332-2742 and the field are distinct.

An environmental difference between the size of the galaxies in clusters and the field can be regarded as a supporting evidence for the minor merger scenario \citep[e.g.][]{Cooperetal2012, Strazzulloetal2013}. It is interesting that we see larger median light-weighted sizes in XMMU~J2235-2557 and XMMXCS~J2215-1738 than in the field, but not in Cl~0332-2742.  In Section~\ref{subsec:Minor mergers and the effect from the environment}, we will discuss this further and explore possible implications together with other results.

\begin{table*}
  \caption{The mean (of the best-fit skew normal distributions) and the median of the mass-normalised size distributions of the KCS clusters and the L14 field sample}
  \centering
  \resizebox{2.1\columnwidth}{!}{%
  \label{tab_sizeMN}
  \begin{tabular}{@{}p{4.37cm}ccccccccc@{}}
  \\
  \hline
  \hline
 
  \multicolumn{10}{c}{Light-weighted size distributions} \\
  Sample               &          $N_{\rm{gal,total}}$    &  $\overline{\log(R_{e\rm{-circ},\rm{MN}})}$\textsuperscript{*}            &                  $\langle \log(R_{e\rm{-circ},\rm{MN}}) \rangle$       & $p_{\rm{KS},R_{e\rm{-circ}}}$  &   $p_{\rm{AD},R_{e\rm{-circ}}}$\textsuperscript{+}  &  $\overline{\log(a_{e,\rm{MN}})}$\textsuperscript{*}               &                  $\langle \log(a_{e,\rm{MN}}) \rangle$            & $p_{\rm{KS},a_e}$      &   $p_{\rm{AD},a_e}$\textsuperscript{+}  \\
  \hline
  XMMU~J2235 +  XCS~J2215                         &    47  &$0.416 \pm 0.055$  &   $0.398 \pm 0.037$  &   0.31  & 0.53  &$0.493 \pm 0.070$  &   $0.486 \pm 0.031$  &  0.45    &  0.86  \\   
  XMMU~J2235 +  XCS~J2215 ($\beta=0.76$) &    47  &$0.332 \pm 0.054$ & $0.401 \pm 0.036$ &   0.33 &  0.14    & $0.457 \pm 0.047$  &  $0.480 \pm 0.031$   &  0.70   &  0.06 \\  
  L14 field ($1.3 < z < 1.5$)                                &   95  & $0.303 \pm 0.032$ &  $0.274 \pm 0.025$ &   0.87 &  0.87   & $0.396 \pm 0.038$  &  $0.394 \pm 0.026$   & 0.96   &  0.79 \\   
  Cl~0332-2742                                                  &    15  &     -    &   $0.152 \pm 0.051$  &   -  &  -     &        -                         &   $0.238 \pm 0.052$  &  -  &  -   \\  
  Cl~0332-2742  ($\beta=0.76$)                         &     15  &    -    &   $0.161 \pm 0.052$   &     -  &  -   &        -                         &   $0.229 \pm 0.053$  &  -  &  - \\  
  L14 field ($1.5 < z < 1.7$)                                 &   131 & $0.243 \pm 0.038$   &   $0.230 \pm 0.027$   &  0.61 &  0.27   &   $0.347 \pm 0.031$   &   $0.320 \pm 0.025$  &  0.43  & 0.18  \\  
  3 KCS clusters                                                  &    62   &$0.351 \pm 0.077$   &  $0.348 \pm 0.036$  &   0.24 & 0.50    &   $0.392 \pm 0.079$   &  $0.448 \pm 0.032$  & 0.53    &  0.22\\  
  3 KCS clusters  ($\beta=0.76$)                         &    62   & $0.240 \pm 0.085$   &  $0.359 \pm 0.035$  &   0.14 & 0.03  &   $0.360 \pm 0.080$   &  $0.457 \pm 0.313$  & 0.13   &  0.004  \\  
   L14 field ($1.3 < z < 1.7$)                                 &    226  & $0.268 \pm 0.012$   &   $0.255 \pm 0.019$  &    0.56 & 0.17   &   $0.376 \pm 0.023$   &   $0.347 \pm 0.018$  & 0.49   &  0.16  \\  
  \hline
  \hline
  \\
   \hline
  \hline
  \multicolumn{10}{c}{Mass-weighted size distributions} \\  
  Sample               &          $N_{\rm{gal,total}}$    &  $\overline{\log(R_{e\rm{-circ},\rm{MN}})}$                 &                  $\langle \log(R_{e\rm{-circ},\rm{MN}}) \rangle$       & $p_{\rm{KS},R_{e\rm{-circ}}}$  &   $p_{\rm{AD},R_{e\rm{-circ}}}$ &  $\overline{\log(a_{e,\rm{MN}})}$               &                  $\langle \log(a_{e,\rm{MN}}) \rangle$            & $p_{\rm{KS},a_e}$      &   $p_{\rm{AD},a_e}$  \\
  \hline
  XMMU~J2235 +  XCS~J2215                       &    36   &   $0.033 \pm 0.089$  &  $0.081  \pm 0.046$    & 0.65 & 0.75    &       -                          &  $0.203  \pm 0.044$ &  -  &   -  \\  
  L14 field ($1.3 < z < 1.5$) ($\beta=0.76$)     &    95   &  -  &  $0.056  \pm 0.034$   & - &  -  &   $0.200 \pm 0.026$  &  $0.179  \pm 0.033$ & 0.95   &  0.22 \\  
  L14 field ($1.3 < z < 1.5$) ($\beta=0.49$)     &    95   &  - &   $0.010  \pm 0.031$     & - & -  &  $0.189 \pm 0.030$ &   $0.136  \pm 0.030$ & -   &  -   \\ 
  Cl~0332-2742                                                &    15   &   -   &   $0.023 \pm 0.044$    &  - &  -  &       -                          &   $0.124 \pm 0.025$  &  -  &   -   \\  
  L14 field ($1.5 < z < 1.7$) ($\beta=0.76$)     &  131    &  - & $0.054 \pm 0.052$  & - &  - &  -   &   $0.198 \pm 0.046$  &  -   &  -    \\
  L14 field ($1.5 < z < 1.7$) ($\beta=0.49$)     &  131    &  - & $0.024 \pm 0.047$  & - &  - &  -   &   $0.127 \pm 0.044$  &  -  &  -  \\  
  3 KCS clusters                                               &  51   &  $0.031 \pm 0.059$ &  $0.037 \pm 0.034$  & 0.94  & 0.86 &  $0.159 \pm 0.121$    &  $0.175 \pm 0.034$  &  0.50  &  0.26   \\  
  L14 field ($1.3 < z < 1.7$)  ($\beta=0.76$)    & 226   &   -  & $0.056 \pm 0.030$  & - & - &   -   & $0.175 \pm 0.034$   &  -  & -    \\  
  L14 field ($1.3 < z < 1.7$)  ($\beta=0.49$)    & 226   &   - & $0.018 \pm 0.028$   & - & - &  -  & $0.181 \pm 0.028$   & -   & -       \\  
  \hline
  \hline
  \multicolumn{10}{p{1.0\textwidth}}{\textsuperscript{*}The uncertainties quoted for $\overline{\log(R_{e\rm{-circ},\rm{MN}})}$ and $\overline{\log(a_{e,\rm{MN}})}$ are computed by bootstrapping.  We repeated the fitting procedure for 1000 times each with a randomly drawn subset of the sample, and the uncertainty is given by the standard deviation of these 1000 measurements.  The uncertainties of the median are estimated as $1.253\sigma/\sqrt{N_{\rm{gal,total}}}$, where $\sigma$ is the standard deviation of the size distributions.} \\
\end{tabular}
}
\end{table*}

\begin{table*}
  \caption{The results of the Kolmogorov-Smirnov and Anderson-Darling test on the size distributions of the KCS clusters and the L14 field sample}
  \centering
  \begin{tabular}{@{}lcccc@{}}
  \label{tab_sizeKS}
  \\
  \hline
  \hline
  \multicolumn{5}{c}{Light-weighted size distributions} \\
  Sample               &          $p_{\rm{KS},R_{e\rm{-circ}}}$  &   $p_{\rm{AD},R_{e\rm{-circ}}}$      & $p_{\rm{KS},a_e}$      &   $p_{\rm{AD},a_e}$                \\
  \hline
  XMMU~J2235 +  XMMXCS~J2215 vs. L14 field\textsuperscript{a}                                        &   0.02   &  0.01    &  0.01  &  0.01    \\  
  XMMU~J2235 +  XMMXCS~J2215 ($\beta=0.76$) vs. L14 field\textsuperscript{a}                &   0.01   &  0.02   &  0.01  &  0.01    \\   
  Cl~0332-2742   vs. L14 field\textsuperscript{b}                                                                       &    0.03   &  0.10   &  0.20  &   0.08  \\   
  Cl~0332-2742  ($\beta=0.76$)   vs. L14 field\textsuperscript{b}                                              &   0.11    &  0.15    &  0.30  &  0.11   \\  
  3 KCS clusters   vs.    L14 field\textsuperscript{c}                                                                    &   0.02    &  0.08    &  0.02 &   0.04   \\  
  3 KCS clusters  ($\beta=0.76$)  vs. L14 field\textsuperscript{c}                                               &   0.01    &  0.09    &  0.02 &  0.03   \\   
  \hline
  \hline
  \\  
   \hline
  \hline
  \multicolumn{5}{c}{Mass-weighted size distributions} \\  
 Sample               &          $p_{\rm{KS},R_{e\rm{-circ}}}$  &   $p_{\rm{AD},R_{e\rm{-circ}}}$      & $p_{\rm{KS},a_e}$      &   $p_{\rm{AD},a_e}$                \\
  \hline
   XMMU~J2235 +  XMMXCS~J2215 vs. L14 field\textsuperscript{a} ($\beta=0.76$)                  &  0.87   & 0.56   &  0.87  &  0.64  \\   
  XMMU~J2235 +  XMMXCS~J2215 vs. L14 field\textsuperscript{a} ($\beta=0.49$)                   &  0.61   & 0.46   &  0.54  &  0.23  \\    
  Cl~0332-2742  vs. L14 field\textsuperscript{b} ($\beta=0.76$)                                                   &  0.12   &  0.07  &  0.13  &  0.09  \\  
  Cl~0332-2742  vs. L14 field\textsuperscript{b} ($\beta=0.49$)                                                   &  0.15   &  0.09  &  0.22  &  0.15  \\  
  3 KCS clusters  vs. L14 field\textsuperscript{c}  ($\beta=0.76$)                                                 &  0.14   &  0.04  &  0.28  &  0.06   \\   
  3 KCS clusters  vs. L14 field\textsuperscript{c}  ($\beta=0.49$)                                                 &  0.12   &  0.06  &  0.15  &  0.05   \\  
  \hline
  \hline
   \multicolumn{4}{p{.475\textwidth}}{\textsuperscript{a}L14 field sample at $1.3 < z < 1.5$. \textsuperscript{b}L14 field sample at $1.5 < z < 1.7$. } \\
   \multicolumn{4}{p{.475\textwidth}}{\textsuperscript{c}L14 field sample at $1.3 < z < 1.7$, corresponds to all three KCS clusters.} \\
   \\
\end{tabular}
\end{table*}

\subsubsection{S\'ersic index distributions in different environments}
\label{subsec:n distributions in different environment}
In this section we compare the S\'ersic index distribution of the KCS sample to the L14 field sample.  

\begin{figure*}
  \centering
  \includegraphics[scale=0.465]{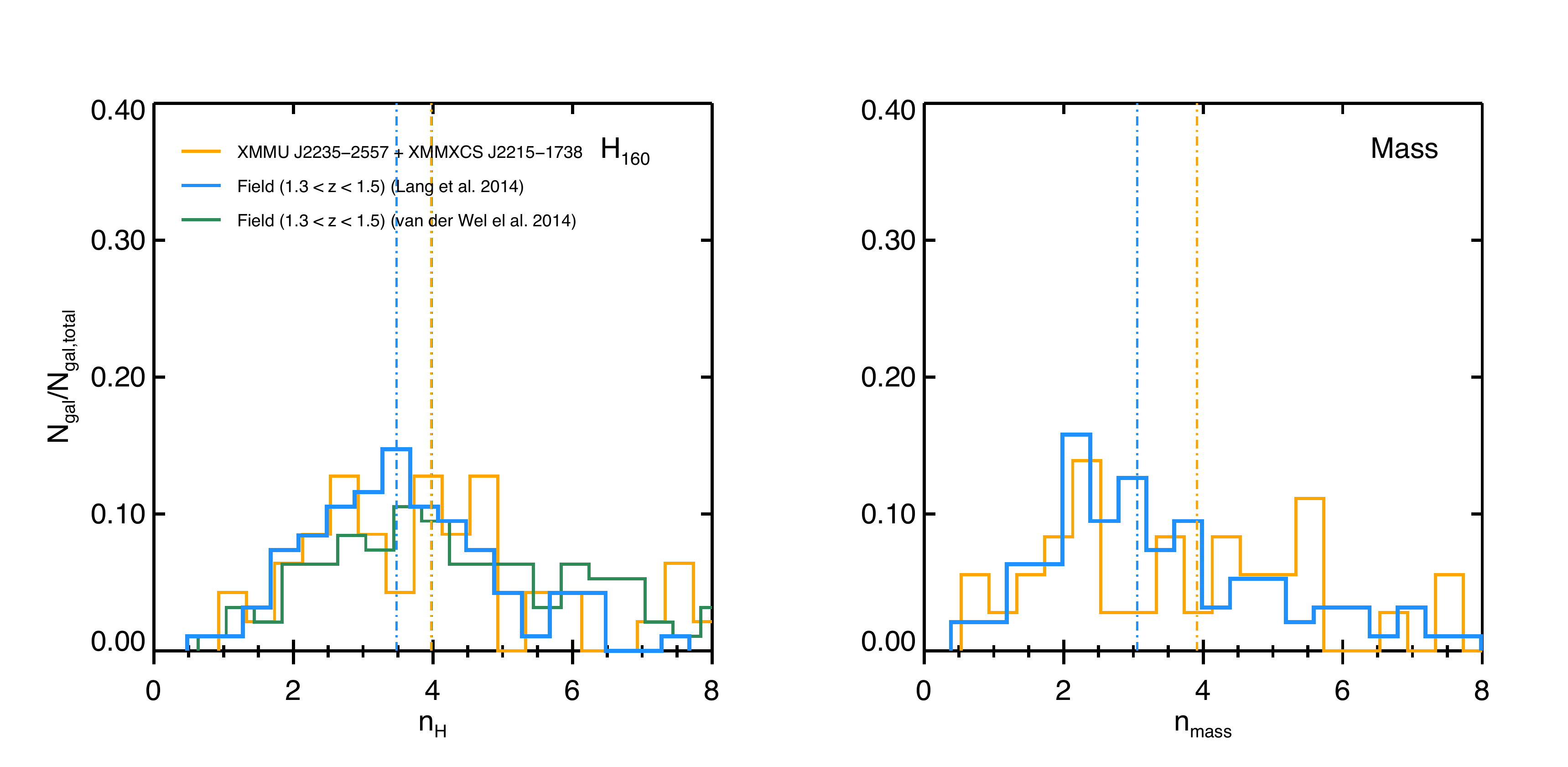}
  \caption[Comparison of the S\'ersic index distributions of XMMU~J2235-2557 and XMMXCS~J2215-1738 with the field]{Comparison of the S\'ersic index distributions  of XMMU~J2235-2557 and XMMXCS~J2215-1738 with the field.  Left:  The light-weight S\'ersic index distributions.  The combined S\'ersic index distribution of XMMU~J2235-2557 $+$ XMMXCS~J2215-1738 is shown in orange.  The S\'ersic index distribution of the L14 field sample with a redshift range of $1.3<z<1.5$ is shown in blue.  The green histogram shows the S\'ersic index distribution of the \citet{vanderWeletal2014} sample with the same redshift and $UVJ$ selection.   Right:  The mass-weighted S\'ersic index distributions.} 
  \label{field_histo_xmm_n}
\end{figure*}

Figure~\ref{field_histo_xmm_n} shows the comparison of the combined light-weighted and mass-weighted S\'ersic index distributions of XMMU~J2235-2557 \& XMMXCS~J2215-1738 with the L14 field sample at $1.3 < z < 1.5$.  The median S\'ersic index of the combined sample is $n=3.98$ ($n=4.01$ for XMMU~J2235-2557, $n=3.45$ for XMMXCS~J2215-1738).  
We found that the L14 sample at this redshift range seems to show a lower median ($n=3.45$) compared to the cluster sample.  Nevertheless, applying the same $UVJ$ selection on the \citet{vanderWeletal2014} sample gives a median of $n=3.98$, which perhaps reflects the large uncertainty of the S\'ersic index measurements.  Similar to the size distributions, we have performed KS and AD tests to evaluate whether the distributions are different.  Overall we find no evidence that the light-weighted S\'ersic index distributions of the clusters are distinct from the field (see Table~\ref{tab_nKS}).

Similarly we found no evidence that the mass-weighted S\'ersic index distributions of the two clusters are distinct from the field, as shown by the KS and AD tests.  The L14 field sample again shows a smaller median ($n_{mass} =3.05$) compared to the combined cluster sample ($n_{mass} =3.52$).  Comparing to the light-weighted distributions, the mass-weighted distributions of XMMU~J2235-2557 shows a larger median ($n_{mass}=4.13$), although XMMXCS~J2215-1738 shows vice versa ($n_{mass}=2.98$). The distributions are more widespread, which is primarily due to the fact that the uncertainties of mass-weighted parameters are $\sim2$ times larger than light-weighted parameters \citep[see][for a description]{Chanetal2016}.

\begin{figure*}
  \centering
  \includegraphics[scale=0.465]{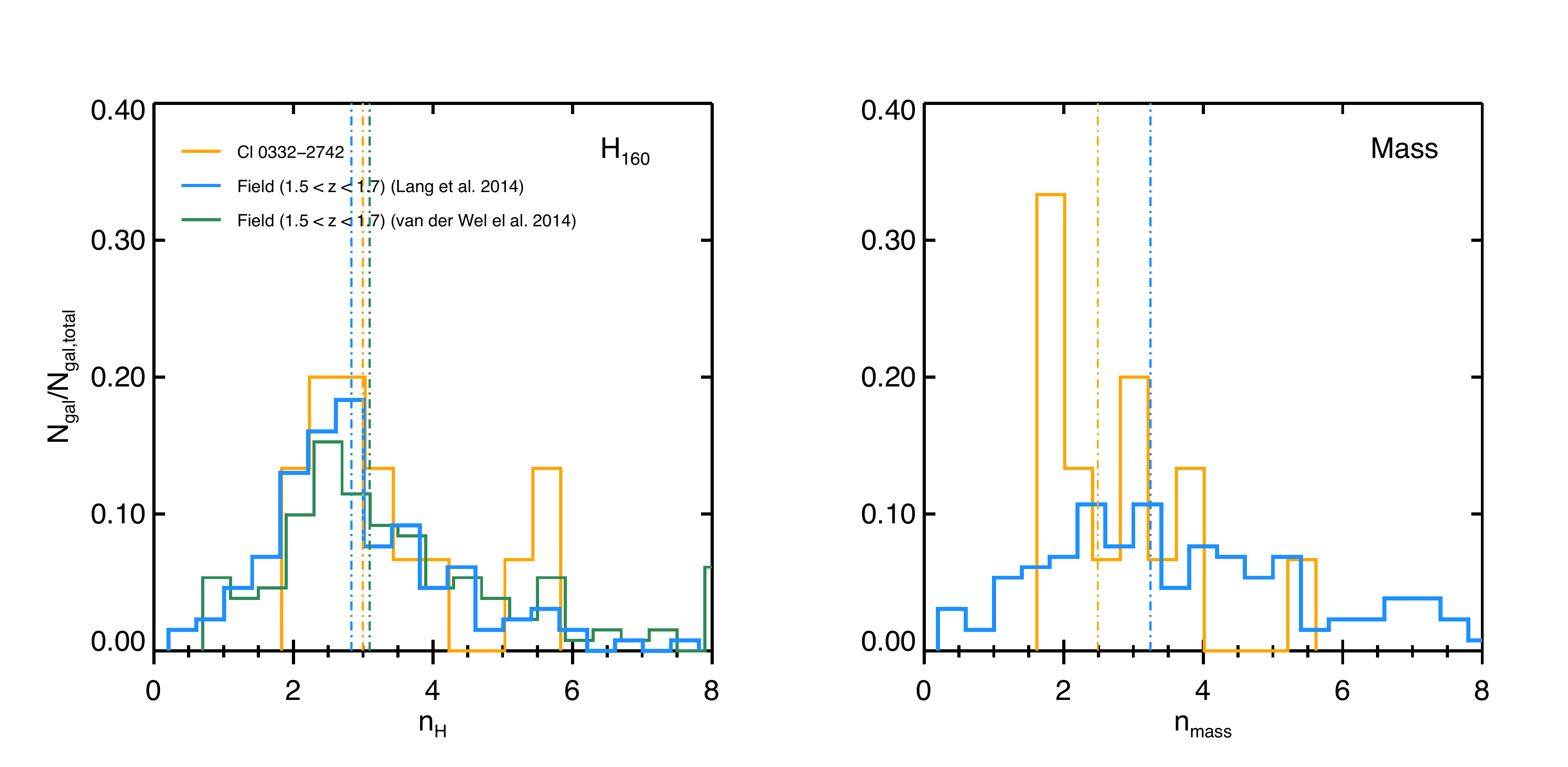}
  \caption[Comparison of the S\'ersic index distributions of Cl~0332-2742 with the field]{Comparison of the S\'ersic index distributions of Cl~0332-2742 with the field.  Left:  The light-weight S\'ersic index distributions.  The S\'ersic index distribution of Cl~0332-2742 is shown in orange.  The S\'ersic index distribution of the L14 field sample with a redshift range of $1.5<z<1.7$ is shown in blue.  The green histogram shows the S\'ersic index distribution of the citet{vanderWeletal2014} sample with the same redshift and $UVJ$ selection.   Right:  The mass-weighted S\'ersic index distributions.} 
  \label{field_histo_cl_n}
\end{figure*}

The light-weighted and mass-weighted S\'ersic index distributions of Cl~0332-2742  with the L14 field sample at $1.5 < z < 1.7$ are shown in Figure~\ref{field_histo_cl_n}. Cl~0332-2742 has a median S\'ersic index of $n=2.99$, which is comparable to the field median of the L14 sample ($n=2.84$) and the \citet{vanderWeletal2014} sample ($n=3.09$) despite the small number statistics.  The median mass-weighted axis ratio of Cl~0332-2742 is $n_{mass}=2.49$, which is smaller than the median of the L14 sample ($n_{mass}=3.17$). 

In summary, the light-weighted S\'ersic indices of the KCS samples are clearly lower than those observed in local passive ellipticals \citep[$n \sim 4 - 6$, e.g.][]{LaBarberaetal2010b}, suggesting that these galaxies are structurally distinct from local ellipticals and have more prominent disky components.  Similar results have been found in previous studies of high-redshift clusters \citep[e.g.][]{Papovichetal2012, Strazzulloetal2013, DeProprisetal2016} and of the field \citep[e.g.][]{Chevanceetal2012,Changetal2013,vanderWeletal2014,Langetal2014}.  From the cluster and field comparisons, we found no evidence that their distributions are distinct at both redshift ranges.  Interestingly, the KS and AD tests do show relatively low $p$-values for the comparison of the light-weighted S\'ersic index distributions of all three clusters combined to the L14 field sample at $1.3 < z < 1.7$. We confirm that this is entirely due to the difference between the S\'ersic index distribution of the L14 sample at different redshifts. For example, a KS test gives a $p$-value of $0.001$ for the comparison between the L14 $1.3 < z < 1.5$ and $1.5 < z < 1.7$ sample. 

\begin{table}
  \caption{The results of the Kolmogorov-Smirnov and Anderson-Darling test on the S\'ersic index distributions of the KCS clusters and the L14 field sample}
  \centering
  \begin{tabular}{@{}lcc@{}}
  \label{tab_nKS}
  \\
  \hline
  \hline
  \multicolumn{3}{c}{Light-weighted S\'ersic index distributions} \\
  Sample               &         $p_{\rm{KS}}$  &   $p_{\rm{AD}}$                    \\
  \hline
  XMMU~J2235 +  XMMXCS~J2215 vs. L14 field\textsuperscript{a}                                          &  0.29      &   0.24   \\  
  Cl~0332-2742   vs. L14 field\textsuperscript{b}                                                                           &   0.73    &   0.46  \\   
  3 KCS clusters   vs.    L14 field\textsuperscript{c}                                                                       &  0.04     &  0.02  \\  
  \hline
  \hline
  \\  
   \hline
  \hline
  \multicolumn{3}{c}{Mass-weighted S\'ersic index distributions} \\  
  Sample               &          $p_{\rm{KS}}$  &   $p_{\rm{AD}}$              \\
  \hline
  XMMU~J2235 +  XMMXCS~J2215 vs. L14 field\textsuperscript{a}                                              &  0.20   &   0.10      \\   
  Cl~0332-2742  vs. L14 field\textsuperscript{b}                                                                              &   0.05    &   0.04       \\  
  3 KCS clusters  vs. L14 field\textsuperscript{c}                                                                            &  0.93     &  0.60 \\   
  \hline
  \hline
   \multicolumn{3}{p{.475\textwidth}}{\textsuperscript{a}L14 field sample at $1.3 < z < 1.5$. \textsuperscript{b}L14 field sample at $1.5 < z < 1.7$. } \\
   \multicolumn{3}{p{.475\textwidth}}{\textsuperscript{c}L14 field sample at $1.3 < z < 1.7$, corresponds to all three KCS clusters.} \\
   \\
\end{tabular}
\end{table}

\subsubsection{Axis ratio distributions in different environments}
\label{subsec:q distributions in different environment}
\begin{figure*}
  \centering
  \includegraphics[scale=0.465]{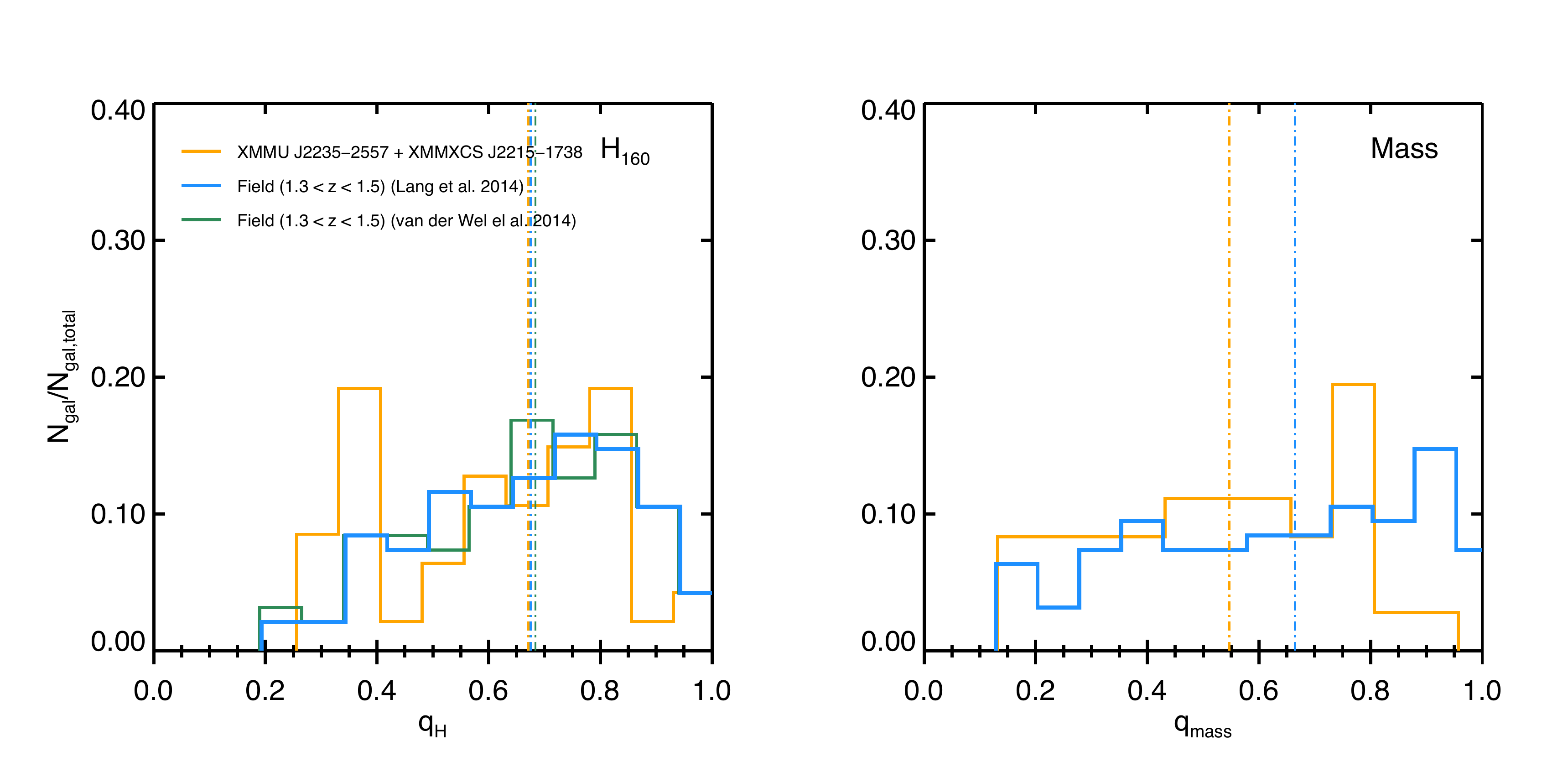}
  \caption[Comparison of the axis ratio distributions of XMMU~J2235-2557 and XMMXCS~J2215-1738 with the field]{Comparison of the axis ratio distributions  of XMMU~J2235-2557 and XMMXCS~J2215-1738 with the field.  Left:  The light-weighted axis ratio distributions.  The combined axis ratio distribution of XMMU~J2235-2557 $+$ XMMXCS~J2215-1738 is shown in orange.  The axis ratio distribution of the L14 field sample with a redshift range of $1.3<z<1.5$ is shown in blue.  The green histogram shows the axis ratio distribution of the \citet{vanderWeletal2014} sample with the same redshift and $UVJ$ selection.   Right:  The mass-weighted axis ratio distributions.} 
  \label{field_histo_xmm_q}
\end{figure*}

We then compare the axis ratio distribution of the KCS sample to the L14 field sample.  Figure~\ref{field_histo_xmm_q} shows the comparison of the combined light-weighted and mass-weighted axis ratio distributions of XMMU~J2235-2557 \& XMMXCS~J2215-1738 with the L14 field sample at $1.3 < z < 1.5$.  The median axis ratio of the combined sample is $q=0.67$ ($q=0.69$ for XMMU~J2235-2557, $q=0.67$ for XMMXCS~J2215-1738), which is very close to the median of the L14 field sample at this redshift ($q = 0.68$).  The median axis ratio distribution of the L14 sample is also consistent with the \citet{vanderWeletal2014} sample ($q=0.68$).  From the distributions, there seems to be an excess of low axis ratio objects ($q < 0.4$) in the two clusters compared to the field sample.  We have checked that these low-$q$ objects do not have very low S\'ersic indices (i.e. they are not contamination from edge-on disks).  This is consistent with the $z\sim1.62$ cluster in \citet{Papovichetal2012} where they found a noticeable population of passive galaxies with low axis ratio ($q \sim 0.2 - 0.3$) with the median $q$ being $\sim 0.6$.  The results of the KS and AD tests can be found in Table~\ref{tab_qKS}.  Both tests present large $p$-values for the comparison.

We found that the mass-weighted axis ratios are on average smaller compared to the light-weighted ones. The median mass-weighted axis ratio of the combined sample is $q_{mass} = 0.55$ ($q_{mass} =0.48$ for XMMU~J2235-2557, $q_{mass} =0.65$ for XMMXCS~J2215-1738).  Similar to the S\'ersic index distributions, the mass-weighted axis ratio distributions are also more widespread.  Similarly, the mass-weighted axis ratios are also found to be smaller in the L14 field sample, with a median of $q_{mass} =0.65$.  The KS and AD tests give a relatively small $p$-value of $\sim0.03$ and $\sim0.04$ respectively.  We found that this is driven by the axis ratio distributions in XMMU~J2235-2557, as seen in the individual cluster KS and AD test results in Table~\ref{tab_qKS}.  However, we note that part of this effect is perhaps due to objects that were discarded in the sample selection (see Section~\ref{subsec:Resolved stellar mass surface density maps and mass-weighted structural parameters}).  For example, if we examine the light-weighted axis ratio of only the objects that have reliable mass-weighted fits, those in XMMU~J2235-2557 have a median of $q=0.62$ (i.e. lower than the whole sample of XMMU~J2235-2557) , while those in XMMXCS~J2215-1738 have $q=0.71$.

\begin{figure*}
  \centering
  \includegraphics[scale=0.465]{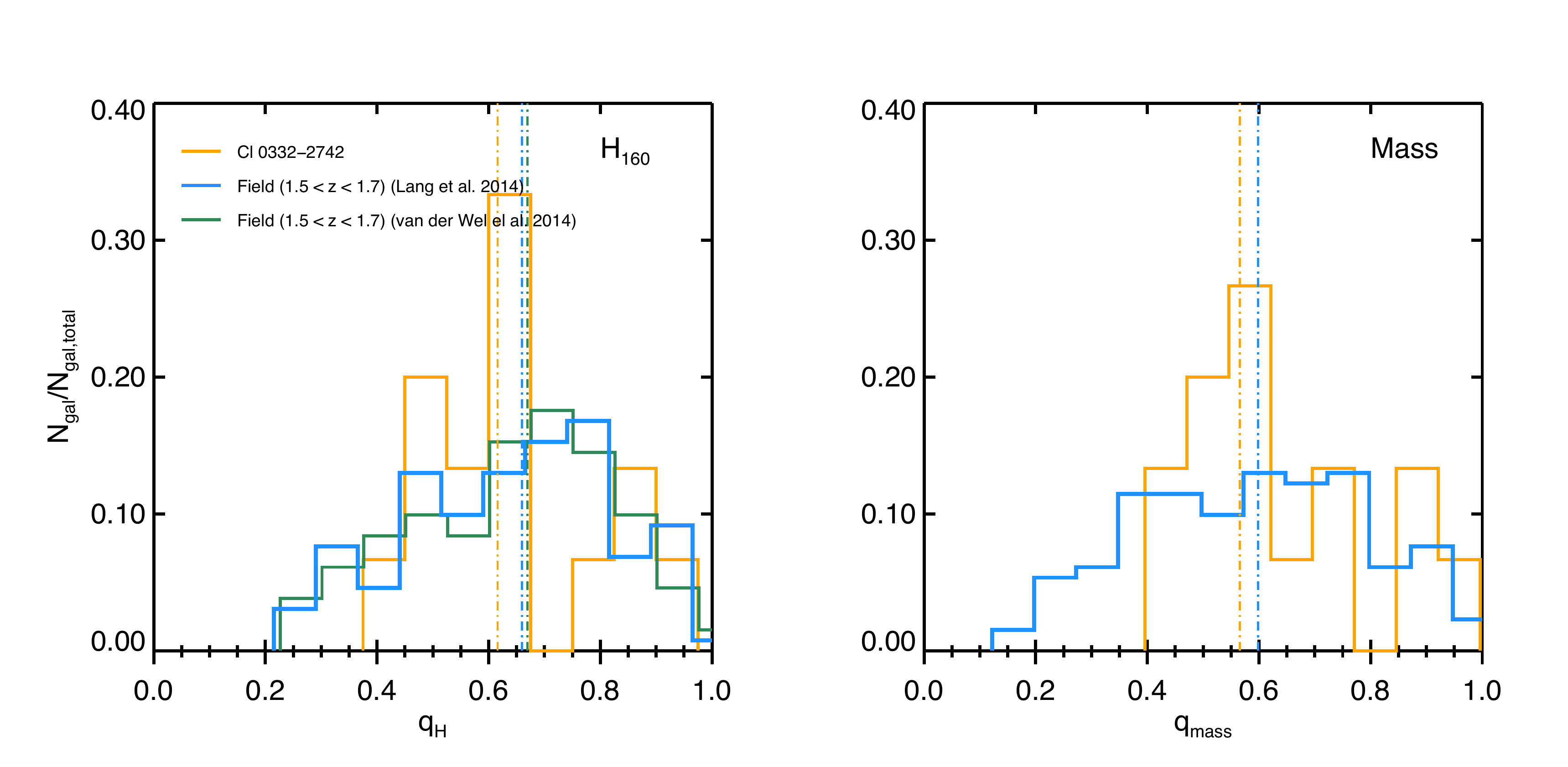}
  \caption[Comparison of the axis ratio distributions of Cl~0332-2742 with the field]{Comparison of the axis ratio distributions of Cl~0332-2742 with the field.  Left:  The light-weight axis ratio distributions.  The axis ratio distribution of Cl~0332-2742 is shown in orange.  The axis ratio distribution of the L14 field sample with a redshift range of $1.5<z<1.7$ is shown in blue.  The green histogram shows the axis ratio distribution of the \citet{vanderWeletal2014} sample with the same redshift and $UVJ$ selection.   Right:  The mass-weighted axis ratio distributions.} 
  \label{field_histo_cl_q}
\end{figure*}

Figure~\ref{field_histo_cl_q} shows the light-weighted and mass-weighted axis ratio distributions of Cl~0332-2742 with the L14 field sample at $1.5 < z < 1.7$.   Cl~0332-2742 has a median axis ratio $q =0.62$, which is smaller than the field median ($q=0.66$).  As in the other two clusters, the median of the mass-weighted axis ratio distribution of Cl~0332-2742 is also smaller than light-weighted one ($q_{\rm{mass}} = 0.57$).  Similarly, we see a similar decrease in mass-weighted axis ratio in the field, with a median of $q_{\rm{mass}} = 0.60$.  Both the KS and AD tests do not suggest the axis ratio distributions of Cl~0332-2742 and the field are distinct.

Overall, we found that the light-weighted axis ratio distributions of the KCS clusters are comparable to the field.  We do not see evidence of cluster galaxies having higher $q$ compared to the field at the same redshift as suggested in \citet{Delayeetal2014}.

\begin{table}
  \caption{The results of the Kolmogorov-Smirnov and Anderson-Darling test on the axis ratio distributions of the KCS clusters and the L14 field sample}
  \centering
  \begin{tabular}{@{}lcc@{}}
  \label{tab_qKS}
  \\
  \hline
  \hline
  \multicolumn{3}{c}{Light-weighted axis ratio distributions} \\
  Sample               &         $p_{\rm{KS}}$  &   $p_{\rm{AD}}$                    \\
  \hline
  XMMU~J2235 vs. L14 field\textsuperscript{a}                                                                           &  0.88      &   0.50    \\ 
  XMMXCS~J2215 vs. L14 field\textsuperscript{a}                                                                      &  0.50      &   0.29    \\  
  XMMU~J2235 +  XMMXCS~J2215 vs. L14 field\textsuperscript{a}                                          &  0.39      &   0.20   \\  
  Cl~0332-2742   vs. L14 field\textsuperscript{b}                                                                           &   0.52    &   0.51  \\   
  3 KCS clusters   vs.    L14 field\textsuperscript{c}                                                                       &  0.59     &  0.52  \\  
  \hline
  \hline
  \\  
   \hline
  \hline
  \multicolumn{3}{c}{Mass-weighted axis ratio distributions} \\  
  Sample               &          $p_{\rm{KS}}$  &   $p_{\rm{AD}}$              \\
  \hline
  XMMU~J2235 vs. L14 field\textsuperscript{a}                                                                             &  0.07      &   0.03    \\ 
  XMMXCS~J2215 vs. L14 field\textsuperscript{a}                                                                        &  0.39      &   0.36    \\  
  XMMU~J2235 +  XMMXCS~J2215 vs. L14 field\textsuperscript{a}                                              &  0.04    &    0.03     \\   
  Cl~0332-2742  vs. L14 field\textsuperscript{b}                                                                              &   0.55    &   0.51       \\  
  3 KCS clusters  vs. L14 field\textsuperscript{c}                                                                            &  0.45     &  0.23 \\   
  \hline
  \hline
   \multicolumn{3}{p{.475\textwidth}}{\textsuperscript{a}L14 field sample at $1.3 < z < 1.5$. \textsuperscript{b}L14 field sample at $1.5 < z < 1.7$. } \\
   \multicolumn{3}{p{.475\textwidth}}{\textsuperscript{c}L14 field sample at $1.3 < z < 1.7$, corresponds to all three KCS clusters.} \\
   \\
\end{tabular}
\end{table}

\subsection{The ratio of mass-weighted to light-weighted sizes in different environment}
\label{subsec:The ratio of mass-weighted to light-weighted sizes in different environment}
As discussed in both section~\ref{subsec:Stellar mass -- mass-weighted size relations} and \ref{subsec:Size distributions in different environment}, the mass-weighted size of our galaxies are generally smaller than their light-weighted sizes, suggesting the mass distributions are more concentrated than the light.  The ratio of the two sizes (hereafter size ratio) can hence be used as a probe of the spatial variation of the mass-to-light ratio, i.e. $M_{*}/L$ gradient within galaxies.

Figure~\ref{field_comparison_ratio_z_circ} shows the comparison of the size ratio in the three KCS clusters with the L14 field sample. Again we have shown both the results of using $R_{e\rm{-circ}}$ and $a_{e}$ as galaxy sizes.  To compare the size ratio at different redshifts we have converted the $H_{160}$ sizes of the L14 field sample into rest-frame $r$-band sizes, assuming the wavelength size dependence of \citet{Kelvinetal2012}, which we found to be consistent with the dependence in the KCS clusters \citep[see][for a detailed discussion on the wavelength-dependence relation of XMMU~J2235-2557]{Chanetal2016}.  Note that the wavelength size dependence itself has a negligible impact on the environment comparison.  For the redshift for which we make our comparison, the $H_{160}$-band is very close to rest-frame $r$-band.

We correct the progenitor bias for the field sample by removing galaxies with ages that are too young (at each redshift) to be descendants of galaxies at the lowest redshift of our cluster sample, at $z=1.39$, similar to Section~\ref{subsubsec: Caveats - Progenitor bias}. The age of the galaxy has to simply be longer than the time difference between $z\sim1.39$ and the redshift it resides.  The ages are taken from the 3D-HST public catalog \citep{Skeltonetal2014} and are derived using the \texttt{FAST} code \citep{Krieketal2009}.

In Figure~\ref{field_comparison_ratio_z_circ} we also plot the size ratio from \citet{Szomoruetal2013}, who derived rest-frame $g$-band and mass-weighted sizes for a mass-selected sample ($\log(M_{*}/M_\odot) > 10.7$) in GOODS-S.  Here we only include their quiescent sample, which was defined to have specific star formation rates $< 0.3/t_{H}$, where $t_H $ is the Hubble time.  We have also converted their sizes into rest-frame $r$-band, assuming the same wavelength dependence as above.  We noticed that there are discrepancies between the size ratio in \citet{Szomoruetal2013} and the L14 field sample.  With a subset of galaxies common to both catalogues, we conclude that the discrepancy comes from the mass-weighted sizes:  their mass-weighted sizes are on average $\sim40\%$ larger than those derived by L14.  This is likely due to the methodology they adopted and the large uncertainites.  \citet{Szomoruetal2013} derived mass-weighted sizes by integrating the 1D mass profiles derived from 1D color profiles, while L14 derived their sizes with 2D S\'ersic fitting to the mass-maps.  Different consideration of the `background mass level' (which arises from the sky background) and neighbouring galaxies may affect the accuracies of the 1D mass-weighted sizes. Understanding this discrepancy is beyond the scope of this paper.  For completeness we kept both for our comparison.  Since only circularised radius is available for the \citet{Szomoruetal2013} sample, we assumed the same size ratio for the $a_{e}$ case (i.e. $q_{mass}/q = 1$).  Note that the L14 sample actually shows $q_{mass}/q \lesssim 1$.

Regardless of using $R_{e\rm{-circ}}$ and $a_{e}$ as galaxy sizes, we found that the median size ratios in XMMU~J2235-2557 and XMMXCS~J2215-1738 are smaller than the \citet{Szomoruetal2013} samples at similar redshifts.  For the L14 field sample, this is only true for XMMXCS~J2215-1738; The offset of XMMU~J2235-2557 and XMMXCS~J2215-1738 from the L14 field sample are $-0.082\pm0.037$ and $-0.158\pm0.046$ dex if $R_{e\rm{-circ}}$ is used, but they are reduced to $-0.005\pm0.042 $ and $-0.098\pm0.045$ dex if $a_{e}$ is used as galaxy size.  On the other hand, the median size ratio in Cl~0332-2742 is comparable to, if not slightly larger than, the field galaxies in both samples ($0.056\pm0.033$ dex to L14 for $R_{e\rm{-circ}}$, $0.030\pm0.030$ dex for $a_{e}$).  

As our result depends on the field sample used for comparison and the choice of $R_{e\rm{-circ}}$ and $a_{e}$, a larger sample is required to confirm whether the size ratios of the clusters are different from the field.  Nevertheless, it is clear that Cl~0332-2742 has the largest size ratio among the three clusters regardless of using $R_{e\rm{-circ}}$ and $a_{e}$ as galaxy sizes. Using $a_{e}$ the size ratio offsets of Cl~0332-2742 to XMMU~J2235-2557 and XMMXCS~J2215-1738 are $-0.094 \pm0.045$ dex, $-0.179 \pm0.047$ dex, and these offsets are even larger if $R_{e\rm{-circ}}$ is used.  We will revisit the possible implications of these offsets in Section~\ref{subsec:Minor mergers and the effect from the environment}.


\begin{figure*}
  \centering
  \includegraphics[scale=0.66]{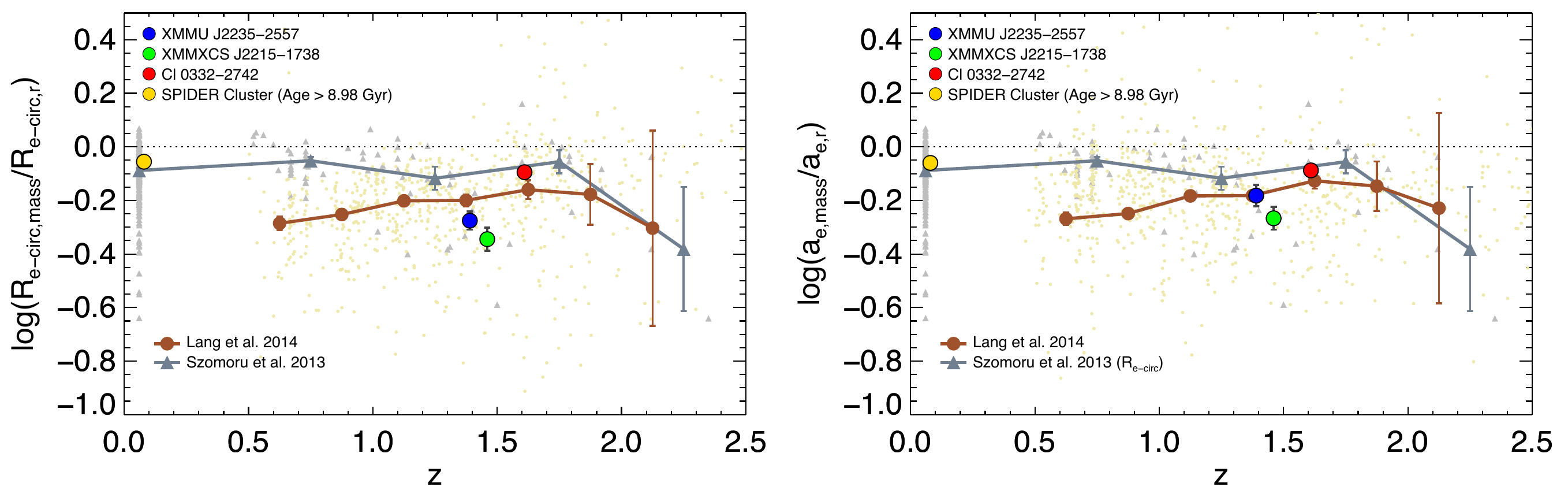}
  \caption[Comparison of the size ratio of the three KCS clusters with the field]{Comparison of the size ratio of the three KCS clusters with the field.  Left:  Size ratios are computed using $R_{e\rm{-circ}}$ as galaxy size.  Right:  Size ratios are computed using $a_{e}$ as galaxy size instead.  The median size ratios of XMMU~J2235-2557, XMMXCS~J2215-1738 and Cl~0332-2742 are shown as red, blue and green circles respectively.  The orange circle corresponds to the median size ratio of the progenitor bias corrected SPIDER cluster sample at $z\sim0$.  The brown circles and line show the binned median size ratio of the progenitor bias corrected L14 field sample across redshift in redshift bin of 0.25, while the light brown dots correspond to individual galaxies in the sample.  The grey triangles and line show the binned median size ratio of the \citet{Szomoruetal2013} field sample across redshift in redshift bin of 0.5, while the light grey triangles represent individual galaxies in their sample. Only $R_{e\rm{-circ}}$ is available for the \citet{Szomoruetal2013} sample. The error bars show the uncertainty of the median in each bin.}
  \label{field_comparison_ratio_z_circ}
\end{figure*}

\subsubsection{The evolution of ratio of mass-weighted to light-weighted sizes}
\label{subsubsec:The evolution of ratio of mass-weighted to light-weighted sizes}
In Figure~\ref{field_comparison_ratio_z_circ} we also plot the median size ratio of a local comparison sample, selected from the Spheroids Panchromatic Investigation in Different Environmental Regions (SPIDER) survey \citep{LaBarberaetal2010a} (hereafter the SPIDER cluster sample).  For the selection, we follow similar criteria as \citet{LaBarberaetal2010a}: a magnitude cut at the 95\% completeness magnitude ($M_{r} \leq -20.55$), a $\chi^{2}$ cut for the $g$-band and $r$-band S\'ersic fit ($\chi^{2} < 2.0$) from the publicly available multi-band structural catalogue, and a seeing cut at $\leq 1.5''$.  On top of that we apply a halo mass cut ($\log(M_{200}/M_{\odot}) \geq 14$) using the group catalogue from \citet{LaBarberaetal2010b}, which gives us a sample of 627 galaxies residing in high density environments. The derivation of the mass-weighted structural parameters of the SPIDER cluster sample is described in \citet{Chanetal2016}.

We correct for the progenitor bias in the SPIDER cluster sample using the same procedure as described in Section~\ref{subsubsec: Caveats - Progenitor bias} and \ref{subsec:The ratio of mass-weighted to light-weighted sizes in different environment} by using the age measurements from \citet{LaBarberaetal2010c}.  The correction for the SPIDER sample is small and mainly affects the size-ratio at the low-mass end, where we expect the progenitor bias to be stronger.

The mass-weighted sizes in the progenitor bias corrected SPIDER cluster sample are on average $\sim$14\% smaller than the $r$-band sizes with a median size ratio of $\langle \log(R_{e\rm{-circ},mass}/R_{e\rm{-circ}}) \rangle = -0.066 \pm 0.013$ ($\langle \log(a_{e,mass}/a_{e}) \rangle = -0.060 \pm 0.015$), which is completely consistent with the result in \citet{Szomoruetal2013} if we restrict our sample to the same mass range and band as theirs.

In \citet{Chanetal2016}, we found that the size ratio of red sequence galaxies in XMMU~J2235-2557 is smaller than those in local clusters, with an observed offset of $\langle \log(R_{e\rm{-circ},mass}^{1.39} / R_{e\rm{-circ}}^{1.39}) - \log(R_{e\rm{-circ},mass}^{0}/ R_{e\rm{-circ}}^{0}) \rangle = -0.188$ dex, suggesting an evolution of $M_{*}/L$ gradient across redshift.  Note that we revised this number due to the additional color-color selection applied in this work and the change is insignificant.  Here, we find that the offset in XMMXCS~J2215-1738 is even larger, with $\langle \log(R_{e\rm{-circ},mass}^{1.46} / R_{e\rm{-circ}}^{1.46}) - \log(R_{e\rm{-circ},mass}^{0}/ R_{e\rm{-circ}}^{0}) \rangle = -0.259$.  On the other hand, Cl~0332-2742 shows almost no offset, with $\langle \log(R_{e\rm{-circ},mass}^{1.61} / R_{e\rm{-circ}}^{1.61}) - \log(R_{e\rm{-circ},mass}^{0} /  R_{e\rm{-circ}}^{0}) \rangle = -0.006$.  Using $a_{e}$ we find similar results, with the offsets for the three clusters being $-0.122$, $-0.207$ and $-0.028$, respectively.

Assuming the cluster galaxies evolve in a self-similar way, the small size ratios in the $z\sim1.39$ and $1.46$ cluster imply an evolution of $M_{*}/L$ gradient with redshift.  Nevertheless, the similarity between the local size ratio and those in Cl~0332-2742 argues against a monotonic evolution of $M_{*}/L$ gradient. Recall that Cl~0332-2742 is a proto-cluster still in its assembly phase, we speculate that the $M_{*}/L$ gradients in high redshift cluster galaxies are related to or originate from physical processes occurring during cluster assembly.

\subsection{The origin of the $M_{*}/L$ gradients}
\label{subsec: The origin of the $M_{*}/L$ gradients}
Before we discuss the possible physical processes responsible for the size ratio (or $M_{*}/L$ gradients), it is worth investigating its origin in terms of the stellar population.  
In this section we focus on the color gradients in the three KCS clusters, as a probe of the spatial distribution of different stellar population in the galaxies.

We follow the method of \citet{Chanetal2016} to derive the observed color gradient along the semi-major axis ($\nabla_{z_{850} - H_{160}} = d (z_{850} - H_{160}) / d \log(a)$).  At redshift $\sim1.5$, $\nabla_{z_{850} - H_{160}}$ roughly corresponds to rest-frame $\nabla_{U-R}$. In short, the color gradients are measured from linear fits to 1D $z_{850}-H_{160}$ color radial profiles, which are derived from PSF-matched elliptical annular photometry.

\begin{figure}
  \centering
  \includegraphics[scale=0.70]{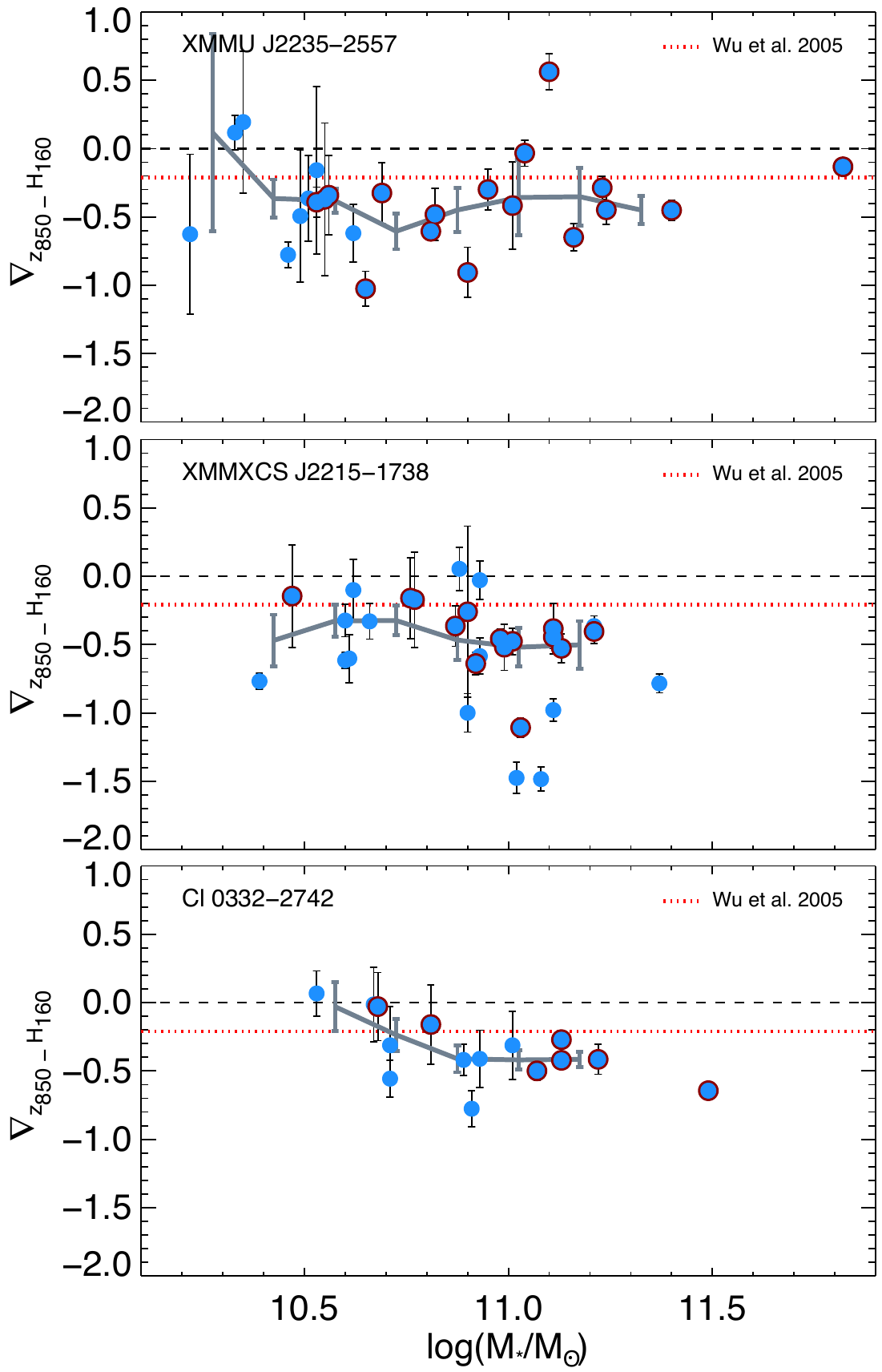}
  \caption[color and $M_{*}/L$ gradients in the KCS clusters]{color gradient $\nabla_{z_{850} - H_{160}}$ in the KCS clusters as a function of stellar mass.  Spectroscopically confirmed objects are circled in dark red.  At redshift $z\sim1.5$, this roughly corresponds to the rest-frame $(U-R)$ color gradient.  The black dashed line in each panel shows the reference zero level.  The red dotted line shows the average local $(U-R)$ gradient from \citet{Wuetal2005}.  The grey line in each panel shows the running median and the error bars show the uncertainty of the median in each bin.  When there is only one point in the bin, the uncertainty of the quantity is plotted instead.  The average $(U-R)$ color gradient of our red sequence sample is $\sim 2$ times steeper than color gradients observed locally.}
  \label{fig_colorgrad}
\end{figure}

Figure~\ref{fig_colorgrad} shows the color gradients $\nabla_{z_{850} - H_{160}}$ of the passive sample as a function of stellar mass.  We find that the majority of the KCS galaxies ($\sim 93 \%$) have negative color gradients.  The median color gradient of the XMMU~J2235-2557 sample is $\langle \nabla_{z_{850} - H_{160}} \rangle = -0.39 \pm 0.07$.  The color gradients are slightly steeper (but still consistent within uncertainties) in XMMXCS~J2215-1738 with a median of $\langle \nabla_{z_{850} - H_{160}} \rangle = -0.46 \pm 0.07$.  Similarly, galaxies in Cl~0332-2742 have negative gradients with a median of $\langle \nabla_{z_{850} - H_{160}} \rangle = -0.41 \pm 0.07$.   Also plotted on Figure~\ref{fig_colorgrad} is the average local $(U-R)$ color gradient from \citet{Wuetal2005}.  Similar to \citet{Chanetal2016}, we find that the average $(U-R)$ color gradient in the three clusters is $\sim 2$ times steeper than color gradients observed locally.

Despite the similarities in the median values of the color gradients, the $M_{*}/L$ gradients of the three clusters do not necessarily share the same values, due to the slope of the $M_{*}/L$-color relations.  Using the $M_{*}/L$-color relations we convert 1D color profiles to 1D $M_{*}/L$ profiles, from which we derive the $\log (M_{*}/L)$ gradients $\nabla_{\log(M/L)}$:  the median values for XMMU~J2235-2557, XMMXCS~J2215-1738 and Cl~0332-2742 are $-0.26 \pm 0.04$,  $-0.27 \pm 0.04$ and $-0.11 \pm 0.02$, respectively.  They are qualitatively consistent with the size ratios in Figure~\ref{field_comparison_ratio_z_circ}.  

Color gradients in passive galaxies are commonly interpreted via either age gradients ($\nabla_{age} = d \log(\text{age}) / d \log(a)$) at fixed metallicity or metallicity gradients ($\nabla_{Z} = d \log(Z) / d \log(a)$) at fixed age.  The age gradients in local passive galaxies are consistent with 0 (or slightly positive) while the average metallicity gradient is found to be of $\nabla_{Z} \approx -0.1 $ to $ -0.3$ \citep[see,][]{Mehlertetal2003, Wuetal2005, LaBarberaetal2009b, Kuntschneretal2010, Oliva-Altamiranoetal2015, Wilkinsonetal2015}.  Previous works on clusters at $z\sim0.4$ \citep{Sagliaetal2000} and local clusters \citep[e.g.][]{TamuraOhta2003} also showed that color gradients may be preferentially produced by radial variation in metallicity rather than age.  At $z > 1$, studies of color gradients are limited \citep[see, for example][]{Gargiuloetal2012, DeProprisetal2015, DeProprisetal2016, Cioccaetal2017}.  In \citet{Chanetal2016} we quantitatively investigated the evolution of color gradients in XMMU~J2235-2557 by modeling the inner (defined to be the color at $0.5 a_{e}$) and outer ($2.0 a_{e}$) regions of the galaxies under different assumptions of the radial variation of stellar population properties, using stellar population models of \citet{BruzualCharlot2003}.  The goal is to evolve the observed $\nabla_{z_{850} - H_{160}}$ assuming different age and metalicity gradients in order to determine which initial conditions match the observed gradient at $z\sim0$.

We repeat the procedure and test cases in \citet{Chanetal2016} for XMMXCS~J2215-1738 and Cl~0332-2742.  Because of the age-metallicity degeneracy, we consider three scenarios with assumptions on the age or metallicity gradients, summarised briefly below:
\begin{itemize}
   \item \textbf{Case I - Pure age-driven gradient evolution} -- In this case we try to use a single age gradient to reproduce the evolution of color gradients. The inner and outer regions of the passive galaxies are assumed to have identical metallicities (i.e. flat metallicity gradients $\nabla_{Z} =0$).
   \item \textbf{Case II - Pure metallicity-driven gradient evolution} -- In this case we assume the evolution of color gradients is solely due to a metallicity gradient.  The stellar population in the inner and outer region are fixed to be coeval (i.e. flat age gradients $\nabla_{age} = 0$).
   \item \textbf{Case III - Age-driven gradient evolution with an assumed metallicity gradient} -- Same as case I, but we also assume a fixed metallicity gradient with $\nabla_Z  = -0.2$, which is the mean value observed in local passive galaxies \citep[e.g.][]{TamuraOhta2003, Wuetal2005, Redaetal2007} as well as in recent simulations \citep[e.g.][]{Hirschmannetal2015}.
\end{itemize}

For each of the cases, scenarios with different assumed metallicity for the inner regions, sub-solar, solar and super-solar ($Z = 0.008, 0.02, 0.05 = 0.4 Z_{\odot},Z_{\odot}, 2.5Z_{\odot}$) are also tested. Assuming metallicities with $Z < 0.008$ or $Z > 0.05$ is unphysical for most galaxies in the sample.

We find that among the three cases an age-driven gradient evolution with a local metallicity gradient (Case III) is the most probable scenario for all three clusters.   Case III reproduces the observed evolution of the color gradients with redshift well in both median and scatter.  On the other hand, an age gradient alone would over-predict the evolution with redshift, causing the evolved gradients to be too shallow, while a metallicity gradient alone could not explain the observation as the evolution it predicts goes in the opposite direction. Details of the test scenarios and the results are shown and discussed in Appendix~\ref{Results of the color gradient test scenarios}.  As found in \citet{Chanetal2016}, among the three metallicities the one with solar metallicity seems to best match the evolution of color gradients for most galaxies in the three clusters.  In this scenario we find a median age gradient of $\langle \nabla_{age} \rangle = -0.32 \pm 0.08$ at $z=1.39$ for XMMU~J2235-2557, $\langle \nabla_{age} \rangle = -0.35 \pm 0.09$ at $z=1.46$ for XMMXCS~J2215-1738 and $\langle \nabla_{age} \rangle = -0.31 \pm 0.10$ at $z=1.61$ for Cl~0332-2742.  The median age difference ($\langle \delta_{age} \rangle$) between the inner and outer regions in each cluster are $0.90 \pm 0.20$, $1.10 \pm 0.20$ and $0.70 \pm 0.38$ Gyr, respectively. 

This indicates that an age gradient is needed to explain the observed evolution of the color gradients, while metallicity gradients probably dominate at $z\sim0$. This conclusion is in partial agreement with other studies \citep{Gargiuloetal2012, DeProprisetal2016}.  Using exponentially declining $\tau$-models with various $\tau$ (instead of SSPs as we did here) will result in shallower age gradients, but the conclusion remains unchanged \citep[see Appendix D in][]{Chanetal2016}.   

A caveat in the above analysis is the effect of dust obscuration.  Although the color gradient is unaffected by global dust obscuration, it can be affected by the radial variation of dust content.  It is not possible to derive reliable dust gradients for passive galaxies with our photometric data \citep[however see][for a method to derive dust gradients for star-forming galaxies with multi-band photometry]{Wangetal2017}.  We have derived an upper bound dust gradient below which our conclusion remains valid.  Our conclusion is robust if the extinction gradient is smaller than $d A_V/d \log(a) \sim 0.40$ or $d E(B-V) /d \log(a) \sim -0.10$, assuming a Calzetti extinction law \citep{Calzettietal2000}.  On the other hand, the values we derived, in particular the age difference between the inner and outer regions are strongly affected by the global dust extinction.  For example, if we adopt the global dust extinction values of the Cl~0332-2742 sample ($\langle A_V \rangle = 0.6$) from the public 3D-HST catalogue \citep{Skeltonetal2014, Momchevaetal2016} and repeat the analysis, the median age difference between the inner and outer region will drastically reduce from $0.70$ to $0.45 \pm 0.18$. 
Note that a dust extinction value of $A_V = 0.6$ is moderately high compared to other field passive galaxies at this redshift range \citep[e.g.][]{Gargiuloetal2012, Bellietal2015}.  This suggests that without an accurate estimation of the dust extinction, one should not over-interpret the age differences derived with this method.

Another caveat is that our analysis is based on a single color gradient.  Recent work by \citet{Cioccaetal2017} measured the rest-frame $UV-U$ gradients (via $\nabla_{i_{775} - z_{850}}$) as well as $\nabla_{z_{850} - H_{160}}$ in XMMU~J2235-2557.  While their $\nabla_{z_{850} - H_{160}}$ gradients are negative and roughly consistent with our work (the discrepancies may result from the depth of data they used and the methodology), they also found positive rest-frame $UV-U$ gradients, which cannot be simultaneously explained by an age, metallicity or dust gradient.  \citet{Cioccaetal2017} suggested that a small amount of central star formation, the presence of a QSO and / or He-rich stars may contribute to the observed UV excess toward the centre of the galaxies.  This should result in a large fraction of emission in the spectra of the sample studied here, which is not the case in our KMOS data \citep{Beifiorietal2017}.  Nevertheless, it is possible that the stellar populations comprise more components than a combination of age and metallicity gradient as we have shown here.
   
\subsection{Minor mergers and the effect from the environment}
\label{subsec:Minor mergers and the effect from the environment}
In the last three sections, we have shown that (1) passive galaxies in XMMU~J2235-2557 and XMMXCS~J2215-1738 show larger median light-weighted sizes compared to the field, but not in the highest redshift cluster Cl~0332-2742, (2) XMMU J2235-2557 and XMMXCS J2215-1738 show smaller size ratios (i.e. steeper $M_{*}/L$ gradients) compared to Cl 0332-2742, and (3) the color gradients in the three clusters are qualitatively consistent with the size ratios and can be explained with a combination of an age gradient and a metallicity gradient. In this section we discuss how these results are consistent with the minor mergers scenario and their implications.

As we discussed in the introduction, dry minor mergers have been shown to be able to explain the observed evolution in size and stellar mass surface density profile of passive galaxies in the field \citep[e.g][]{Trujilloetal2011, Pateletal2013, vanderWeletal2014}.  During a minor merger, the less massive satellite galaxy gets tidally disrupted and is accreted to the main (more massive) galaxy.  The accreted mass assembles predominately in the outer part of the galaxy and induces an increase in size but only minimally increases the mass, as demonstrated in simulations \citep[e.g][]{Naabetal2009, Hilzetal2013, RodriguezGomezetal2016}.  A prediction of this scenario is that there should be an environmental dependence on sizes, due to the difference in merger rates in different environments. 

Traditionally merger activity is believed to be suppressed in virialized clusters because of their high velocity dispersion, which results in high relative velocities between cluster members \citep[e.g.][]{Conroyetal2007}. The exception being mergers of satellites onto the BCG due to dynamical friction \citep[e.g][]{Burkeetal2013, Burkeetal2015}.  However, this does not apply to young forming clusters (and group structures) at high redshift; young clusters are expected to be extreme merger-rich environments, as they are still in an active assembly phase and have lower local velocity dispersion \citep[see discussion in e.g.][]{Lotzetal2013, Newmanetal2014}.  Various simulations have also shown that such mergers are common in forming clusters \citep[e.g.][]{DeLuciaetal2007b, Jianetal2012, Lackneretal2012}, galaxy merger rates may also be enhanced (before and) during merging clusters or cluster-group mergers \citep[e.g.][]{Vijayaraghavanetal2013}.

Hence, the fact that we see larger median light-weighted sizes in the passive galaxies in XMMU~J2235-2557 and XMMXCS~J2215-1738 compared to the field is consistent with the prediction of the minor merger scenario.  The mass-weighted sizes and color gradients (or $M_{*}/L$ gradients) in the three clusters are also consistent with the minor merger picture.  Figure~\ref{field_histo_xmm} shows that there is almost no size difference between cluster and field for XMMU~J2235-2557 and XMMXCS~J2215-1738 if mass-weighted sizes are used.  This again indicates that the driver of the difference in light-weighted sizes between different environments is minor mergers rather than major mergers or adiabatic expansion, as the accreted stars predominately stayed in the outskirt of the galaxies and hence are not affecting the mass distribution (and mass-weighted size) to a large extent.  Similarly, the accretion will result in the negative age gradients we see, given that the stars accreted from the minor mergers are relatively young, as suggested in \citet{Chanetal2016}.  This is perhaps not surprising for the massive galaxies we are considering here, as the passive fraction of satellite galaxies is found to increase sharply with mass \citep[see the discussion in][]{Fossatietal2016}.

The effect of having larger sizes in clusters than the field presumably due to an epoch of enhanced merger rate (i.e. the ``accelerated structural evolution'') is seen observationally in various works \citep[e.g.][]{Cooperetal2012, Zirmetal2012, Lanietal2013, Strazzulloetal2013, Jorgensenetal2013, Delayeetal2014}, although a consensus has not been reached \cite[see e.g.][]{Retturaetal2010, Raichooretal2012, Newmanetal2014}.  The fact that we see larger light-weighted sizes in XMMU~J2235-2557 and XMMXCS~J2215-1738 than in the field, but not in Cl~0332-2742, indicates that this merger enhancement is the strongest when the various groups are infalling and merging to form a single cluster.  Galaxies in Cl~0332-2742 have similar structural properties to the field galaxies as they have not yet undergone or completed this phase.  Although a larger sample is needed to confirm the correlation between the dynamical state of clusters and the environmental effects on galaxy sizes, we suggest that this may be a possible explanation to the discrepancies on the environmental effect in previous works.  For example, while \citet{Newmanetal2014} have found no evidence for a difference between the sizes in the passive galaxies in the cluster JKCS 041 and the field at $z\sim1.8$, recent work by \citet{Prichardetal2017} has shown that JKCS 041 is still in formation and comprises two merging groups of galaxies extending eastward and towards the southwest of the cluster with different stellar ages.  We also note that the size difference between cluster and field \textit{due to this epoch of enhanced merger rate} may be somewhat short-lived, as the merger rate will soon decline as the cluster virializes.  To illustrate this, consider a hypothetical scenario in which a cluster just ended its epoch of enhanced merger rate and has a passive population with an average size that is $\sim30\%$ larger than the field at $z\sim1.5$ similar to our findings, the size difference between this cluster and the field would have vanished by $z\sim1.05$, assuming the field galaxy population has a size evolution as in \citet{vanderWeletal2014} and ignoring any subsequent size growth in the cluster population.  Note that from the observations we do see evidence of subsequent size growth in the cluster population (see Section~\ref{subsubsec: Caveats - Progenitor bias}), but if the subsequent size growth rate in clusters is much slower than the field \citep[which is supported by the decreasing size difference between cluster and field with redshift in, e.g.][]{Cooperetal2012, Lanietal2013, Delayeetal2014}, the size difference between a given cluster and the field may only be able to be detected for a short period of time.

While the minor merger scenario can provide a qualitative explanation for the results we see, one should not treat it as the sole process responsible for the evolution of passive galaxies all the way to $z\sim0$.  As the cluster becomes virialized over time mergers are more disfavoured (except those to the BCG), other environmental processes are likely to dominate at lower redshift.  The two main types of cluster ETG observed at low redshifts \citep[the fast and slow rotators, see][for a review]{Cappellari2016} could be formed through distinct scenarios.  Massive ($\geq 2 \times 10^{11} M_{*}$) slow rotators are preferentially found near the centres of clusters; they were quenched early and assembled their mass continually through repeated dry mergers \citep[e.g.][and reference therein]{Cappellari2013,Scottetal2014}.  Fast rotators in clusters are presumably formed from spiral galaxies quenched via internal processes \citep[due to AGN or supernovae feedback or formation of bulge, e.g.][]{Boweretal2006, Hopkinsetal2007, Dekeletal2009} and various environmental processes after they enter the cluster halo, including ram-pressure gas stripping as they pass through the hot intracluster medium \citep{GunnGott1972} and the suppression of ongoing gas accretion onto galaxies due to the halo \citep[e.g.][]{Larson1980}.  Gravitational interactions between these galaxies and the members of the cluster (or the halo itself) also play a role in the subsequent evolution \citep[galaxy harassment, e.g.][]{Mooreetal1996, Mooreetal1998}.  It is also possible that during cluster assembly these tidal interactions can affect the size of the galaxies.  A similar two-channel evolution has also been suggested from photometric studies of high redshift galaxies \citep[the fast and slow assembly,][]{Huertascompanyetal2015}.

Another issue that remains unclear is whether minor mergers in forming clusters are sufficient to account for all the differences we see.  Currently the merger rates in high-redshift clusters are not well constrained \citep[see, e.g.][]{Lotzetal2013}, although the importance of mergers in growing the cluster red sequence have been established in various works \citep[e.g.][]{Papovichetal2012, Rudnicketal2012}.  After the assembly stage, minor mergers are still likely to happen when group-scale structures are being accreted into (virialized) clusters.  Whether this can fully explain the size evolution in clusters again requires a stringent constraint on the minor merger rate, which is out of the scope of this paper.  In the field, it has been shown that the observed minor merger rate may not be sufficient to explain the observed size evolution at $z >1$ \citep[e.g][]{Newmanetal2012, Manetal2016}, hinting that additional mechanisms may be in place.

\section{Summary and Conclusions}
\label{sec:Conclusion}
In this paper we have presented the structural properties of a sample of red sequence galaxies in dense environments at $1.39 < z < 1.61$ from the KMOS Cluster Survey (KCS).  With \textit{HST}/ACS and WFC3 imaging we derive light-weighted structural parameters for individual galaxies through 2D S\'ersic fitting.  In addition, we derive mass-weighted structural parameters for these galaxies from their resolved stellar mass surface density maps constructed with an empirical $M_{*}/L$ -color relation and the $z_{850}$ and $H_{160}$ images.  We then compare these quantities to the field to investigate the effect of environment on the structural properties.  Our results can be summarised as follows:  

\begin{itemize}
  
  \item The $H_{160}$ band sizes of the passive galaxies in the KCS clusters are on average smaller than that expected from the local mass-size relation by \citet{Bernardietal2014} and \citet{Cappellarietal2013b} at the same rest-frame wavelength.  Comparing to \citet{Bernardietal2014}, the sizes are on average $\sim42\%$, $\sim55\%$ and $\sim69\%$ smaller than local passive galaxies of the same mass for XMMU~J2235-2557, XMMXCS~J2215-1738 and Cl~0332-2742, respectively.  We have also shown the progenitor bias can reduce, but not fully eliminate, the observed difference between the KCS clusters and the local sample.
  
  \item The slopes $\beta$ of the stellar mass -- light weighted size relation of the three KCS clusters are consistent with each other within the uncertainties.  The derived slope for the full sample $\beta = 0.79 \pm 0.14$ as well as the values for individual clusters are consistent with the values derived from the field population at a similar redshift range.
    
  \item Using the mass-weighted sizes of the galaxies, we study the stellar mass -- mass-weighted size relation of the KCS clusters.   We find that the mass-weighted sizes are $\sim45\%$,  $\sim55\%$ and $\sim20\%$ smaller compared to the light-weighted ones for XMMU~J2235-2557, XMMXCS~J2215-1738 and Cl~0332-2742, respectively, which tell us that the stellar mass distributions are more concentrated than the light.  The derived slope of mass -- mass-weighted size relation for the full sample is $\beta = 0.49 \pm 0.13$.
  
   \item Comparing the mass-normalised size distribution of the KCS galaxies to a field sample in a similar redshift range, we find that the median size of the combined light-weighted mass-normalised size distribution of XMMU~J2235-2557 and XMMXCS~J2215-1738 is offset to larger sizes.  We confirm that this is true if either circularized effective radii ($\sim33\%$ larger) or semi-major axes ($\sim24\%$ larger) are used as galaxy sizes.  This observed offset is reduced if mass-weighted mass-normalised sizes are used for the comparison.  In contrast, the size distribution of Cl~0332-2742 has comparable, if not on average smaller, light-weighted and mass-weighted sizes compared to the field galaxies.  
   
   \item The median ratios of mass-weighted to light-weighted size in XMMU J2235-2557 and XMMXCS J2215-1738 are smaller than Cl~0332-2742.  The median size ratios in XMMU~J2235-2557 and XMMXCS~J2215-1738 are also smaller than the \citet{Szomoruetal2013} samples at similar redshifts.  But for the L14 field sample, this is only true for XMMXCS~J2215-1738.  The offset of XMMU~J2235-2557 and XMMXCS~J2215-1738 from the L14 field sample are $-0.082\pm0.037$ and $-0.158\pm0.046$ dex if $R_{e\rm{-circ}}$ is used, but they are reduced to $-0.005\pm0.042 $ and $-0.098\pm0.045$ dex if $a_{e}$ is used as galaxy size. On the other hand, the median size ratio in Cl~0332-2742 is comparable to, if not larger than, the field galaxies (an offset of $0.056\pm0.033$ dex for $R_{e\rm{-circ}}$, $0.030\pm0.030$ dex for $a_{e}$).
  
   \item Comparing the KCS cluster galaxies to the progenitor bias corrected local SPIDER cluster sample, we find an offset in the ratio of mass-weighted to light-weighted sizes for XMMU~J2235-2557 and XMMXCS~J2215-1738. On the other hand, Cl~0332-2742 shows almost no offset with the SPIDER cluster sample.  We attribute the difference seen in XMMU~J2235-2557 and XMMXCS~J2215-1738 to an evolution of the $M_{*}/L$ gradient over redshift.

   \item From a case study on the color gradient $\nabla_{z_{850}-H_{160}}$ of the three clusters, we find that an age gradient is needed at high redshift to explain the observed evolution of the color gradients, suggesting that the $M_{*}/L$ gradient is a combination of an age gradient and a metallicity gradient in the stellar population.  The presence of an age gradient is consistent with the picture of recent accretion of young stars from minor mergers.
   
   \end{itemize}
 
    We attribute the larger median light-weighted size in XMMU~J2235-2557 and XMMXCS~J2215-1738 compared to the field to an enhanced dry minor merger rate during cluster assembly.    As a (proto)cluster still not yet fully assembled and comprising several group structures, it is likely that Cl~0332-2742 still has not undergone this phase, hence shows similar structural properties to the field and larger mass-weighted size to light-weighted size ratio compared to the other two clusters.
    
    Our measurements of the size ratio and mass-weighted size provide further evidence to support the idea of accelerated structural evolution proposed by previous works comparing the light-weighted sizes between clusters and the field \citep[e.g.][]{Lanietal2013, Delayeetal2014}. This enhanced merger rate in forming clusters is also seen in simulations \citep[e.g.][]{DeLuciaetal2007b, Jianetal2012, Lackneretal2012}.  While this picture can qualitatively explain our results, an important question that remains unclear is whether the rate of minor mergers can sufficiently explain the environmental differences and size evolution. Despite a few works attempt to quantify the minor merger rate in the field \citep[e.g.][]{Newmanetal2012, Manetal2016}, the merger rate in clusters remains poorly constrained.
    
    There are certain limitations to our studies, as our sample is based on photometric (and partial spectroscopic) selection on three clusters within the \textit{HST}/WFC3 FOV.  Increasing the number of members in a single cluster would reduce the measurement uncertainty in the slope of the mass-size relations we have, while increasing the number of clusters with a range of environment and dynamical states would allow for a more precise measurement of the structural differences between clusters and field passive galaxies, and perhaps pinpoint the epoch where environment influences their structural properties.

\acknowledgments

The authors would like to thank Philipp Lang and Stijn Wuyts for providing the structural parameter catalogue of their field sample for the comparison between clusters and field.  The authors would also like to thank Francesco La Barbera for providing the SPIDER catalogues for the comparison with local passive galaxies, Chris Lidman for providing HAWK-I images for XMMU~J2235-2557 and XMMXCS~J2215-1738 for the construction of our photometry catalogue, and Masao Hayashi for providing the catalogue of [O II] emitters in XMMXCS~J2215-1738.  We would also like to thank Matteo Fossati and Fabrizio Finozzi for useful discussions.  We thank the anonymous referee and the editor for their comments that improved the quality of the manuscript. 

J.C.C. Chan acknowledges the support of the Deutsche Zentrum f\"ur Luft- und Raumfahrt (DLR) via Project ID 50OR1513.  David J. Wilman acknowledges the support of the Deutsche Forschungsgemeinschaft via Project ID 3871/1-1 and ID 3871/1-2.  This work was supported by the UK Science and Technology Facilities Council through the `Astrophysics at Oxford' grant ST/K00106X/1.  Roger L. Davies acknowledges travel and computer grants from Christ Church, Oxford and support from the Oxford Hintze Centre for Astrophysical Surveys which is funded by the Hintze Family Charitable Foundation.  LJP is supported by a Hintze Scholarship, awarded by the Oxford Hintze Centre for Astrophysical Surveys, which is funded through generous support from the Hintze Family Charitable Foundation.
Michele Cappellari acknowledges support from a Royal Society University Research Fellowship.  John P. Stott gratefully acknowledges support from a Hintze Research Fellowship. 




\bibliographystyle{aasjournal}
\bibliography{jchan_kcsii_sizecg_apj_arxiv_submission} 


\appendix

\section{A. Size comparison with van de Wel et al. 2014 and Lang et al. 2014}
\label{Mass-weighted size comparison with Lang et al. 2014}
\begin{figure}
\centering
  \includegraphics[scale=0.5]{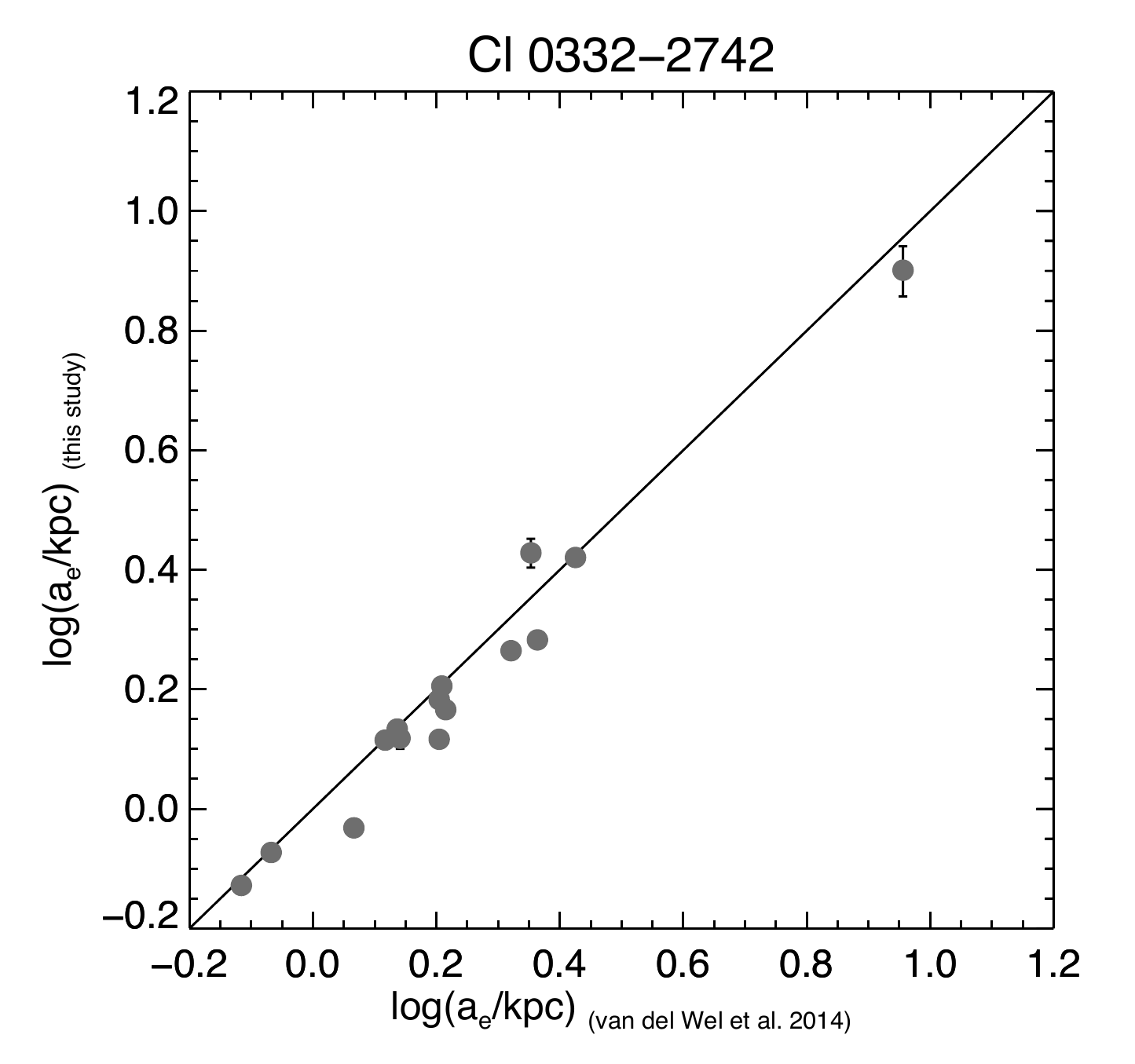}
  \caption[Comparison of light-weighted sizes with sizes from \citet{vanderWeletal2014}]{Comparison of light-weighted sizes with sizes from \citet{vanderWeletal2014}.  The solid black line is the one-to-one relation.  The error bars represent the $1\sigma$ uncertainties of the light-weighted sizes.}
  \label{fig_lightsize_comparison}
\end{figure}

In this section we compare our derived light-weighted and mass-weighted sizes with the literature.  Being currently the largest study on the mass -- size relation, \citet{vanderWeletal2014} derived light-weighted sizes for both passive and star-forming galaxies using the imaging data from CANDELS \citep{Groginetal2011, Koekemoeretal2011} and 3D-HST \citep{Brammeretal2012, Skeltonetal2014} for all five CANDELS fields, including the GOODS-S field where Cl~0332-2742 resides.  Figure~\ref{fig_lightsize_comparison} shows a comparison of our $H_{160}$ sizes of the 15 galaxies in Cl~0332-2742 to those sizes derived by \citet{vanderWeletal2014} in the same band.  Overall, the sizes are consistent with each other.  The entire sample has a median difference and $1\sigma$ dispersion of $-0.029 \pm 0.046$ dex ($\sim10\%$ scatter).  The light-weighted sizes from \citet{Langetal2014} are also generally consistent with those from \citet{vanderWeletal2014}.

\citet{Langetal2014} also derived stellar mass maps for a mass-selected sample ($\log(M_{*}/M_\odot) > 10$) in all five CANDELS fields.  They fitted both two-dimensional S\'ersic models and two-component (i.e., bulge + disk) decomposition to the mass maps.  Below we compare the mass-weighted sizes derived from our mass maps using the $M_{*}/L$-color relation to those derived from resolved SED fitting from \citet{Langetal2014}.

\begin{figure}
\centering
  \includegraphics[scale=0.5]{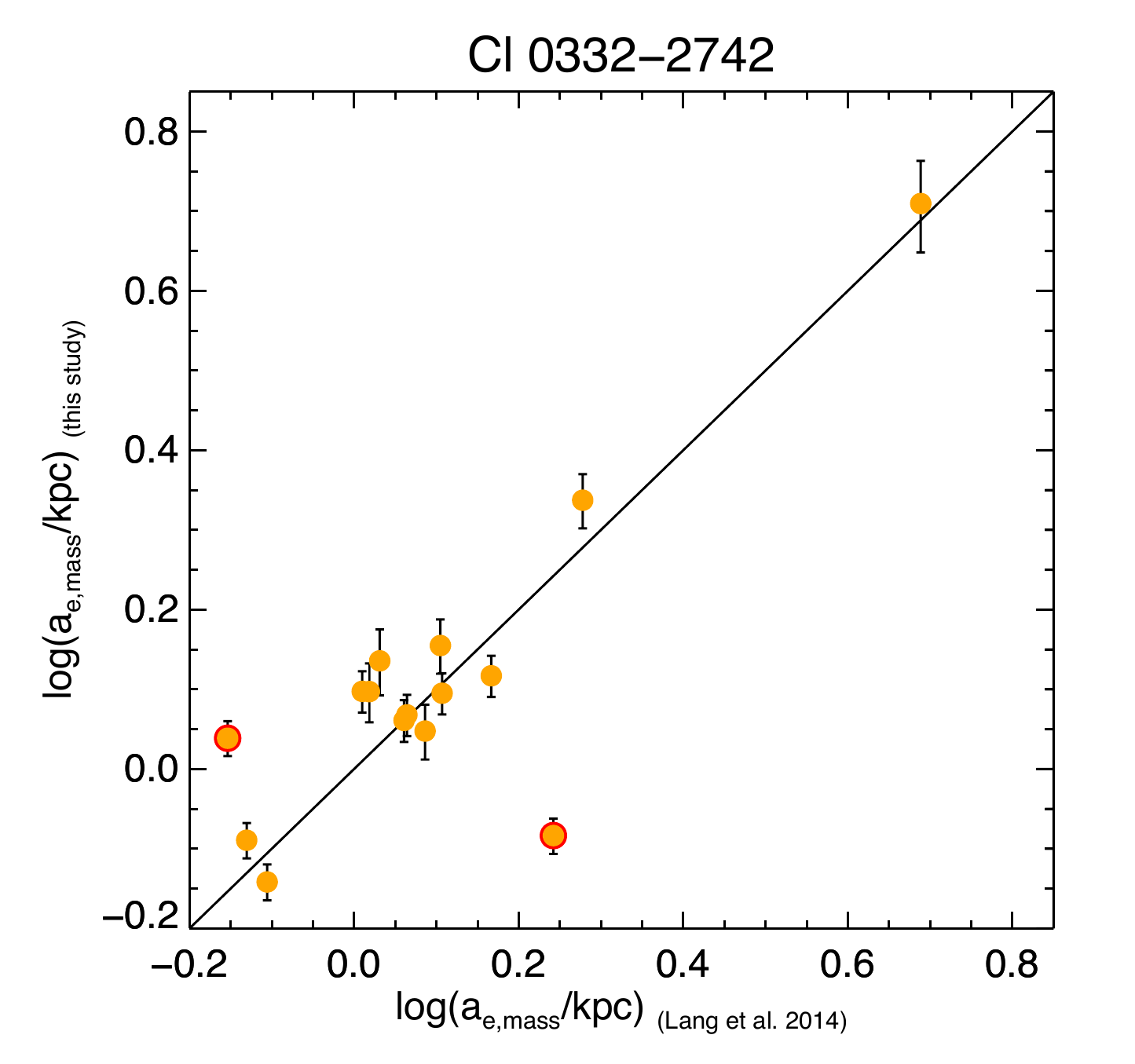}
  \caption[Comparison of mass-weighted sizes derived using $M_{*}/L$ - color relation with resolved SED fitting from \citet{Langetal2014}]{Comparison of mass-weighted sizes derived using $M_{*}/L$ - color relation with resolved SED fitting from \citet{Langetal2014}.  The solid black line is the one-to-one relation.  The error bars represent the $1\sigma$ uncertainties of our mass-weighted sizes. The two outliners (ID 25989, 25338) are marked in red. See text for details.}
  \label{fig_masssize_comparison}
\end{figure}

Figure~\ref{fig_masssize_comparison} shows a direct mass-weighted size comparison of the 15 galaxies in Cl~0332-2742 (P. Lang, private communication).   Overall, the sizes from the two methods are very consistent with each other.  The entire sample has a median difference and $1\sigma$ scatter of $0.012 \pm 0.081$ dex ($\sim20\%$ scatter).   We have inspected the mass maps of the two objects that deviate from the one-to-one relation, marked in red in Figure~\ref{fig_masssize_comparison}.  Galaxy ID 25989 has a close neighbouring object, which we treated as a separate object and fitted simultaneously.  On the other hand,  the close neighbour is not deblended in the 3D-HST catalogue (and hence not fitted in \citet{Langetal2014}).  As a result, our size of this object appears to be much smaller.   For galaxy ID 25338, there is a giant diffuse halo component in the galaxy which may make the sky level determination problematic and results in the size difference. 

Hence we conclude that the methods we used produced compatible results with those from the literature, at least at $z\sim1.61$.

\ifx
\begin{table}
  \centering
  \noindent
  \caption{Best-fit parameters of the stellar mass -- size relations of the three KCS clusters using semi-major axes as galaxy sizes}
  \label{tab_bestfit_sma}
  \begin{tabular}{@{}p{2.45cm}lccc@{}}
  \\
  \multicolumn{5}{c}{XMMU~J2235-2557} \\  
  \hline
  \hline
  Case &  Mass range &  $\alpha \pm \Delta \alpha$  & $\beta \pm \Delta \beta$ & $\epsilon$ \\
  \hline
  Light-weighted (C) & $10.5 \leq M_{*} \leq 11.5$              & $ 0.529  \pm 0.046 $  &  $ 0.660  \pm 0.155  $   & 0.163 \\  
  Mass-weighted (C) & $10.5 \leq M_{*} \leq 11.5$              & $ 0.320  \pm 0.058 $  &  $ 0.465  \pm 0.192  $   & 0.197 \\  
  \hline
  \\
  \multicolumn{5}{c}{XMMXCS~J2215-1738} \\  
  \hline
  \hline
  Case &  Mass range &  $\alpha \pm \Delta \alpha$  & $\beta \pm \Delta \beta$ & $\epsilon$ \\
  \hline
  Light-weighted (C) & $10.5 \leq M_{*} \leq 11.5$              & $ 0.446  \pm 0.038 $  &  $ 0.909  \pm 0.202  $   & 0.134 \\  
  Mass-weighted (C) & $10.5 \leq M_{*} \leq 11.5$              & $ 0.104  \pm 0.044 $  &  $ 0.605  \pm 0.241  $   & 0.120 \\  
   \hline
  \\
  \multicolumn{5}{c}{Cl~0332-2742} \\  
  \hline
  \hline
  Case &  Mass range &  $\alpha \pm \Delta \alpha$  & $\beta \pm \Delta \beta$ & $\epsilon$ \\
  \hline
  Light-weighted (C) & $10.5 \leq M_{*} \leq 11.5$              & $ 0.308  \pm 0.057 $  &  $ 0.631  \pm 0.228  $   & 0.140 \\  
  Mass-weighted (C) & $10.5 \leq M_{*} \leq 11.5$              & $ 0.113  \pm 0.042 $  &  $ 0.440  \pm 0.241  $   & 0.101 \\  
   \hline
   \hline
\end{tabular}
\end{table}

\begin{table*}
  \caption{The mean (of the best-fit skew normal distributions) and the median of the mass-normalised size distributions of the KCS clusters and the L14 field sample using semi-major axes as galaxy sizes}
  \centering
  \label{tab_sizeMN_sma}
  \begin{tabular}{lcccc@{}}
  \\
  \hline
  \hline
  
  \multicolumn{5}{c}{Light-weighted size distributions} \\
  Sample               &           $\overline{\log(a_{e,\rm{MN}})}$*                 &                  $\langle \log(a_{e,\rm{MN}}) \rangle$            & $p_{\rm{KS}}$      &   $p_{\rm{AD}}$      \\
  \hline
  XMMU~J2235 +  XCS~J2215                          &   $0.493 \pm 0.070$  &   $0.486 \pm 0.031$  &  0.45    &  0.86  \\  
  XMMU~J2235 +  XCS~J2215 ($\beta=0.76$)  &   $0.457 \pm 0.047$  &  $0.480 \pm 0.031$   &  0.70   &  0.06 \\   
  L14 field ($1.3 < z < 1.5$)                                 &   $0.396 \pm 0.038$  &  $0.394 \pm 0.026$   & 0.96   &  0.79 \\  
  Cl~0332-2742                                                    &        -                         &   $0.238 \pm 0.052$  &  -  &  -  \\   
  Cl~0332-2742  ($\beta=0.76$)                           &        -                         &   $0.229 \pm 0.053$  &  -  &  -  \\   
  L14 field ($1.5 < z < 1.7$)                                  &   $0.347 \pm 0.031$   &   $0.320 \pm 0.025$  &  0.43  & 0.18   \\   
  3 KCS clusters                                                    &   $0.392 \pm 0.079$   &  $0.448 \pm 0.032$  & 0.53    &  0.22   \\  
  3 KCS clusters  ($\beta=0.76$)                          &   $0.360 \pm 0.080$   &  $0.457 \pm 0.313$  & 0.13   &  0.004  \\   
  L14 field ($1.3 < z < 1.7$)                                   &   $0.376 \pm 0.023$   &   $0.347 \pm 0.018$  & 0.49   &  0.16   \\  
  \hline
  \hline
  \\
   \hline
  \hline
  \multicolumn{5}{c}{Mass-weighted size distributions} \\  
  Sample               &           $\overline{\log(a_{e,\rm{MN}})}$                 &                  $\langle \log(a_{e,\rm{MN}}) \rangle$            & $p_{\rm{KS}}$      &   $p_{\rm{AD}}$            \\
  \hline
  XMMU~J2235 +  XCS~J2215                            &       -                          &  $0.203  \pm 0.044$ &  -  &   -  \\  
  L14 field ($1.3 < z < 1.5$) ($\beta=0.76$)          &   $0.200 \pm 0.026$  &  $0.179  \pm 0.033$ & 0.95   &  0.22  \\  
  L14 field ($1.3 < z < 1.5$) ($\beta=0.49$)           &  $0.189 \pm 0.030$ &   $0.136  \pm 0.030$ & -   &  -    \\  
  Cl~0332-2742                                                      &       -                          &   $0.124 \pm 0.025$  &  -  &   -  \\  
  L14 field ($1.5 < z < 1.7$) ($\beta=0.76$)           &  -   &   $0.198 \pm 0.046$  &  -   &  -    \\  
  L14 field ($1.5 < z < 1.7$) ($\beta=0.49$)           &  -   &   $0.127 \pm 0.044$  &  -  &  -     \\  
  3 KCS clusters                                                     &  $0.159 \pm 0.121$    &  $0.175 \pm 0.034$  &  0.50  &  0.26      \\  
  L14 field ($1.3 < z < 1.7$)  ($\beta=0.76$)          &   -   & $0.175 \pm 0.034$   &  -  & -        \\  
  L14 field ($1.3 < z < 1.7$)  ($\beta=0.49$)          &  -  & $0.181 \pm 0.028$   & -   & -        \\  
  \hline
  \hline
  \multicolumn{4}{p{.7\textwidth}}{\textsuperscript{*}The uncertainties quoted for $\overline{\log(R_{e\rm{-circ},\rm{MN}})}$ are computed by bootstrapping.  We repeated the fitting procedure for 1000 times each with a randomly drawn subset of the sample, and the uncertainty is given by the standard deviation of these 1000 measurements.  The uncertainties of the median are estimated as $1.253\sigma/\sqrt{N_{\rm{gal,total}}}$, where $\sigma$ is the standard deviation of the size distributions.} \\
\end{tabular}
\end{table*}

\begin{figure*}
  \centering
  \includegraphics[scale=0.75]{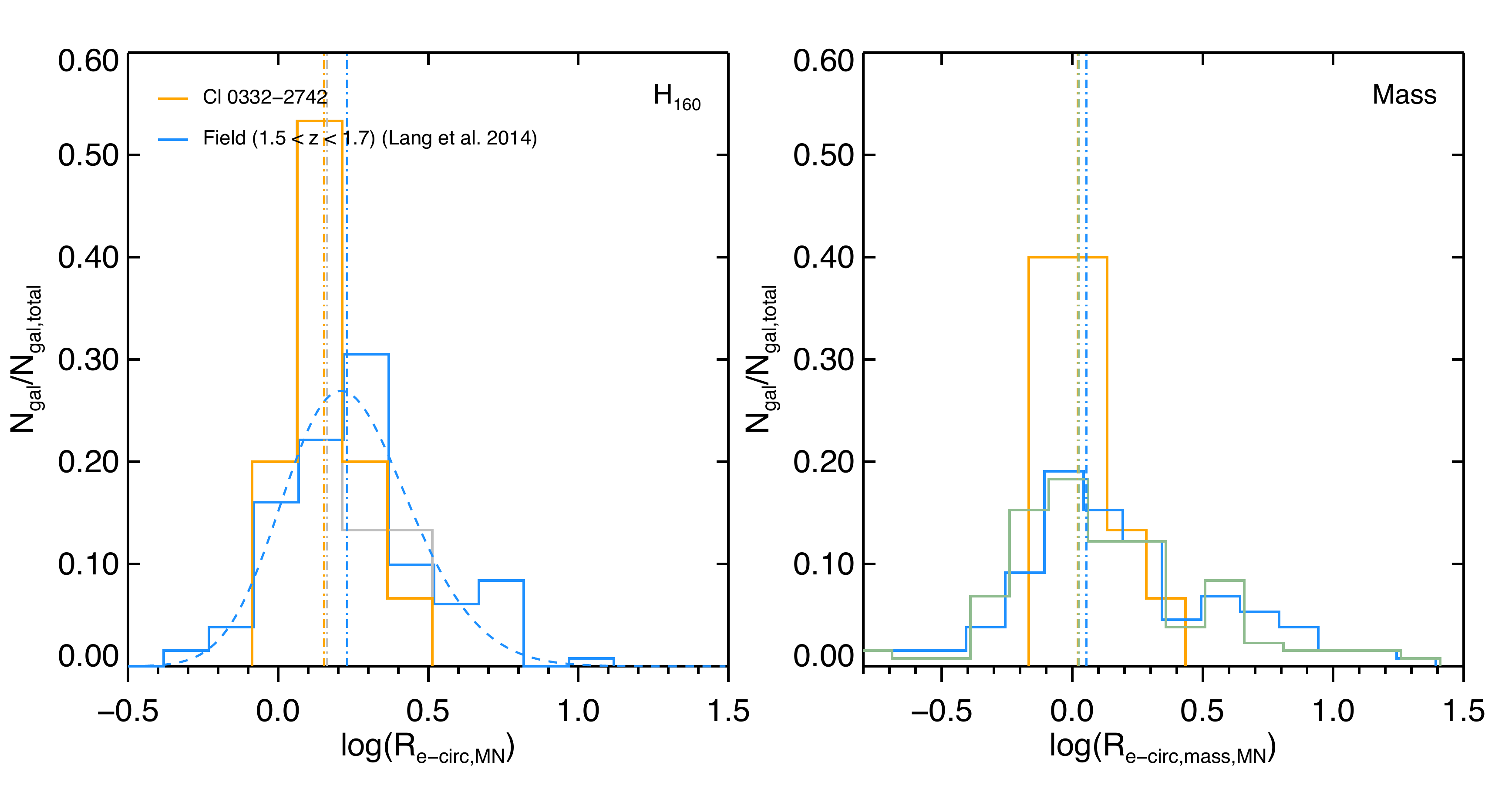}
  \caption[Comparison of the size ratio of the three KCS clusters with the field]{Comparison of the size ratio of the three KCS clusters with the field.  The median size ratios of XMMU~J2235-2557, XMMXCS~J2215-1738 and Cl~0332-2742 are shown as red, blue and green circles respectively.  The orange circle corresponds to the median size ratio of the progenitor bias corrected SPIDER cluster sample at $z\sim0$.  The brown circles and line show the binned median size ratio of the progenitor bias corrected L14 field sample across redshift in redshift bin of 0.25, while the light brown dots correspond to individual galaxies in the sample.  The grey triangles and line show the binned median size ratio of the \citet{Szomoruetal2013} field sample across redshift in redshift bin of 0.5, while the light grey triangles represent individual galaxies in their sample.  The error bars show the uncertainty of the median in each bin.}
  \label{field_comparison_ratio_z_circ}
\end{figure*}

\begin{figure*}
  \centering
  \includegraphics[scale=0.65]{./fig_ratio_sma}
  \caption[Comparison of the size ratio of the three KCS clusters with the field]{Comparison of the size ratio of the three KCS clusters with the field.  The median size ratios of XMMU~J2235-2557, XMMXCS~J2215-1738 and Cl~0332-2742 are shown as red, blue and green circles respectively. Same as Figure~\ref{field_comparison_ratio_z_circ}, but using semi-major axis ($a_e$) as galaxy size instead of the circularized radius.}
  \label{field_comparison_ratio_z_sma}
\end{figure*}
\fi

\section{B. The fitted parameters of mass -- size relations using semi-major axes}
\label{The fitted parameters of mass -- size relations using semi-major axes}
Table~\ref{tab_bestfit_sma2} and~\ref{tab_bestfitmass_sma2} show the fitted parameters stellar mass -- size relations of the three clusters using semi-major axes ($a_e$) as galaxy size.

\begin{table}
  \centering
  \noindent
  \caption{Best-fit parameters of the stellar mass -- light-weighted size relations of the three KCS clusters}
  \label{tab_bestfit_sma2}
  \begin{tabular}{@{}p{0.3cm}lccc@{}}
  \\
  \multicolumn{5}{c}{XMMU~J2235-2557\textsuperscript{$\star$}} \\  
  \hline
  \hline
  \multicolumn{5}{l}{Stellar mass -- light-weighted size relation} \\
  \hline
  Case &  Mass range &  $\alpha \pm \Delta \alpha$  & $\beta \pm \Delta \beta$ & $\epsilon$ \\
  \hline
A & $10.0 \leq M_{*} \leq 11.5$              & $ 0.534 \pm  0.057 $ &  $ 0.548  \pm 0.144 $  &  0.212  \\    
C & $10.5 \leq M_{*} \leq 11.5$              & $ 0.529  \pm 0.046 $  &  $ 0.660  \pm 0.155  $   & 0.163  \\  
D & $10.5 \leq M_{*} \leq 11.5$ (spec)   & $ 0.534  \pm 0.051 $  &  $ 0.584  \pm  0.195 $   & 0.179  \\  
  \hline
  \multicolumn{5}{p{.45\textwidth}}{\textsuperscript{$\star$} Since all the spectroscopic members in XMMU~J2235-2557 are $\log(M_{*}/M_\odot) \geq 10.5$, case B is identical to case D.} \\
  \\
  \multicolumn{5}{c}{XMMXCS~J2215-1738} \\  
  \hline
  \hline
  Case &  Mass range &  $\alpha \pm \Delta \alpha$  & $\beta \pm \Delta \beta$ & $\epsilon$ \\
  \hline
  A & $10.0 \leq M_{*} \leq 11.5$              & $ 0.444 \pm 0.044$  &  $0.720 \pm 0.179$  &  0.172 \\    
  B & $10.0 \leq M_{*} \leq 11.5$ (spec)  & $ 0.396 \pm 0.057$   &  $1.174 \pm 0.309$  &  0.129 \\    
  C & $10.5 \leq M_{*} \leq 11.5$              & $ 0.446 \pm  0.038 $  &  $0.909 \pm  0.202 $   & 0.134 \\     
  D & $10.5 \leq M_{*} \leq 11.5$ (spec)  & $  0.401 \pm  0.068 $  &  $1.596 \pm  0.662 $   & 0.125 \\    
  \hline
  \\
  \multicolumn{5}{c}{Cl~0332-2742*} \\  
  \hline
  \hline
  Case &  Mass range &  $\alpha \pm \Delta \alpha$  & $\beta \pm \Delta \beta$ & $\epsilon$ \\
  \hline
  C & $10.5 \leq M_{*} \leq 11.5$               & $ 0.228  \pm 0.057 $  &  $ 0.631  \pm 0.308  $   & 0.140 \\     
  D & $10.5 \leq M_{*} \leq 11.5$ (spec)    & $ 0.152 \pm 0.106  $   &  $1.012 \pm 0.638   $  &  0.176 \\      
   \hline
   \hline
    \multicolumn{5}{p{.45\textwidth}}{\textsuperscript{*} Since all the selected galaxies in Cl~0332-2742 have $\log(M_{*}/M_\odot) \geq 10.5$, only case C \& D are applicable.} \\
\end{tabular}
\end{table}

\begin{table}
  \centering
  \noindent
  \caption{Best-fit parameters of the stellar mass -- mass-weighted size relations of the three KCS clusters}
  \label{tab_bestfitmass_sma2}
  \begin{tabular}{@{}p{0.3cm}lccc@{}}
  \\
  \multicolumn{5}{c}{XMMU~J2235-2557\textsuperscript{$\star$}} \\  
  \hline
  \hline
  \multicolumn{5}{l}{Stellar mass -- mass-weighted size relation} \\
  \hline
  Case &  Mass range &  $\alpha \pm \Delta \alpha$  & $\beta \pm \Delta \beta$ & $\epsilon$ \\
  \hline
 A & $10.0 \leq M_{*} \leq 11.5$              & $ 0.335 \pm 0.074 $  &  $ 0.325 \pm 0.190 $  &  0.246 \\        
 C & $10.5 \leq M_{*} \leq 11.5$              & $ 0.320  \pm 0.058 $  &  $ 0.465  \pm 0.192  $   & 0.197  \\       
 D & $10.5 \leq M_{*} \leq 11.5$ (spec)   & $ 0.321  \pm 0.065 $  &  $ 0.422  \pm 0.233  $  & 0.213   \\          
  \hline
  \multicolumn{5}{p{.45\textwidth}}{\textsuperscript{$\star$} Since all the spectroscopic members in XMMU~J2235-2557 are $\log(M_{*}/M_\odot) \geq 10.5$, case B is identical to case D.} \\
  \\
  \multicolumn{5}{c}{XMMXCS~J2215-1738} \\  
  \hline
  \hline
   Case &  Mass range &  $\alpha \pm \Delta \alpha$  & $\beta \pm \Delta \beta$ & $\epsilon$ \\
   \hline
 A & $10.0 \leq M_{*} \leq 11.5$              & $  0.107 \pm 0.046 $  &  $ 0.386 \pm 0.206 $  &  0.143 \\          
 B & $10.0 \leq M_{*} \leq 11.5$ (spec)  & $   0.121 \pm 0.084 $  &  $ 1.060 \pm 0.917 $  &  0.170 \\          
 C & $10.5 \leq M_{*} \leq 11.5$              & $  0.104 \pm 0.044 $  &  $ 0.605 \pm 0.240  $  & 0.120 \\           
 D & $10.5 \leq M_{*} \leq 11.5$ (spec)   & $  0.128 \pm 0.083 $  &  $ 1.058 \pm 0.881  $  & 0.166 \\          
   \hline
  \\
  \multicolumn{5}{c}{Cl~0332-2742*} \\  
  \hline
  \hline
   Case &  Mass range &  $\alpha \pm \Delta \alpha$  & $\beta \pm \Delta \beta$ & $\epsilon$ \\
   \hline
 C & $10.5 \leq M_{*} \leq 11.5$              & $ 0.113  \pm  0.042$  &  $0.440 \pm 0.241$  &  0.101 \\          
 D & $10.5 \leq M_{*} \leq 11.5$ (spec)   & $ 0.072  \pm  0.103$  &  $0.753 \pm 0.647$  &  0.182 \\          
   \hline
   \hline
 \multicolumn{5}{p{.45\textwidth}}{\textsuperscript{*} Since all the selected galaxies in Cl~0332-2742 have $\log(M_{*}/M_\odot) \geq 10.5$, only case C \& D are applicable.} \\
\end{tabular}
\end{table}

\section{C. Details and results of the color gradient test scenarios}
\label{Results of the color gradient test scenarios}
In this section, we expand the discussion in Section~\ref{subsec: The origin of the $M_{*}/L$ gradients} on the details and results of our three test scenarios to evolve the observed color gradients \citep[see][for additional details]{Chanetal2016}.

\begin{figure}
  \centering
  \includegraphics[scale=0.70]{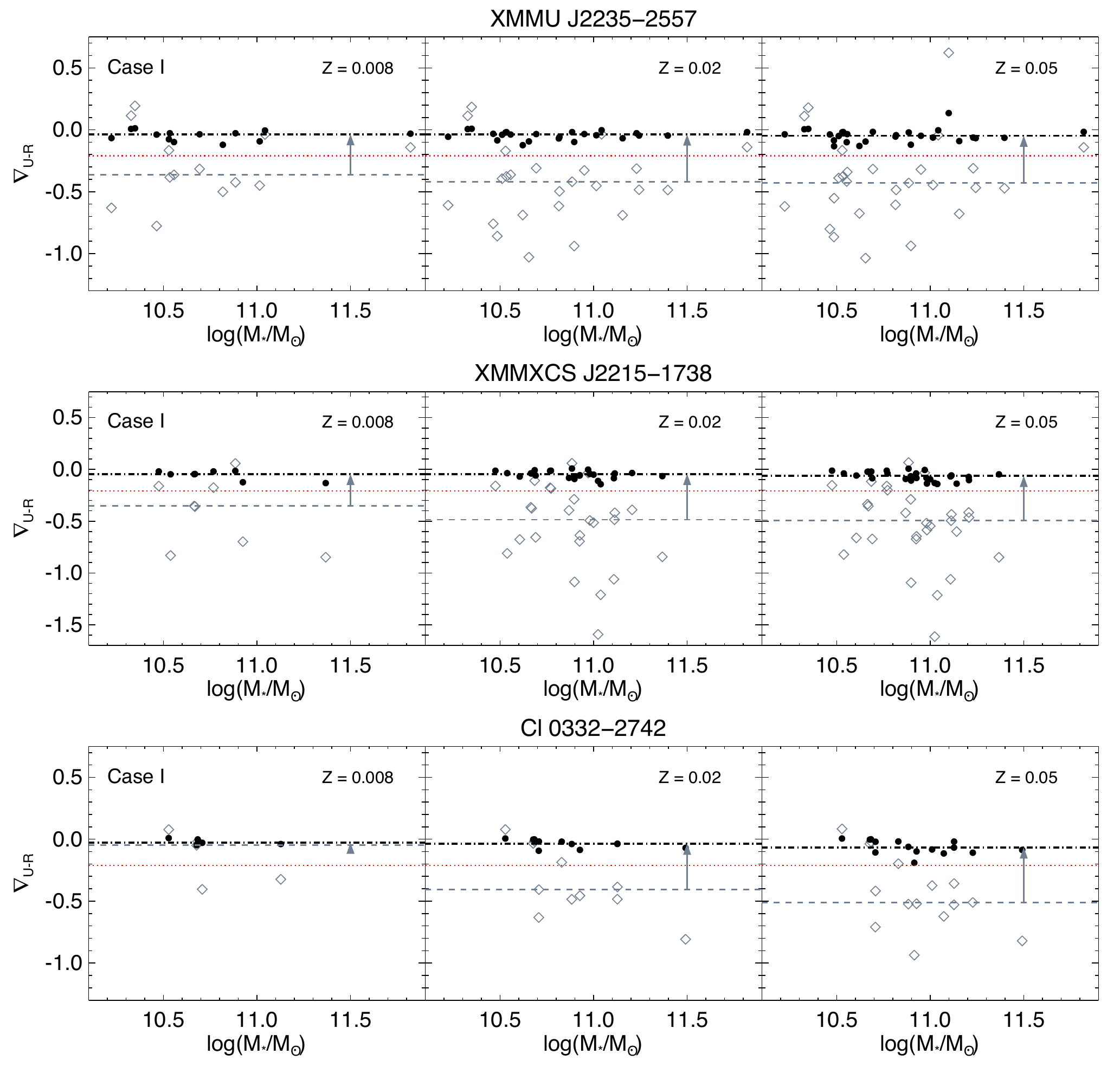}
  \caption[Evolution of colour gradients over redshift in case I (Pure age-driven gradient evolution) for the KCS clusters]{Evolution of colour gradients over redshift in case I (Pure age-driven gradient evolution) for the KCS clusters.   From top row to bottom:  The evolution of XMMU~J2235-2557, XMMXCS~J2215-1738 and Cl~0332-2742 respectively.  Left to right: The sub-solar ($Z=0.008$), solar ($Z=0.02$) and super-solar ($Z=0.05$) metallicity scenarios in each cluster.   Grey diamonds correspond to the $(U-R)$ gradient at the cluster redshifts, with the median plotted as the grey dashed line.  Black circles indicate the predicted $(U-R)$ gradient at redshift 0 of the same galaxy, and the black dot-dashed line indicate the median.  Their masses remain unchanged as we do not consider any mass growth over the period.  The grey arrow in each panel shows the direction of evolution of the median gradient.  The red dotted line corresponds to the observed $(U-R)$ gradient at redshift 0 by \citet{Wuetal2005}.}
  \label{fig_gradevo_c1}
\end{figure}

In Figure~\ref{fig_gradevo_c1}, we show the evolution of the rest-frame $(U-R)$ colour gradient from the cluster redshifts to $z=0$ under the assumption of pure age gradient (Case I) for the three KCS clusters.  Although the colour gradients evolve in the correct direction, the median gradients of the evolved sample at $z=0$ in all three clusters are very close to zero (i.e. flat colour profiles), and hence are too shallow compared to the local colour gradients. For example, the median evolved colour gradients of the solar metallicity scenarios ($Z=0.02$) are $-0.038 \pm 0.009$,  $-0.044 \pm 0.009$ and $-0.036 \pm 0.013$ for XMMU~J2235-2557,  XMMXCS~J2215-1738 and Cl~0332-2742, respectively.

Under a different assumed metallicity for the inner region, occasionally the age (or the metallicity for the following case II) determination for some galaxies results in an unphysical value.  
Galaxies that have deduced ages that are too old for the cluster (e.g. $> 4.48$ Gyr for $z=1.39$ or $> 13.46$ Gyr for $z=0$ within $1\sigma$ uncertainty) are discarded to avoid drawing incorrect conclusions.  They may simply be unphysical to be modelled with a particular metallicity or have a more complicated star formation history, which cannot be well-represented by SSPs.

Under the assumption of sub-solar metallicity, only 13 out of 27 galaxies in XMMU~J2235-2557 have a physically meaningful age.  On the other hand, most of the galaxies are retained if we assume a solar (24 out of 27) or super-solar metallicity (27 out of 27).  Similarly, 8, 25 and 28 out of 29 galaxies in XMMXCS~J2215-1738 are retained in each metallicity scenario ($Z=0.008,0.02,0.05$).  For Cl~0332-2742,  5, 11 and 15 out of 15 galaxies are retained.  We conclude that in the reasonable range of metallicity that we covered, a pure age-driven gradient is not able to match the observed evolution of colour gradients.

The reason behind the rapid evolution is the flattening of the SSP colour-age relation over time \citep[see Figure~D1 in][]{Chanetal2016}.  Since we assume identical metallicities for both inner and outer regions, the inner and outer regions of an individual galaxy lie on the same colour-age relation.  Take the solar metallicity $Z=0.02$ case in Figure~D1 as an example, the $U-R$ colour increases sharply from 0 to 4 Gyr but flattens after, hence the $(U-R)$ gradient evolves to almost zero at redshift 0.

\begin{figure}
  \centering
  \includegraphics[scale=0.75]{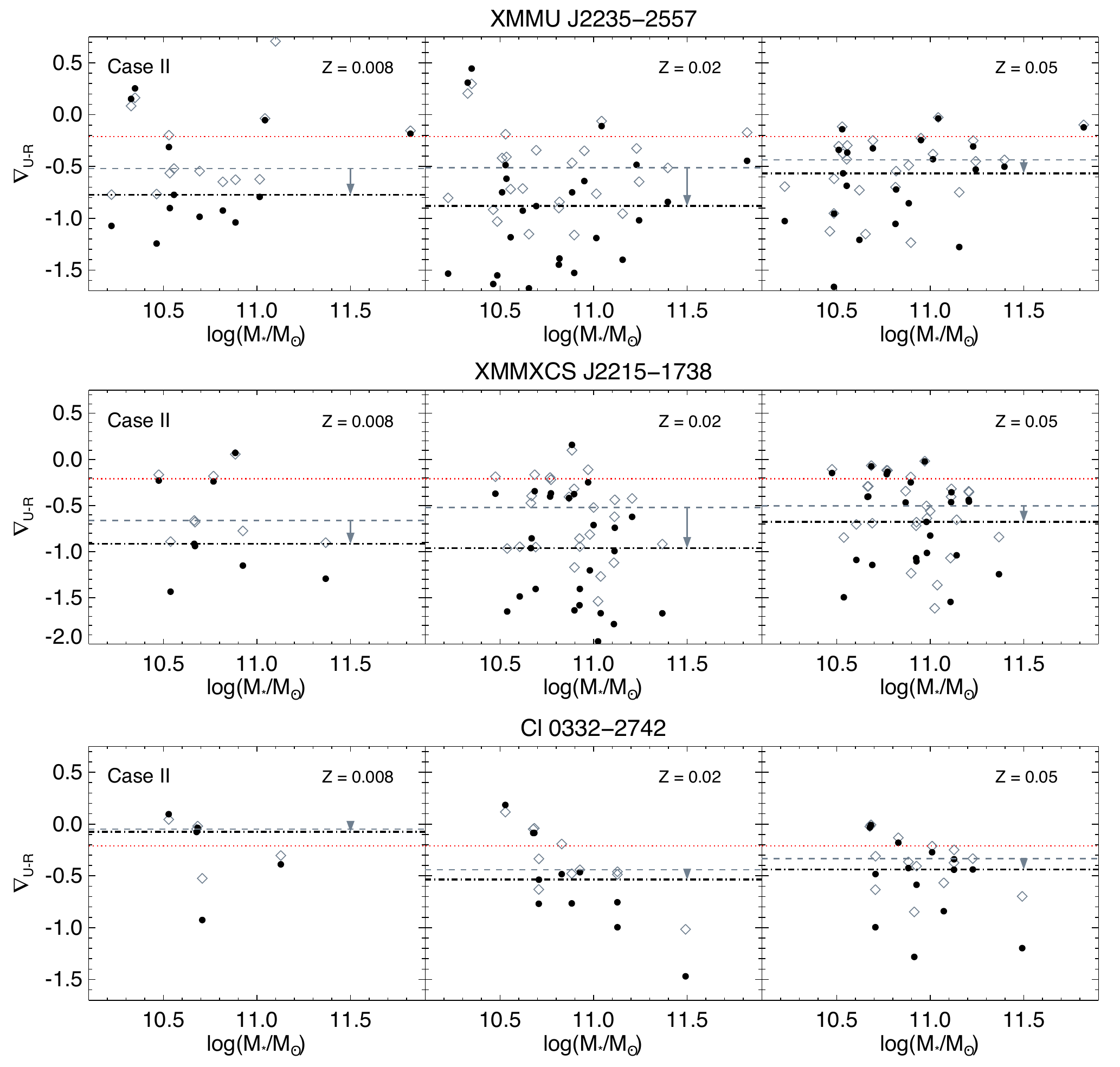}
  \caption[Evolution of colour gradients over redshift in case II (Pure metallicity-driven gradient evolution) for the KCS clusters]{Evolution of colour gradients over redshift in case II (Pure metallicity-driven gradient evolution) for the KCS clusters.   From top row to bottom:  The evolution of XMMU~J2235-2557, XMMXCS~J2215-1738 and Cl~0332-2742 respectively.  Left to right: The sub-solar ($Z=0.008$), solar ($Z=0.02$) and super-solar ($Z=0.05$) metallicity scenarios in each cluster.   Grey diamonds correspond to the $(U-R)$ gradient at the cluster redshifts, with the median plotted as the grey dashed line.  Black circles indicate the predicted $(U-R)$ gradient at redshift 0 of the same galaxy, and the black dot-dashed line indicate the median.  Their masses remain unchanged as we do not consider any mass growth over the period.  The grey arrow in each panel shows the direction of evolution of the median gradient.  The red dotted line corresponds to the observed $(U-R)$ gradient at redshift 0 by \citet{Wuetal2005}.}
  \label{fig_gradevo_c2}
\end{figure}

Instead of using a flat metallicity gradient as case I,  Figure~\ref{fig_gradevo_c2} shows the evolution of the $(U-R)$ gradient under the assumption of pure metallicity-driven gradient (Case II), i.e. a flat age gradient $\nabla_{age} = 0$.  Similar to case I, galaxies that have unphysical ages or metallicities are discarded. 14, 24 and 24 out of 27 galaxies in XMMU~J2235-2557 are retained in each metallicity scenario respectively.  For XMMXCS~J2215-1738,  8, 25 and 27 out of 29 galaxies are retained.  5, 11 and 14 out of 15 galaxies are retained for Cl~0332-2742.

From Figure~\ref{fig_gradevo_c2}, we can see that the median gradients of the evolved sample are even steeper compared to the one at the cluster redshifts.  This is true for all metallicity scenarios.  It is therefore clear that a pure metallicity-driven gradient fails to reproduce the observed gradient.  As an example, the median gradients of the evolved sample of the solar metallicity scenarios are $-0.88 \pm 0.15$,  $-0.96 \pm 0.16$ and $-0.54 \pm 0.19$ for XMMU~J2235-2557,  XMMXCS~J2215-1738 and Cl~0332-2742, respectively.  Of course in reality metallicity in individual galaxies differs, but mixing galaxies with different metallicity within our metallicity range would not change this conclusion.

\begin{figure}
  \centering
  \includegraphics[scale=0.75]{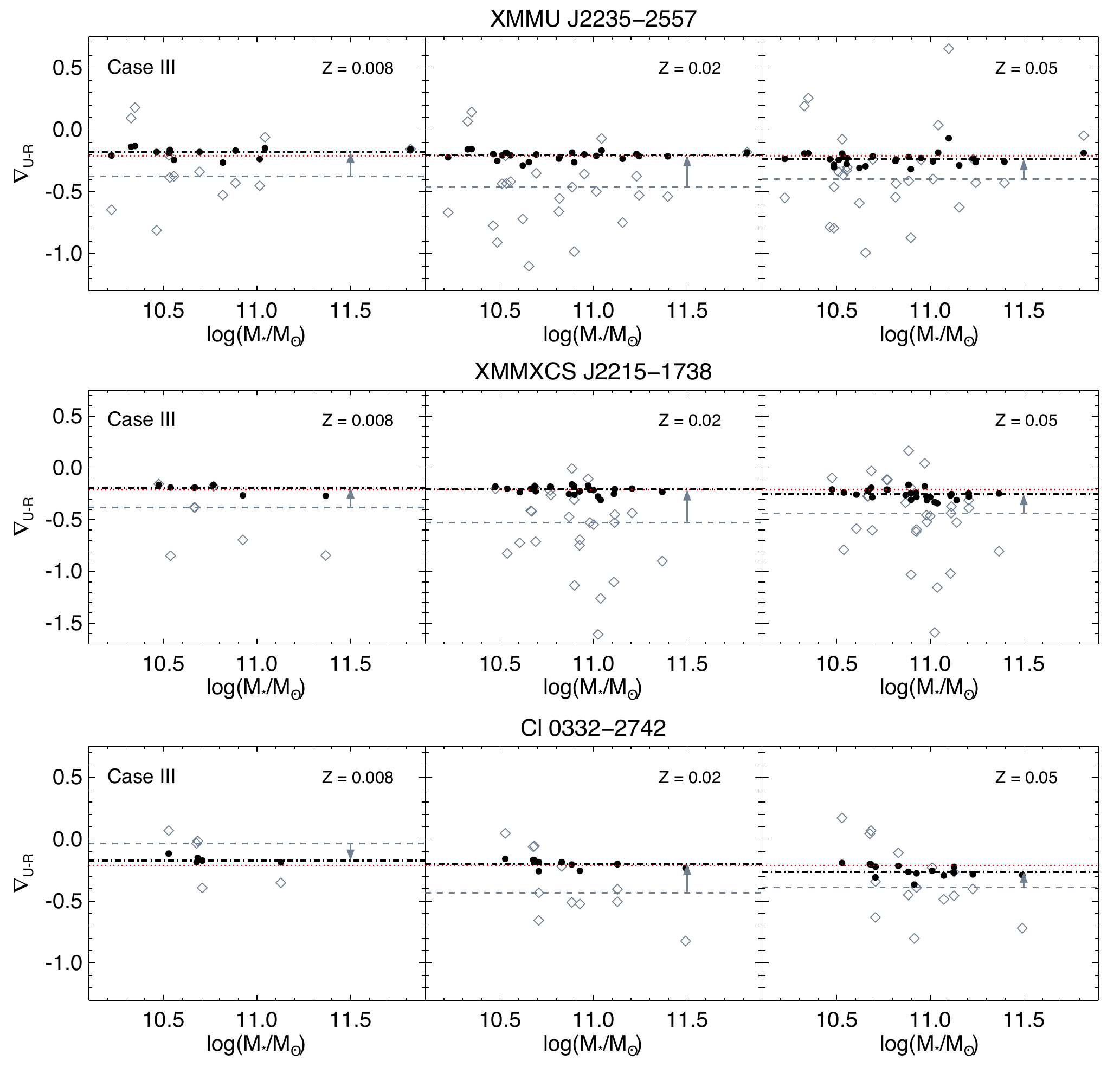}
  \caption[Evolution of colour gradients over redshift in case III (Age-driven gradient evolution with assumed metallicity gradient) for the KCS clusters]{Evolution of colour gradients over redshift in case III (Age-driven gradient evolution with assumed metallicity gradient) for the KCS clusters.   From top row to bottom:  The evolution of XMMU~J2235-2557, XMMXCS~J2215-1738 and Cl~0332-2742, respectively.  Left to right: The sub-solar ($Z=0.008$), solar ($Z=0.02$) and super-solar ($Z=0.05$) metallicity scenarios in each cluster.   Grey diamonds correspond to the $(U-R)$ gradient at the cluster redshifts, with the median plotted as the grey dashed line.  Black circles indicate the predicted $(U-R)$ gradient at redshift 0 of the same galaxy, and the black dot-dashed line indicate the median.  Their masses remain unchanged as we do not consider any mass growth over the period.  The grey arrow in each panel shows the direction of evolution of the median gradient.  The red dotted line corresponds to the observed $(U-R)$ gradient at redshift 0 by \citet{Wuetal2005}. Due to a larger number of discarded objects, the initial median color gradient of Cl~0332-2742 in the sub-solar metallicity scenario is drastically different from the other two scenarios, which results in an opposite direction of evolution.}
  \label{fig_gradevo_c3}
\end{figure}

Figure~\ref{fig_gradevo_c3} shows the results of Case III, the evolution of the $(U-R)$ gradient with a metallicity gradient as observed in local passive galaxies: $\nabla_{Z} = -0.2$.  Again,  galaxies with unphysical ages are rejected. For XMMU~J2235-2557, 13, 24 and 27 out of 27 galaxies are retained in each metallicity scenario respectively.  7, 25 and 28 out of 29 galaxies in XMMXCS~J2215-1738 are retained, and 5, 11 and 14 out of 15 galaxies are retained in Cl~0332-2742.

The solar metallicity scenario works reasonably well for the majority of the sample in all three clusters with evolved median gradients of $\nabla_{U-R} = -0.21 \pm 0.01$, $-0.21 \pm 0.01$ and $-0.20 \pm 0.01$ for XMMU~J2235-2557, XMMXCS~J2215-1738 and Cl~0332-2742, respectively,  which is in close agreement with the observed value in the local universe by \citet{Wuetal2005}.

Despite a number of objects have to be discarded due to unphysical ages (especially in XMMXCS~J2215-1738 and Cl~0332-2742), the median gradient as well as the individual gradients of the evolved samples in the sub-solar metallicity scenario are also close to but slightly smaller than the observed local value.  Note that the initial median color gradient of Cl~0332-2742 in the sub-solar metallicity scenario is drastically different from the other two scenarios due to a large number of discarded objects.  Assuming super-solar metallicity for the inner regions, on the other hand, predicts gradients that are slightly too steep.

Besides the median values of the colour gradients, the 1$\sigma$ scatter of the evolved colour gradients is also in excellent agreement to the local value by \citet{Wuetal2005} ($\sigma_{U-R} = 0.04$).  For example, for the solar metallicity scenario the scatter reduces from $\sigma_{U-R} = 0.32,  0.36, 0.26$ at the cluster redshifts to $0.034, 0.035, 0.035$ at $z=0$ for the three clusters respectively.

\section{D. Photometry, Structural parameters and color gradients of the red sequence galaxies in the KCS clusters}
\label{app:Photometry, Structural parameters and color gradients of the red sequence galaxies in the KCS clusters}
The photometric catalogues of the red sequence objects in the three clusters are provided in Table~\ref{tab_photopara_xmmu}, Table~\ref{tab_photopara_xmmxcs} and Table~\ref{tab_photopara_cl}, respectively.  In this paper, we have revised the sample of XMMU~J2235-2557 with the additional color-color selection and new redshift information from recent KCS observations. Table~\ref{tab_photopara_xmmu} comprises the new $z_{850} - J_{125}$ and the $J_{125} - H_{160}$ color we used for the $UVR$ selection.  It also comprises more updated spectroscopic member information compared to Table F1 in \citet{Chanetal2016}.  There are 3 objects that are in the red sequence sample of \citet{Chanetal2016} but are excluded in Table~\ref{tab_photopara_xmmu}, as they are confirmed spectroscopically as interlopers with new redshift information.  We also updated the coordinates of these galaxies to those from our new cross-match to the HAWK-I data (see Section~\ref{sec:Construction of photometric catalogues}).  The $H_{160}$ magnitudes and $z_{850}-H_{160}$ colors are identical to those in Table F1 in \citet{Chanetal2016} and are provided for completeness.

The best-fitting light-weighted and mass-weighted structural parameters as well as their corresponding uncertainties of XMMXCS~J2215-1738 and Cl~0332-2742 are provided in Table~\ref{tab_strpara_xmmxcs} and Table~\ref{tab_strpara_cl}, respectively.  For the structural parameters of XMMU~J2235-2557 the interested reader can refer to \citet{Chanetal2016}. 


\ifx true false

\begin{table}
\centering 
\caption{Photometry of the red sequence galaxies in XMMU~J2235-2557}
 (To be updated)
   \label{tab_photopara_xmmu}
\begin{tabular}{@{}lccccccccc@{}}
  \hline
  \hline
  ID   &  ID & R.A.\textsuperscript{a} & Decl. &  $H_{160}$\textsuperscript{b}  &   $z_{850}-H_{160}$\textsuperscript{c} &   $z_{850}-J_{125}$\textsuperscript{d}  & $J_{125}-H_{160}$\textsuperscript{d}  &   $UVR$\textsuperscript{e}   &   Spec-z\textsuperscript{f} \\
     &  (C16)                    & (J2000) & (J2000) & (\texttt{MAG\_AUTO}) & (AB mag) & (AB mag) & (AB mag) & passive &  member \\
   \hline
     &  36 &  338.829482 &  -25.974178 & $  21.105 \pm  0.004$ & $  1.765 \pm  0.010$ & $  1.414 \pm  0.009$ & $  0.390 \pm  0.005$ &  1 &  1 \\
     &  77 &  338.818075 &  -25.972015 & $  22.351 \pm  0.013$ & $  1.844 \pm  0.037$ & $  1.147 \pm  0.035$ & $  0.680 \pm  0.019$ &  0 &  0 \\
     &  148 &  338.833366 &  -25.967277 & $  21.755 \pm  0.008$ & $  1.527 \pm  0.024$ & $  1.052 \pm  0.023$ & $  0.527 \pm  0.014$ &  0 &  0 \\
     &  159 &  338.825351 &  -25.968396 & $  21.965 \pm  0.011$ & $  1.259 \pm  0.015$ & $  0.834 \pm  0.015$ & $  0.478 \pm  0.010$ &  0 &  1 \\
     &  170 &  338.836760 &  -25.961046 & $  19.558 \pm  0.003$ & $  1.956 \pm  0.009$ & $  1.542 \pm  0.008$ & $  0.451 \pm  0.004$ &  1 &  1 \\
     &  198 &  338.824435 &  -25.936956 & $  22.339 \pm  0.024$ & $  1.377 \pm  0.025$ & $  1.017 \pm  0.026$ & $  0.401 \pm  0.032$ &  1 &  0 \\
\end{tabular}
\end{table} 

\begin{table}

\centering 
\caption{Photometry of the red sequence galaxies in XMMXCS~J2215-1738}
 (To be updated)
   \label{tab_photopara_xmmxcs}
  \begin{tabular}{@{}lcccccccc@{}}
  \hline
  \hline
  ID\textsuperscript & R.A. & Decl. &  $H_{160}$\textsuperscript{a}  &   $z_{850}-H_{160}$\textsuperscript{b} &   $z_{850}-J_{125}$\textsuperscript{c}  & $J_{125}-H_{160}$\textsuperscript{c}  &   $UVR$\textsuperscript{d}   &   Spec-$z$\textsuperscript{e} \\
                  & (J2000) & (J2000) & (\texttt{MAG\_AUTO}) & (AB mag) & (AB mag) & (AB mag) & passive &  member \\
   \hline
  1 &  333.9776786 &  -17.6486591 & $  22.142 \pm  0.005$ & $  2.199 \pm  0.016$ & $  1.560 \pm  0.015$ & $  0.721 \pm  0.005$ &  0 &  0 \\
  317 &  334.0024952 &  -17.6434686 & $  21.816 \pm  0.003$ & $  1.817 \pm  0.010$ & $  1.436 \pm  0.008$ & $  0.402 \pm  0.003$ &  1 &  0 \\
  321 &  334.0006392 &  -17.6435969 & $  22.249 \pm  0.004$ & $  1.815 \pm  0.013$ & $  1.025 \pm  0.010$ & $  0.766 \pm  0.005$ &  0 &  0 \\
  342 &  334.002421 &  -17.643257 & $  21.523 \pm  0.003$ & $  2.216 \pm  0.011$ & $  1.789 \pm  0.009$ & $  0.444 \pm  0.003$ &  1 &  0 \\
  498 &  333.9926561 &  -17.6395221 & $  22.067 \pm  0.004$ & $  1.802 \pm  0.011$ & $  1.213 \pm  0.009$ & $  0.605 \pm  0.004$ &  0 &  0 \\
  500 &  333.9846986 &  -17.6361219 & $  22.364 \pm  0.006$ & $  1.489 \pm  0.013$ & $  0.868 \pm  0.010$ & $  0.580 \pm  0.006$ &  0 &  0 \\
\end{tabular}
\end{table}

\fi

\begin{table}
\centering 
\caption{Photometry of the red sequence galaxies in XMMU~J2235-2557}
   \label{tab_photopara_xmmu}
  \begin{tabular}{@{}lcccccccc@{}}
  \hline
  \hline
      ID\textsuperscript{a} & R.A.\textsuperscript{b} & Decl. &  $H_{160}$\textsuperscript{c}  &   $z_{850}-H_{160}$\textsuperscript{d} &   $z_{850}-J_{125}$\textsuperscript{e}  & $J_{125}-H_{160}$\textsuperscript{e}  &   $UVR$\textsuperscript{f}   &   Spec-z\textsuperscript{g} \\
                   & (J2000) & (J2000) & (\texttt{MAG\_AUTO}) & (AB mag) & (AB mag) & (AB mag) & passive &  member \\
   \hline
       36 &  338.829552 &  -25.974256 & $  21.105 \pm  0.004$ & $  1.765 \pm  0.010$ & $  1.414 \pm  0.009$ & $  0.390 \pm  0.005$ &  1 &  1 \\
       77 &  338.818124 &  -25.972095 & $  22.351 \pm  0.013$ & $  1.844 \pm  0.037$ & $  1.147 \pm  0.035$ & $  0.680 \pm  0.019$ &  0 &  0 \\
       148 &  338.833442 & -25.967342 & $  21.755 \pm  0.008$ & $  1.527 \pm  0.024$ & $  1.052 \pm  0.023$ & $  0.527 \pm  0.014$ &  0 &  0 \\
       159 &  338.825410 & -25.968464 & $  21.965 \pm  0.011$ & $  1.259 \pm  0.015$ & $  0.834 \pm  0.015$ & $  0.478 \pm  0.010$ &  0 &  1 \\
       170 &  338.836838 & -25.961102 & $  19.558 \pm  0.003$ & $  1.956 \pm  0.009$ & $  1.542 \pm  0.008$ & $  0.451 \pm  0.004$ &  1 &  1 \\
       198 &  338.824493 & -25.936966 & $  22.339 \pm  0.024$ & $  1.377 \pm  0.025$ & $  1.017 \pm  0.026$ & $  0.401 \pm  0.032$ &  1 &  0 \\
\end{tabular}
\end{table} 

\begin{table}

\centering 
\caption{Photometry of the red sequence galaxies in XMMXCS~J2215-1738}
   \label{tab_photopara_xmmxcs}
  \begin{tabular}{@{}lcccccccc@{}}
  \hline
  \hline
  ID\textsuperscript{a} & R.A. & Decl. &  $H_{160}$\textsuperscript{b}  &   $z_{850}-H_{160}$\textsuperscript{c} &   $z_{850}-J_{125}$\textsuperscript{d}  & $J_{125}-H_{160}$\textsuperscript{d}  &   $UVR$\textsuperscript{e}   &   Spec-$z$\textsuperscript{f} \\
                  & (J2000) & (J2000) & (\texttt{MAG\_AUTO}) & (AB mag) & (AB mag) & (AB mag) & passive &  member \\
   \hline
  412 &  333.977679 &  -17.648659 & $  22.142 \pm  0.005$ & $  2.199 \pm  0.016$ & $  1.560 \pm  0.015$ & $  0.721 \pm  0.005$ &  0 &  0 \\
  561 &  334.000639 &  -17.643597 & $  22.249 \pm  0.004$ & $  1.815 \pm  0.013$ & $  1.025 \pm  0.010$ & $  0.766 \pm  0.005$ &  0 &  0 \\
  562 &  334.002495 &  -17.643469 & $  21.816 \pm  0.003$ & $  1.817 \pm  0.010$ & $  1.436 \pm  0.008$ & $  0.402 \pm  0.003$ &  1 &  0 \\
  575 &  334.002421 &  -17.643257 & $  21.523 \pm  0.003$ & $  2.216 \pm  0.011$ & $  1.789 \pm  0.009$ & $  0.444 \pm  0.003$ &  1 &  0 \\
  692 &  333.992656 &  -17.639522 & $  22.067 \pm  0.004$ & $  1.802 \pm  0.011$ & $  1.213 \pm  0.009$ & $  0.605 \pm  0.004$ &  0 &  0 \\
  710 &  333.983600 &  -17.639072 & $  21.874 \pm  0.003$ & $  2.020 \pm  0.010$ & $  1.580 \pm  0.008$ & $  0.441 \pm  0.003$ &  1 &  1 \\
\end{tabular}
\end{table}

\begin{table}

\centering 
\caption{Photometry of the red sequence galaxies in Cl~0332-2742}
   \label{tab_photopara_cl}
  \begin{tabular}{@{}lccccccccc@{}}
  \hline
  \hline
  ID\textsuperscript{a} & R.A. & Decl. &  $i_{814}$\textsuperscript{b}  &   $i_{814}-J_{125}$\textsuperscript{b} &   $z_{850}-H_{160}$\textsuperscript{c} &   $U-V$\textsuperscript{d}  & $V-J$\textsuperscript{d}  &   $UVJ$\textsuperscript{e}   &   Spec-$z$\textsuperscript{f} \\
                    & (J2000) & (J2000) & (total)    &  (AB mag) & (AB mag) & (AB mag) & (AB mag) & passive &  member  \\
   \hline
  11827 &  53.044943 &  -27.774395 & $  21.168 \pm  0.007$ & $  2.520 \pm  0.017$ & $  2.418 \pm  0.017$ & $  2.046$ & $  1.334$ &  1 &  1 \\
  12177 &  53.052200 &  -27.774770 & $  21.682 \pm  0.005$ & $  2.422 \pm  0.011$ & $  2.173 \pm  0.014$ & $  1.924$ & $  1.145$ &  1 &  1 \\
  12347 &  53.043829 &  -27.774707 & $  22.188 \pm  0.007$ & $  2.351 \pm  0.015$ & $  2.059 \pm  0.015$ & $  1.764$ & $  1.143$ &  1 &  0 \\
  12412 &  53.049043 &  -27.774496 & $  23.478 \pm  0.018$ & $  1.960 \pm  0.032$ & $  1.689 \pm  0.031$ & $  1.432$ & $  1.493$ &  0 &  0 \\
  13096 &  53.046607 &  -27.772185 & $  23.063 \pm  0.013$ & $  2.315 \pm  0.030$ & $  2.045 \pm  0.025$ & $  1.811$ & $  1.095$ &  1 &  0 \\
  19839 &  53.030910 &  -27.738183 & $  22.595 \pm  0.009$ & $  2.203 \pm  0.026$ & $  1.980 \pm  0.020$ & $  1.620$ & $  1.077$ &  1 &  0 \\
\end{tabular}
\end{table}

\ifx true false
\begin{turnpage}
\begin{center}
\begin{table}

\centering 
\caption{Structural parameters and color gradients of the red sequence galaxies in XMMU~J2235-2557}
    \label{tab_strpara_xmmu}
}
\end{table}
\end{center}
\end{turnpage}

\fi

\begin{turnpage}
\begin{center}
\begin{table}
\centering 
\caption{Structural parameters and color gradients of the red sequence galaxies in XMMXCS~J2215-1738}
    \label{tab_strpara_xmmxcs}
} 
\end{table}

%

\end{center}
\end{turnpage}


\begin{turnpage}
\begin{center}
\begin{table}

\centering 
\caption{Structural parameters and color gradients of the red sequence galaxies in Cl~0332-2742}
    \label{tab_strpara_cl}
   \resizebox{1.325\textwidth}{!}{%
  \begin{tabular}{@{}lccccccccccccc@{}}
  \hline
  \hline
  ID\textsuperscript & $\log M_{*}\textsuperscript{a} $ & $\nabla_{z_{850}-H_{160}}$\textsuperscript{b} & $\nabla_{\log(M/L)}$\textsuperscript{b} &  $a_{e}$ & $n$ & $q$ & $R_{e\rm{-circ}}$ & $a_{e,mass}$ & $n_{mass}$ & $q_{mass}$ & $R_{e\rm{-circ}-mass}$ \\ 
    &  ($M_\odot$) &  &  & (kpc) &  &  & (kpc) & (kpc) &  & &(kpc) \\ 
  \hline
  11827 & $  11.49 \pm  0.07$ & $  -0.644 \pm  0.034$ & $  -0.192 \pm  0.047$ & $  7.97 \pm  0.77$ & $  2.39 \pm  0.25$ & $  0.60 \pm  0.02$ & $  6.17 \pm  0.61$ & $  5.12 \pm  0.67$ & $  1.99 \pm  0.30$ & $  0.60 \pm  0.04$ & $  3.98 \pm  0.53$ \\ 
  12177 & $  11.13 \pm  0.12$ & $  -0.422 \pm  0.049$ & $  -0.138 \pm  0.044$ & $  2.63 \pm  0.08$ & $  1.84 \pm  0.11$ & $  0.55 \pm  0.01$ & $  1.95 \pm  0.06$ & $  2.17 \pm  0.17$ & $  1.79 \pm  0.17$ & $  0.53 \pm  0.02$ & $  1.58 \pm  0.13$ \\ 
  12347 & $  10.89 \pm  0.12$ & $  -0.419 \pm  0.114$ & $  -0.108 \pm  0.113$ & $  1.52 \pm  0.04$ & $  3.41 \pm  0.18$ & $  0.62 \pm  0.01$ & $  1.20 \pm  0.03$ & $  1.24 \pm  0.07$ & $  2.92 \pm  0.25$ & $  0.56 \pm  0.02$ & $  0.93 \pm  0.06$ \\ 
  13096 & $  10.53 \pm  0.12$ & $  0.067 \pm  0.165$ & $  0.001 \pm  0.147$ & $  1.31 \pm  0.05$ & $  2.86 \pm  0.21$ & $  0.67 \pm  0.02$ & $  1.07 \pm  0.04$ & $  1.25 \pm  0.11$ & $  2.94 \pm  0.31$ & $  0.69 \pm  0.03$ & $  1.04 \pm  0.09$ \\ 
  19839 & $  10.67 \pm  0.13$ & $  -0.013 \pm  0.273$ & $  -0.012 \pm  0.223$ & $  1.31 \pm  0.04$ & $  5.16 \pm  0.32$ & $  0.48 \pm  0.01$ & $  0.91 \pm  0.03$ & $  1.15 \pm  0.07$ & $  3.62 \pm  0.33$ & $  0.40 \pm  0.02$ & $  0.72 \pm  0.05$ \\ 
  21613 & $  10.91 \pm  0.07$ & $  -0.776 \pm  0.133$ & $  -0.172 \pm  0.129$ & $  2.68 \pm  0.15$ & $  5.76 \pm  0.47$ & $  0.52 \pm  0.02$ & $  1.94 \pm  0.11$ & $  1.37 \pm  0.13$ & $  3.28 \pm  0.42$ & $  0.50 \pm  0.03$ & $  0.97 \pm  0.10$ \\ 
  21853 & $  11.13 \pm  0.13$ & $  -0.271 \pm  0.055$ & $  -0.097 \pm  0.061$ & $  1.61 \pm  0.04$ & $  2.99 \pm  0.16$ & $  0.95 \pm  0.02$ & $  1.56 \pm  0.04$ & $  1.31 \pm  0.08$ & $  2.10 \pm  0.18$ & $  0.92 \pm  0.04$ & $  1.26 \pm  0.08$ \\ 
  22281 & $  10.68 \pm  0.12$ & $  -0.029 \pm  0.249$ & $  -0.014 \pm  0.239$ & $  0.74 \pm  0.02$ & $  2.62 \pm  0.15$ & $  0.90 \pm  0.02$ & $  0.71 \pm  0.02$ & $  0.72 \pm  0.04$ & $  1.71 \pm  0.15$ & $  0.86 \pm  0.03$ & $  0.67 \pm  0.04$ \\ 
  22777 & $  10.71 \pm  0.12$ & $  -0.556 \pm  0.134$ & $  -0.138 \pm  0.128$ & $  1.30 \pm  0.04$ & $  1.83 \pm  0.12$ & $  0.64 \pm  0.01$ & $  1.04 \pm  0.03$ & $  1.12 \pm  0.09$ & $  1.62 \pm  0.16$ & $  0.70 \pm  0.03$ & $  0.94 \pm  0.08$ \\ 
  24147 & $  10.93 \pm  0.12$ & $  -0.411 \pm  0.210$ & $  -0.128 \pm  0.174$ & $  1.47 \pm  0.03$ & $  2.28 \pm  0.12$ & $  0.53 \pm  0.01$ & $  1.06 \pm  0.03$ & $  1.17 \pm  0.07$ & $  2.49 \pm  0.22$ & $  0.47 \pm  0.02$ & $  0.81 \pm  0.05$ \\ 
  24517 & $  11.22 \pm  0.10$ & $  -0.415 \pm  0.109$ & $  -0.101 \pm  0.112$ & $  1.84 \pm  0.04$ & $  5.73 \pm  0.30$ & $  0.77 \pm  0.01$ & $  1.61 \pm  0.04$ & $  1.25 \pm  0.07$ & $  5.60 \pm  0.48$ & $  0.76 \pm  0.03$ & $  1.09 \pm  0.07$ \\ 
  24882 & $  10.81 \pm  0.12$ & $  -0.160 \pm  0.290$ & $  -0.051 \pm  0.235$ & $  0.85 \pm  0.02$ & $  3.47 \pm  0.18$ & $  0.56 \pm  0.01$ & $  0.63 \pm  0.01$ & $  0.81 \pm  0.04$ & $  3.86 \pm  0.32$ & $  0.56 \pm  0.02$ & $  0.61 \pm  0.03$ \\ 
  25338 & $  11.01 \pm  0.10$ & $  -0.312 \pm  0.248$ & $  -0.108 \pm  0.199$ & $  1.36 \pm  0.03$ & $  3.34 \pm  0.17$ & $  0.38 \pm  0.01$ & $  0.83 \pm  0.02$ & $  1.09 \pm  0.06$ & $  3.05 \pm  0.25$ & $  0.40 \pm  0.01$ & $  0.69 \pm  0.04$ \\ 
  25972 & $  11.07 \pm  0.10$ & $  -0.499 \pm  0.069$ & $  -0.141 \pm  0.077$ & $  1.92 \pm  0.06$ & $  2.73 \pm  0.17$ & $  0.89 \pm  0.02$ & $  1.81 \pm  0.06$ & $  1.43 \pm  0.11$ & $  1.61 \pm  0.15$ & $  0.91 \pm  0.04$ & $  1.36 \pm  0.11$ \\ 
  25989 & $  10.71 \pm  0.13$ & $  -0.312 \pm  0.284$ & $  -0.090 \pm  0.239$ & $  0.93 \pm  0.02$ & $  3.88 \pm  0.22$ & $  0.63 \pm  0.01$ & $  0.74 \pm  0.02$ & $  0.82 \pm  0.04$ & $  2.24 \pm  0.19$ & $  0.57 \pm  0.02$ & $  0.62 \pm  0.03$ \\ 
  \hline
  \multicolumn{12}{l}{\textsuperscript{a} Total stellar masses are estimated using the $M_{*}/L$-color relation, the $z_{850}-H_{160}$ aperture colors and the total luminosity $L_{H_{160}}$ from the best-fit S\'ersic models.} \\  
  \multicolumn{12}{p{1.0\textwidth}}{\textsuperscript{b} color gradients $\nabla_{z_{850}-H_{160}}$ and $M_{*}/L$ gradients $\nabla_{\log(M/L)}$ are defined as $d(z_{850} - H_{160}) / d \log(a)$ and $d(\log(M/L)) / d \log(a)$ respectively.  The gradient is fitted in the radial range of PSF HWHM $< a < 3.5~a_{e}$.}  \\
  \end{tabular}}
\end{table}

%
\end{center}
\end{turnpage}



\end{document}